\documentclass[final,5p,times,twocolumn]{elsarticle}

\usepackage{amssymb}

\usepackage{amsmath}

\usepackage{lineno}
\usepackage{color}
\usepackage{multirow, caption, makecell, tabularx, booktabs}
\usepackage{fancyhdr}

\usepackage[colorlinks=true,linkcolor=red]{hyperref} 

\usepackage{comment}

\usepackage{orcidlink}

\usepackage{enumitem}
\usepackage{setspace}
\newcommand{\ltwo}[1]{\subsection{#1}}
\newcommand{\lthree}[1]{\subsubsection{#1}}

\journal{Nuclear Instruments and Methods in Physics Research Section A}

\begin{document}

\begin{frontmatter}



\title{The SHMS 11 GeV/c Spectrometer in Hall C at Jefferson Lab}


\author[CUA]{S.~Ali}
\author[NCAT]{A.~Ahmidouch}
\author[Regina]{G.R.~Ambrose}
\author[ANSL,JLAB]{A.~Asaturyan}
\author[MSS,ODU]{C.~Ayerbe Gayoso}
\author[JLAB]{J.~Benesch}
\author[CUA,JLAB]{V.~Berdnikov\orcidlink{0000-0003-1603-4320}}
\author[MSS]{H.~Bhatt\orcidlink{0000-0003-0087-5387}}
\author[MSS]{D.~Bhetuwal}
\author[Hampton,VT]{D.~Biswas}
\author[JLAB]{P.~Brindza}
\author[JAZ]{M.~Bukhari}
\author[JMU,JHAPL]{M.~Burton}
\author[JLAB]{R.~Carlini}
\author[CUA]{M.~Carmignotto}
\author[JLAB,Hampton]{M.~E.~Christy}
\author[UVA]{C.~Cotton}
\author[CUA]{J.~Crafts}
\author[UVA]{D.~Day\orcidlink{0000-0001-7126-8934}}
\author[NCAT]{S.~Danagoulian}
\author[UIUC]{A.~Dittmann}
\author[Hampton,LBNL]{D.~H.~Dongwi}
\author[TEM,ANL,NMSU]{B.~Duran}
\author[MSS]{D.~Dutta}
\author[JLAB]{R.~Ent}
\author[JLAB]{H.~Fenker}
\author[JLAB]{M.~Fowler}
\author[JLAB]{D.~Gaskell\orcidlink{0000-0001-5463-4867}}
\author[Regina]{A.~Hamdi\orcidlink{0000-0001-7099-9452}}
\author[Regina]{N.~Heinrich\orcidlink{0009-0005-8720-9329}}
\author[JLAB]{W.~Henry}
\author[CUA]{N.~Hlavin}
\author[CUA,JLAB]{T.~Horn\orcidlink{0000-0003-1925-9541}}
\author[Regina]{G.M.~Huber\orcidlink{0000-0002-5658-1065}}
\author[USC]{Y.~Ilieva}
\author[JMU,LLNL]{J.~Jarrell}
\author[TEM]{S.~Jia}
\author[JLAB]{M.~K.~Jones\corref{cor1}\orcidlink{0000-0002-7089-6311}}
\author[Regina]{M.~Junaid\orcidlink{0000-0002-5853-4326}}
\author[MSS]{M.~L.~Kabir}
\author[VUU]{N.~Kalantarians}
\author[MSS]{A.~Karki\orcidlink{0000-0001-7650-2646}}
\author[Regina,York]{S.J.D.~Kay\orcidlink{0000-0002-8855-3034}}
\author[JLAB]{C.~E.~Keppel}
\author[Regina]{V.~Kumar}
\author[JLAB]{S.~Lassiter}
\author[Regina,MSS,SBU]{W.B.~Li\orcidlink{0000-0002-8108-8045}}
\author[JLAB]{D.~Mack}
\author[JLAB]{S.~Malace}
\author[Hampton]{J.~McMahon}
\author[CUA]{A.~Mkrtchyan}
\author[ANSL]{H.~Mkrtchyan}
\author[Hampton,CNU]{P.~Monaghan}
\author[CUA]{C.~Morean}
\author[JLAB]{P.~Nadel-Turonski}
\author[JMU]{G.~Niculescu}
\author[JMU]{M.~I.~Niculescu}
\author[Hampton,MSS]{A.~Nadeeshani\orcidlink{0000-0002-5736-8274}}
\author[JLAB]{E.~Pooser}
\author[FIU]{A.~Ramos}
\author[FIU]{J.~Reinhold}
\author[JLAB]{B.~Sawatzky}
\author[JLAB]{H.~Szumila-Vance}
\author[ANSL]{V.~Tadevosyan}
\author[CUA]{R.L.~Trotta}
\author[Regina]{A.~Usman}
\author[FIU,CUA]{C.~Yero\orcidlink{0000-0003-2822-7373}}
\author[UVA]{M.~Yurov}
\author[ANSL]{S.~Zhamkochyan}
\author[JLAB]{S.~A.~Wood\orcidlink{0000-0002-1909-5287}}
\author[UVA]{J.~Zhang}

\address[CUA]{The Catholic University of America, Washington, DC 20064, USA}
\address[NCAT]{North Carolina A\&T State University, North Carolina 27411, USA}
\address[Regina]{University of Regina, Regina, Saskatchewan S4S~0A2, Canada}
\address[ANSL]{A.~I.~Alikhanyan National Science Laboratory, Yerevan 0036, Armenia}
\address[ODU]{Old Dominion University, Norfolk, Virginia, USA}
\address[MSS]{Mississippi State University, Mississippi State, Mississippi 39762, USA}
\address[JLAB]{Thomas Jefferson National Accelerator Facility, Newport News, Virginia 23606, USA}
\address[Hampton]{Hampton University, Hampton, Virginia 23668, USA}
\address[VT]{Virginia Tech, Blacksburg, VA 24061, USA}
\address[JAZ]{Jazan University, Jazan 45142, Saudi Arabia }
\address[JMU]{James Madison University, Harrisonburg, Virginia 22801, USA}
\address[JHAPL]{Johns Hopkins University Applied Physics Laboratory, Laurel, Maryland 20723, USA}
\address[UVA]{University of Virginia, Charlottesville, Virginia 22904, USA}
\address[UIUC]{University of Illinois, Urbana-Champaign, Illinois, USA}
\address[LBNL]{Lawrence Berkeley National Laboratory, Berkeley, California 94720, USA}
\address[LLNL]{Lawrence Livermore National Laboratory, Livermore, California 94551, USA}
\address[TEM]{Temple University, Philadelphia, PA, USA}
\address[ANL]{ Argonne National Laboratory, Lemont, IL, USA}
\address[NMSU]{New Mexico State University, Las Cruces, NM 88001, USA}
\address[USC]{University of South Carolina, Columbia, South Carolina 29208, USA}
\address[VUU]{Virginia Union University, Richmond, Virginia 23220, USA}
\address[York]{University of York, Heslington, York, Y010~5DD, UK}
\address[SBU]{Stony Brook University, Stony Brook, New York 11794, USA}
\address[CNU]{Christopher Newport University, Newport News, Virginia, 23606, USA}
\address[FIU]{Florida International University, University Park, Florida 33199, USA}
\cortext[cor1]{jones@jlab.org}

\begin{abstract}
The {\it Super High Momentum Spectrometer} (SHMS) has been built for Hall C at the Thomas Jefferson National Accelerator Facility (Jefferson Lab). With a momentum capability reaching 11\,GeV/$c$, the SHMS provides measurements of charged particles produced in electron-scattering experiments using the maximum available beam energy from the upgraded Jefferson Lab accelerator. The SHMS is an ion-optics magnetic spectrometer comprised of a series of new superconducting magnets which transport charged particles through an array of triggering, tracking, and particle-identification detectors that measure momentum, energy, angle and position in order to allow kinematic reconstruction of the events back to their origin at the scattering target. The detector system is protected from background radiation by a sophisticated shielding enclosure. The entire spectrometer is mounted on a rotating support structure which permits measurements to be taken with a large acceptance over laboratory scattering angles from $5.5^{\circ}$ to $40^{\circ}$, thus allowing a wide range of low cross-section experiments to be conducted. These experiments complement and extend the previous Hall C research program to higher energies.

\end{abstract}

\begin{keyword}
Magnetic spectrometer \sep
Electron scattering \sep
Tracking detectors \sep
Particle identification \sep
Electron calorimetry \sep
Radiation shielding.



\end{keyword}

\end{frontmatter}


\section{Introduction}

\subsection{Jefferson Lab Overview}
\begin{figure}[tbhp]
  \includegraphics[width=\columnwidth]{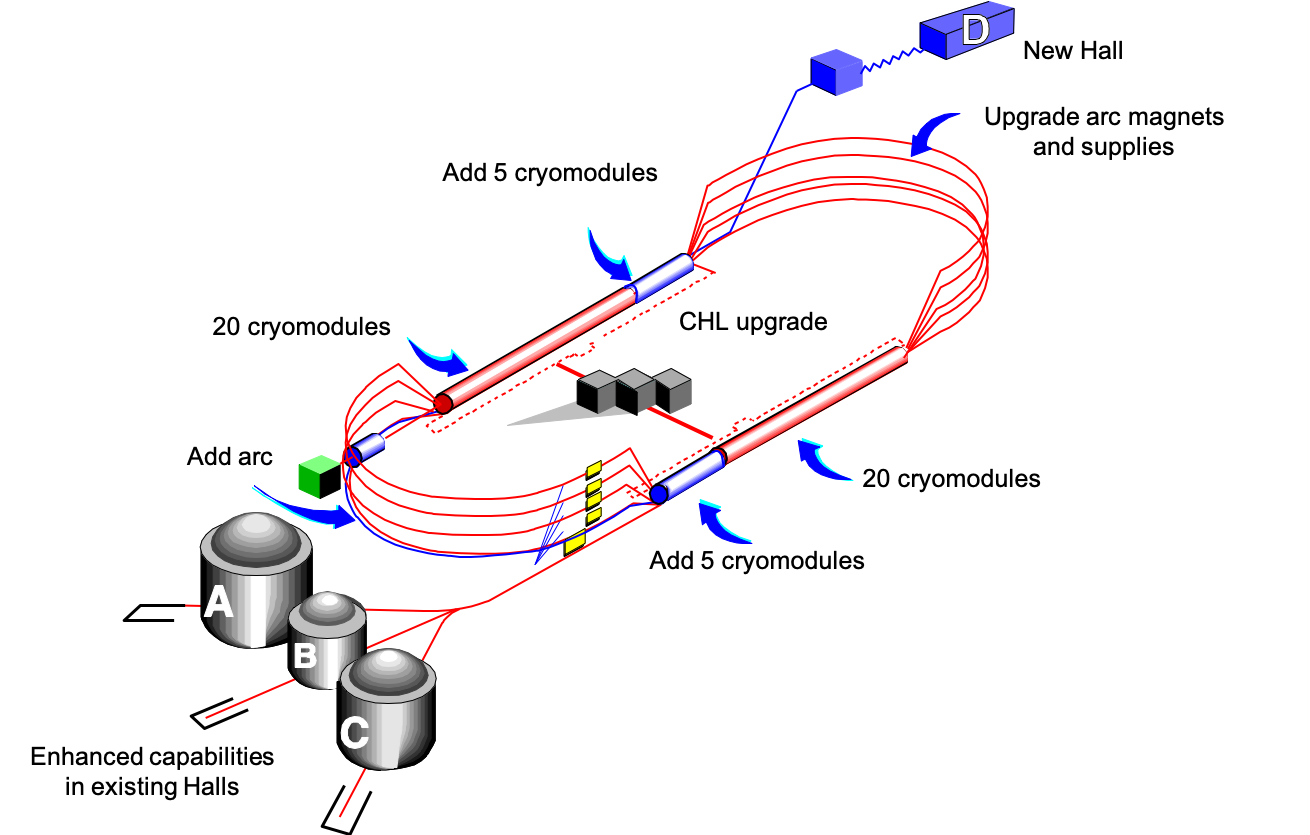}
  \caption{Schematic of hall and accelerator improvements as part of the Jefferson Lab 12~GeV Upgrade.
  \label{fig:jlab_12gev}}
\end{figure}

The Continuous Electron Beam Accelerator Facility at Thomas Jefferson National Accelerator Facility (Jefferson Lab) provides high energy electron beams for fundamental nuclear physics experiments. Originally planned for maximum electron beam energies of 4\,GeV, the accelerator operated at energies of up to 6\,GeV starting in 2000. An upgrade of the facility was completed in 2017, enabling beam delivery at a maximum energy of 12\,GeV to the new experimental Hall D, and 11\,GeV to the existing Halls, A, B, and C \cite{PhysRevAccelBeams.27.084802}. (Fig. \ref{fig:jlab_12gev}.)

The electron beam at Jefferson Lab operates at high duty cycle, with beam repetition rates of 249.5 or 499\,MHz delivered to the experimental halls. High beam polarization ($>80$\%) is also routinely available.

In the 6\,GeV era, Halls A, B, and C executed a large program of experiments focusing primarily on elucidating the quark-gluon structure of nucleons and nuclei. Experimental Hall B made use of a large acceptance spectrometer capable of detecting multi-body final states over a large region of kinematic phase space in one setting. Halls A and C made use of magnetic focusing spectrometers. In Hall A, the two High Resolution Spectrometers (HRS) emphasized excellent momentum resolution. In Hall C, the Short Orbit Spectrometer (SOS) facilitated the detection of short-lived final states (pions and kaons) at modest momentum while the High Momentum Spectrometer (HMS) was capable of detecting particles up the maximum beam energy at Jefferson Lab.

As part of the 12\,GeV upgrade at Jefferson Lab, a new experimental facility, Hall D, was built to search for gluonic excitations in the meson spectrum using a photon beam produced via coherent bremsstrahlung. The GlueX experiment in Hall D began commissioning in 2014 and has taken production-quality data since 2016.

The existing Halls A, B, and C were also upgraded as part of the 12\,GeV upgrade. The Hall A beamline and beam polarimeters were upgraded to accommodate operation at 11\,GeV. Hall A has made use of the existing HRS spectrometers in its early 12\,GeV era experiments (which began initial data-taking in 2014) and has also installed specialized, dedicated equipment for recent measurements. Experimental Hall B replaced its large acceptance CLAS spectrometer with the new CLAS-12 spectrometer. This new spectrometer retains the key features of large acceptance and robust particle identification over a large momentum range but with more emphasis on particle detection in the forward direction, required due to the higher beam energies. Finally, Hall C replaced its Short Orbit Spectrometer with the new Super-High Momentum Spectrometer (SHMS). 
The design of this new spectrometer was 
guided by experience from the 6\,GeV program, with the main goal for the new spectrometer to serve as an optimal partner for the HMS in coincidence experiments.



\subsection{Hall C Experimental Program at 6 GeV}

\begin{figure}[tbhp]
\begin{center}
  \includegraphics[width=0.9\columnwidth]{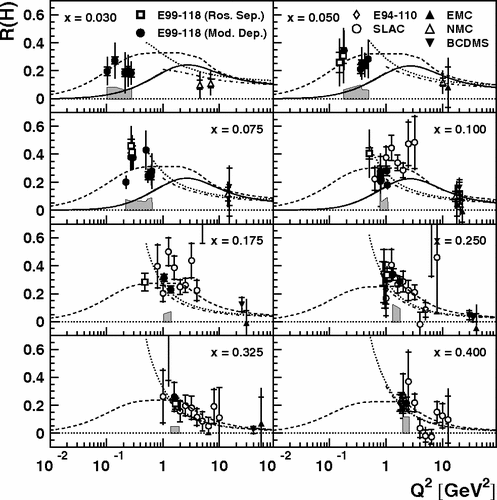}
  \caption{Measurement of $R=\frac{\sigma_L}{\sigma_T}$ at low $Q^2$. The extraction of $R$ requires precise L-T separations with excellent control of point-to-point systematic uncertainties. Figure from \cite{PhysRevLett.98.142301}. \label{fig:inclusive_lts}}
  \end{center}
\end{figure}

The HMS and SOS spectrometers in Hall C enabled the execution of a diverse program of experiments. 
\begin{figure}[tbhp]
\begin{center}
  \includegraphics[width=\columnwidth]{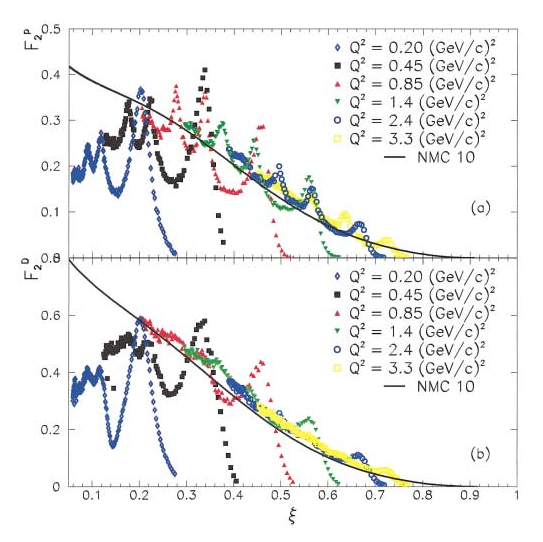}
  \caption{Inclusive $F_2$ structure functions measured in the resonance region compared to a DIS fit. When plotted vs. the Nachtmann variable $\xi$, the DIS fit agrees, on average, with the resonance region data, demonstrating quark-hadron duality~\cite{PhysRevLett.85.1186}.
  \label{fig:duality}}
  \end{center}
\end{figure}
The well-understood acceptance of both spectrometers, in tandem with excellent kinematic reproducibility allowed the extraction of precise cross sections. A particular strength was the control of point-to-point systematic uncertainties, which allowed high precision Rosenbluth, or L-T, separations. Examples of inclusive cross section measurements, using primarily the HMS, are shown in Figs.~\ref{fig:duality} and \ref{fig:inclusive_lts}.

\begin{figure}[tbhp]
\begin{center}
  \includegraphics[width=\columnwidth]{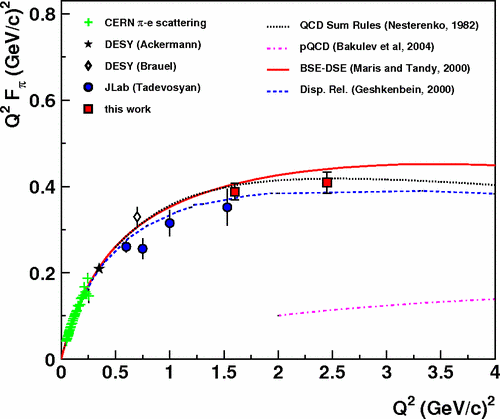}
  \caption{Measurements of the charged pion form factor in Hall C (6 GeV era). Extraction of the pion form factor requires a precise L-T separation, as well as detection of the charged pion at small forward angles. Figure from~\cite{Horn:2006tm}. \label{fig:pionff}}
  \end{center}
\end{figure}

\begin{figure}[tbhp]
\begin{center}
  \includegraphics[width=\columnwidth]{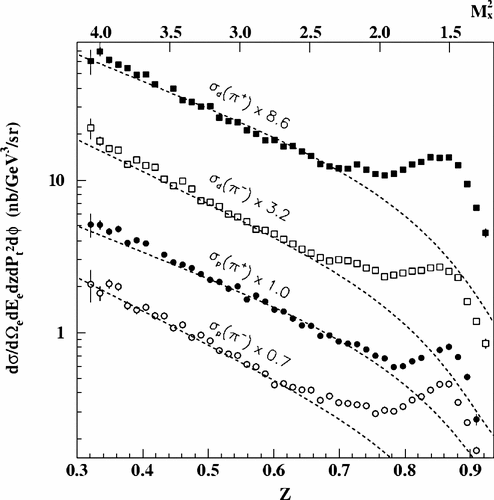}
  \caption{Cross sections for semi-inclusive $\pi^+$ and $\pi^-$ production from hydrogen and deuterium. The cross sections are compared to a parameterization that uses fragmentation functions fit to high energy $e^+e^-$ collisions. Figure from~\cite{Navasardyan:2006gv}. \label{fig:meson_duality}}
  \end{center}
\end{figure}

In addition, the relatively small minimum angle of 10.5 degrees accessible with the HMS allowed the execution of pion electroproduction experiments where, in many cases, the pion is emitted in the forward direction. This allowed the successful execution of a program of measurements of the pion form factor~\cite{Volmer:2000ek, Horn:2006tm}, which also incorporated precise L-T separations, as well measurements of charged pion production in Semi-inclusive Deep Inelastic Scattering (SIDIS)~\cite{Navasardyan:2006gv} (see Figs.~\ref{fig:pionff} and \ref{fig:meson_duality}).

\begin{figure}[tbhp]
\begin{center}
  \includegraphics[width=\columnwidth]{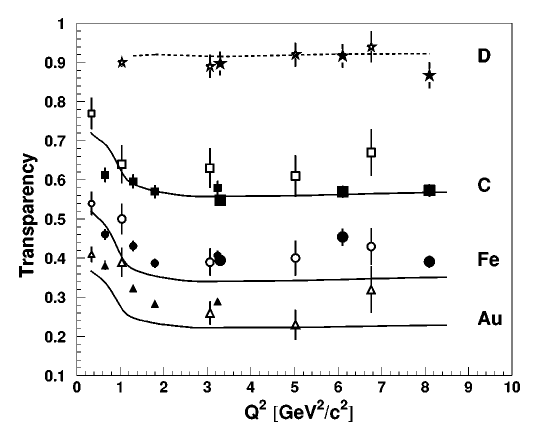}
  \caption{Measurement of transparency for $(e,e'p)$.  Solid points are from (6\,GeV era) Hall C measurements~\cite{Abbott:1997bc, Garrow:2001di}. At the largest $Q^2$, the HMS momentum is $>5$~GeV/c. Figure from~\cite{Garrow:2001di}. \label{fig:color_transparency}}
  \end{center}
\end{figure}

\begin{figure}[tbhp]
\begin{center}
  \includegraphics[width=\linewidth]{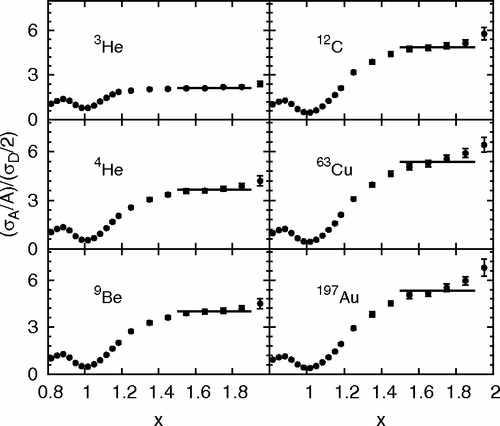}
  \caption{\label{fig:src}Measurements of cross section ratios for nuclear targets relative to deuterium at $x>1$. The size of the ratio is proportional to the relative contributions of 2-nucleon Short Range Correlations to the nuclear wave function. These measurements required high momentum in the HMS. Figure from~\cite{Fomin:2011ng}.}
  \end{center}
\end{figure}

The momentum reach of the HMS, up to the maximum beam energy of 6\,GeV,  enabled measurements of the $A(e,e'p)$ process to large $Q^2$~\cite{Abbott:1997bc, Garrow:2001di}. This allowed a search for signs of color transparency (Fig.~\ref{fig:color_transparency}) as well as measurements of inclusive electron scattering at $x>1$ to access contributions of ``superfast'' quarks to inelastic structure functions~\cite{Fomin:2010ei} and measure the relative contributions of Short Range Correlations (SRCs) in the nuclear wave function~\cite{Fomin:2011ng} (Fig.~\ref{fig:src}).

The experiments noted above are just a sample of the $\sim$30 ``standard equipment'' experiments that were executed in the 6\,GeV era in Hall C. Other experiments include measurements of exclusive kaon production, resonance ($\Delta$, S$_{11}$) production, color transparency via pion electroproduction, and numerous other inclusive electron scattering measurements using hydrogen and deuterium, as well as heavier nuclear targets. In some cases, the HMS was paired with dedicated equipment for special measurements. Examples of this include measurement of the ratio of  proton elastic form factors ($G_E/G_M$) to large $Q^2$, as well as measurements using a dynamically polarized NH$_3$ target.

\subsection{Hall C 12 GeV Program}

The new, Super-High Momentum Spectrometer was designed to build on the experimental capabilities exploited during the Hall C program at lower energies. Notably, this includes:
\begin{enumerate}
\item{Excellent kinematic control and reproducibility.}
\item{Thorough understanding of spectrometer acceptance.}
\item{Small angle capability (down to 5.5 degrees) for detection of forward mesons.}
\item{Central momentum up to (nearly) the maximum beam energy accessible in Hall C.}
\item{In-plane and out-of-plane acceptance well matched to the existing HMS to facilitate experiments detecting two particles in coincidence.}
\end{enumerate}

Several ``commissioning'' experiments were chosen for the first year of 12\,GeV running in Hall C to exercise the above requirements as much as possible. These experiments ran in 2018 and will be discussed briefly below.

\begin{figure}[htbp]
\begin{center}
  \includegraphics[width=0.8\columnwidth]{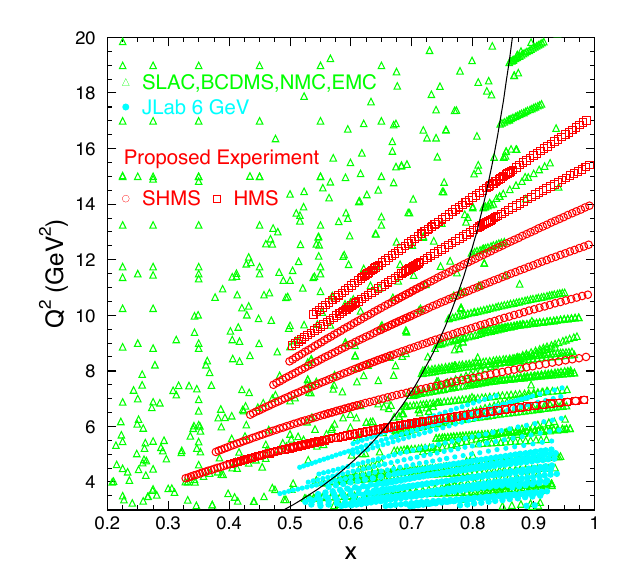}
  \caption{Kinematic coverage of $F_2$ measurements from experiment E12-10-002~\cite{E12-10-002}, which measured inclusive electron scattering cross sections as part of Hall C's 12 GeV commissioning experiments. 
  \label{fig:f2_coverage}}
  \end{center}
\end{figure}

The first commissioning experiment was a measurement of inclusive electron scattering cross sections from hydrogen and deuterium~\cite{E12-10-002} (see Fig.~\ref{fig:f2_coverage}). Such a cross section experiment is an excellent testing ground for understanding of the spectrometer acceptance, while not pushing the SHMS performance in other areas. Some settings for this experiment were chosen to allow simultaneous measurement with the well-understood HMS to provide a cross check. In addition, some time was devoted to the measurement of inclusive cross section ratios for nuclear targets relative to deuterium~\cite{E12-10-008}. These ratios are well-measured for certain nuclei and serve as another straightforward verification of the spectrometer acceptance due to the need to compare yields from extended (10~cm long) targets to shorter, solid targets (mm scale).  These measurements resulted in the first extraction of the EMC Effect in $^{10}$B and $^{11}$B~\cite{HallC:2022utd}.

\begin{figure}[tbhp]
\centering
  \includegraphics[width=0.8\columnwidth]{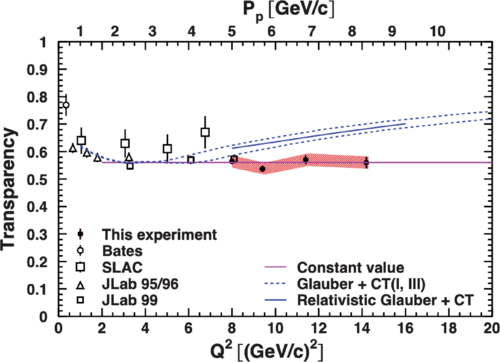}
  \caption{Results from experiment E12-06-107, a measurement of color transparency to large $Q^2$~\cite{HallC:2020ijh}.   This was the first coincidence measurement in the 12\,GeV era in Hall C. \label{fig:ct_results}}
\end{figure}

\begin{figure}[tbhp]
    \centering  
  \includegraphics[width=0.9\columnwidth]{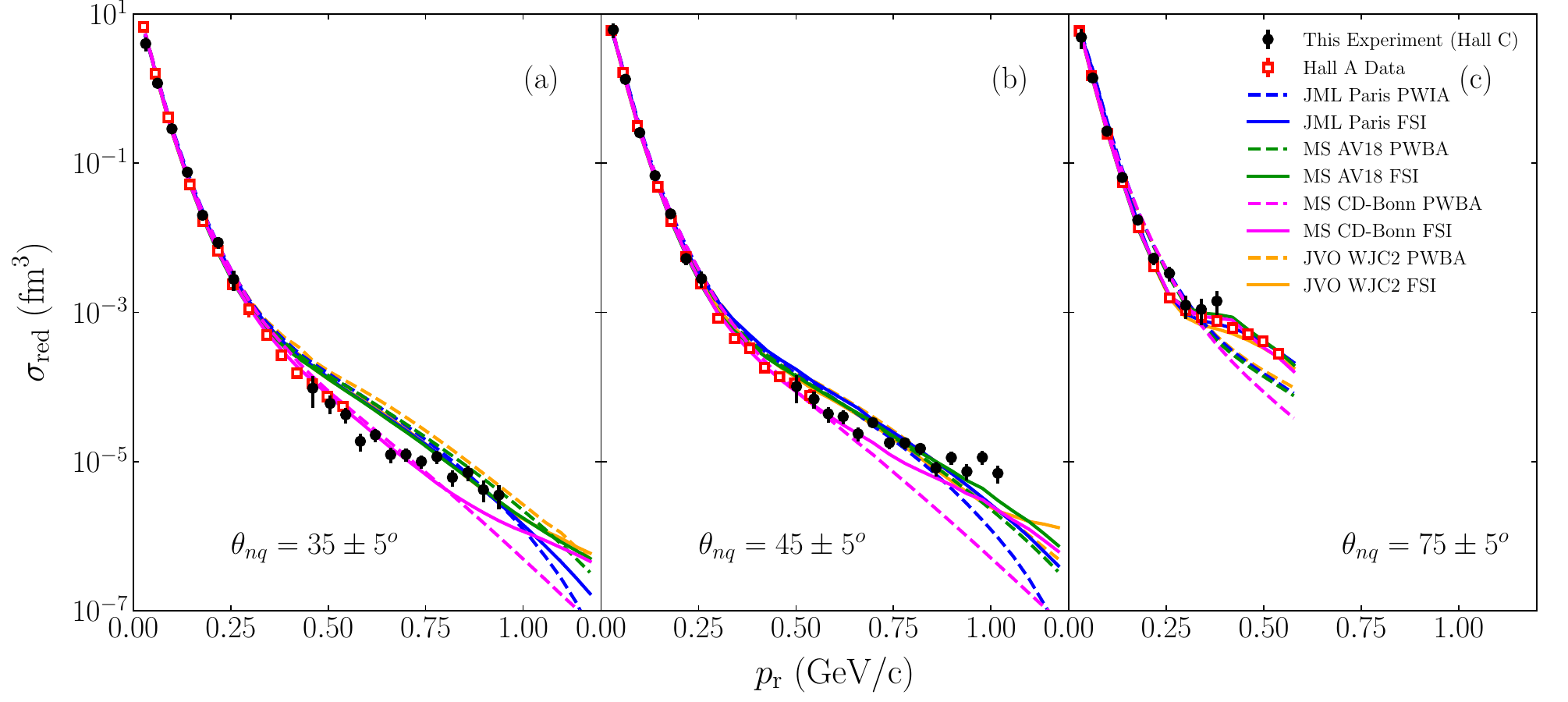}
  \caption{Results from experiment E12-10-003, a measurement of deuteron electro-disintegration at large missing momentum~\cite{jr:CYero_PRL2020}. This too was a 12 GeV era commissioning experiment in Hall C. \label{fig:deuteron_results}}
\end{figure}

An extension of the 6~GeV color transparency experiments to larger $Q^2$~\cite{E12-06-107} served as an excellent first experiment with which to exercise the SHMS in coincidence mode. In this $A(e,e'p)$ experiment, there are few random coincidences, so isolating the coincidence reaction is straightforward. This experiment, as well as a measurement of deuteron electro-disintegration~\cite{E12-10-003}, also tested the high momentum capabilities of the SHMS, exceeding $8.5$\,GeV/$c$ in these experiments. Although the maximum central momentum of the SHMS is almost 11 GeV, the momentum of $8.5$\,GeV/$c$ was already sufficient to learn about the performance of the superconducting magnets and spectrometer optics when pushed to a significant fraction of the spectrometer's ultimate capabilities. In addition, the body of elastic $H(e,e'p)$ data acquired from both these initial coincidence experiments provided constraints on the experiment kinematics, testing the possible variation of, \textit{e.g.} the spectrometer pointing or central momentum for various settings. Results from the color transparency and deuteron electro-disintegration experiments are shown in Figs.~\ref{fig:ct_results} and \ref{fig:deuteron_results}.

A set of meson electroproduction experiments followed the initial commissioning experiments and further exercised the SHMS capabilities. Two of the experiments
measured charged pion electroproduction in semi-inclusive deep inelastic scattering, SIDIS~\cite{E12-09-017, E12-09-002}. The SHMS was used at central angles smaller than 7$^\circ$ for the SIDIS running,
a fact which contributed to relatively high singles rates in the SHMS. Both experiments aimed to make precise measurements of $\pi^+/\pi^-$ ratios, so control of rate dependent systematic effects was a key challenge. The third experiment~\cite{E12-09-011} measured exclusive cross sections for $K^+$ production above the resonance region, extracting the longitudinal and transverse cross sections via a Rosenbluth separation. In this case, the experimental uncertainties were expected to be dominated by statistics rather than systematics, so this served as an excellent candidate for a first L-T separation.
In common with the charged pion SIDIS experiments, the kaon experiment required use of the SHMS at small angles and so faced 
high singles rates.

The ``commissioning'' and ``year-1'' experiments described above give a sense of the SHMS capabilities important for the overall physics program. Since then, a variety of experiments have been completed in Hall C. These include measurements of $J/\Psi$ photoproduction~\cite{Duran:2022xag}, virtual Compton scattering~\cite{Li:2022sqg}, exclusive charged pion electroproduction to extract the pion form factor and for cross section scaling tests~\cite{E12-19-006}, inclusive electron scattering from polarized $^3$He to extract $A_1^n$ and $d_2^n$~\cite{E12-06-110, E12-06-121}, and exclusive and inclusive scattering from nuclei to make measurements of short range correlations and the EMC Effect~\cite{E12-17-005, E12-06-105,E12-10-008}. In the future, additional L-T separations in inclusive scattering (to measure $R=\frac{\sigma_L}{\sigma_T}$ from hydrogen, deuterium, and several nuclei) and semi-inclusive reactions (to make the first precise measurement of $R$ for the SIDIS reaction) are also planned.  While not all future Hall C experiments will make use of the SHMS, it is a key component of its 12\,GeV experimental program.

\subsection{Contents of the Following Sections}

Specifications for the SHMS are given in Sec.\,\ref{sec:specs}. 
SHMS magnetic optics and shielding are discussed in Sec.\,\ref{sec:optics} and Sec.\,\ref{sec:shielding}, respectively.  
Detector system details are presented in Sec.\,\ref{sec:scint}-- \ref{sec:shower}. Event-triggering schemes and the data-acquisition system appear in Sec.\,~\ref{sec:trigdaq}, while software is briefly overviewed in Sec.\,~\ref{sec:software}. Some examples of the overall performance of these SHMS subsystems working in concert are shown in Sec.\,\ref{sec:performance}, followed by a short conclusion in Sec.\,\ref{sec:conclude}. 


\section{Specifications for the upgraded Hall-C Spectrometer complex}
\label{sec:specs}

The physics outlined in the previous section can be accessed only if the Hall~C spectrometer system is capable of providing the necessary measurements with precision, rate, and trigger capabilities consistent with those physics goals. Originally, Hall C offered the 7.4\,GeV/c High Momentum Spectrometer (HMS) and its lower-momentum (1.8\,GeV/c) partner, the Short-Orbit Spectrometer (SOS). These two devices were utilized independently by some experiments and in coincidence by others. The performance specifications for the SHMS were drafted such that the SHMS-HMS pair would provide similar complementary functions in the higher-momentum regime. That is, the SHMS was developed as a general-purpose spectrometer with properties similar to the existing HMS, but with a higher maximum momentum capability (11\,GeV/c). The 11\,GeV/c limit of the SHMS was selected because the accelerator constrained maximum beam energy to any of the first generation endstations (A, B, C) is 11\,GeV, and hadrons at small angles in $(e,e'h)$ experiments may approach the beam momentum. Table~\ref{tab_detsys_SHMS_spec} summarizes the demonstrated performance of the HMS and the design specifications for the SHMS.

With the higher beam energies in use at Jefferson Lab after the 12\,GeV upgrade, scattered electrons and secondary particles are boosted to more forward directions. Thus the SHMS acceptance is made to extend down to a 5.5$^\circ$ scattering angle, and needs to cover angles no higher than 40$^\circ$. Nevertheless, high energies generally lead to smaller cross sections. Therefore precision experiments can be performed only if a spectrometer provides large overall acceptance, high rate capability, and precise momentum measurement. As shown in Table~\ref{tab_detsys_SHMS_spec}, the SHMS design includes a momentum bite even larger than the HMS, and achieves an angular acceptance within a factor of two of its lower-momentum partner. The combination of dispersive optics and precision tracking provides excellent momentum resolution. Triggering, data-acquisition, particle identification, and rate handling capability are the same or better than those of the HMS. This performance is achieved not only through the use of faster, modern electronics, but also by innovative radiation shielding that reduces the background seen by the detectors.

\begin{table*}[bt]
\begin{center}
\begin{tabular}{|p{2.2in}|p{1.2in}|p{1.1in}|}
\hline
{\textbf{\textit{Parameter}}} & \makecell[c]{\textbf{\textit{HMS}}}        & \makecell[c]{\textbf{\textit{SHMS}}}         \\
{\textbf{\textit{  }}}        & \makecell[c]{\textbf{\textit{Performance}}}& \makecell[c]{\textbf{\textit{Specification}}}\\
\hline
{\raggedright Range of Central Momentum} & {\raggedright 0.4 to 7.4\,GeV/c}& {\raggedright 2 to 11\,GeV/c}\\
\hline
{\raggedright Momentum Acceptance } & {\raggedright \ensuremath{\pm}$10\%$} & {\raggedright -10\% to +22\%}\\
\hline
{\raggedright Momentum Resolution} & {\raggedright 0.1\% -- 0.15\%} & {\raggedright 0.03\% -- 0.08\%}\\
\hline
{\raggedright Scattering Angle Range} & {\raggedright 10.5$^\circ$ to 90$^\circ$} &{\raggedright 5.5$^\circ$ to 40$^\circ$}\\
\hline
{\raggedright Target Length Accepted at 90$^\circ$(HMS)/40$^{\circ}$ (SHMS)} & {\raggedright 10\,cm} & {\raggedright 25\,cm}\\
\hline
{\raggedright Horizontal Angle Acceptance }  & {\raggedright \ensuremath{\pm}32 mrad} & {\raggedright \ensuremath{\pm}18 mrad}\\
\hline
{\raggedright Vertical Angle Acceptance }  & {\raggedright \ensuremath{\pm}85\,mrad} & {\raggedright \ensuremath{\pm}45\,mrad}\\
\hline
{\raggedright Solid Angle Acceptance }            & {\raggedright 8.1\,msr}  & {\raggedright 4\,msr}\\
\hline
{\raggedright Horizontal Angle Resolution}  & {\raggedright 0.8\,mrad} & {\raggedright 0.5 -- 1.2\,mrad}\\
\hline
{\raggedright Vertical Angle Resolution}  & {\raggedright 1.0\,mrad} & {\raggedright 0.3 -- 1.1\,mrad}\\
\hline
  {\raggedright Target resolution ($y_\mathit{tar}$)}  & {\raggedright 0.3\,cm} & {\raggedright 0.1 - 0.3\,cm}\\
\hline
{\raggedright Maximum Event Rate}  & {\raggedright 4--5\,kHz} & {\raggedright 4--5\,kHz}\\
\hline
{\raggedright Max. Flux within Acceptance }   & {\raggedright $\sim 5$\,MHz} & {\raggedright $\sim5$\,MHz}\\
\hline
{\raggedright e/h Discrimination }  & {\raggedright $>$1000:1 at\\98\% efficiency} & {\raggedright  $>$1000:1 at\\98\% efficiency}\\
\hline
{\raggedright $\pi$/K Discrimination}  & {\raggedright 100:1 at\\95\% efficiency} & {\raggedright 100:1 at\\95\% efficiency}\\
\hline
\end{tabular}
\caption{Demonstrated performance of the HMS and design specifications for the SHMS. Resolutions are quoted at 1 sigma. 
    \label{tab_detsys_SHMS_spec}}
\end{center}
\end{table*}

\section{Design and Development of the SHMS Optics and Infrastructure}
\label{sec:sysdev}

\begin{figure}[hbtp]
\begin{centering}
\includegraphics[width=0.7\linewidth]{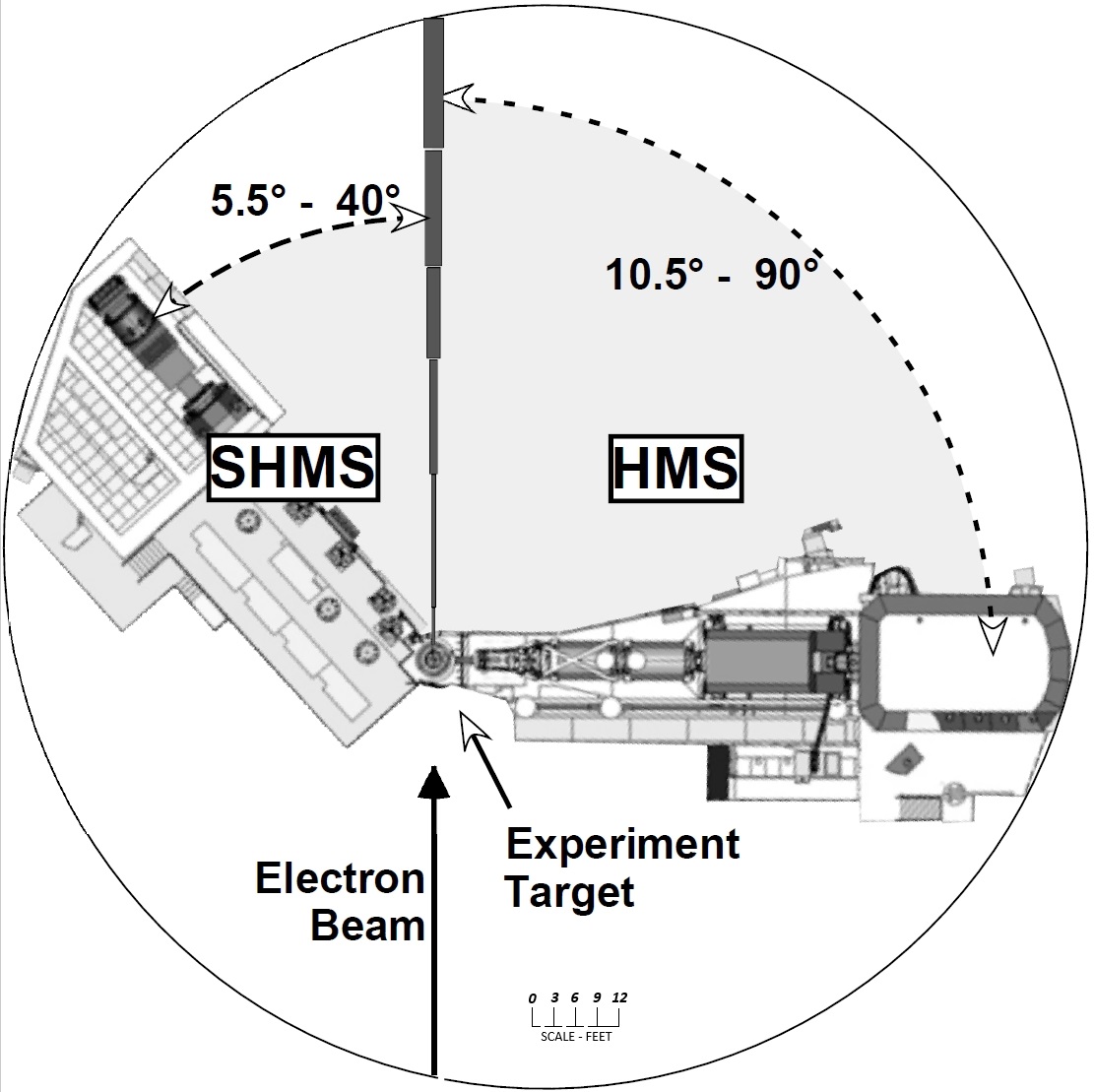}
\caption{Simplified plan view of Hall C showing the footprints of the SHMS and HMS. The SHMS occupies the smaller side of Hall~C, where the smaller, low-momentum Short-Orbit Spectrometer (SOS) was previously located. }
\label{fig:hallcPlanView}
\end{centering}
\end{figure}

The entire spectrometer is carried on a steel support structure which can rotate through an arc on the left side of the beam-line in Hall~C. Like the HMS carriage, it is secured to a central pivot so that it rotates around a vertical axis that intersects the electron beam-line at the experimental target. This is shown in Fig.~\ref{fig:hallcPlanView}.

Acceptance at the smallest scattering angles is enabled by a horizontal-bending dipole as the first element in the magnetic optical system. This small deflection moves the subsequent pieces of the SHMS farther from the beamline, relaxing the size constraints on the other magnetic elements (described in Section~\ref{sec:optics}) and shielding (Section~\ref{sec:shielding}). The shielded enclosure is itself a technically-optimized combination of concrete, lead, boron, and plastic. It surrounds the detectors and the electronics of the control and data-acquisition systems.

\subsection{Magnetic Optics}
\label{sec:optics}

The SHMS consists of five magnets used to determine the momentum, angles and position of particles scattered from the target using their angle and position measurements in the SHMS drift chambers. The first magnet is a dipole which bends the incident particles in the horizontal plane. A quadrupole triplet provides a point-to-point focus. To optimize acceptance in the vertical scattering plane, the first quadrupole focuses in the vertical while the second and third quadrupoles defocus and focus in the vertical, respectively. A vertical-bending dipole magnet follows the last quadrupole and disperses particles with different momenta across the focal plane. In point-to-point optics, all particles with the same momentum will be displaced by the same vertical distance at the focal plane.

\subsubsection{The Magnets and Vacuum Channel}

A specially-designed, horizontal-bend dipole (HB) precedes the first quadrupole. Its purpose is to provide an initial 3$^\circ$ separation between scattered particles and the electron beam so that particles scattered at small angles can be accepted.
\begin{figure}[hbtp]
\begin{centering}
\includegraphics[width=0.9\linewidth]{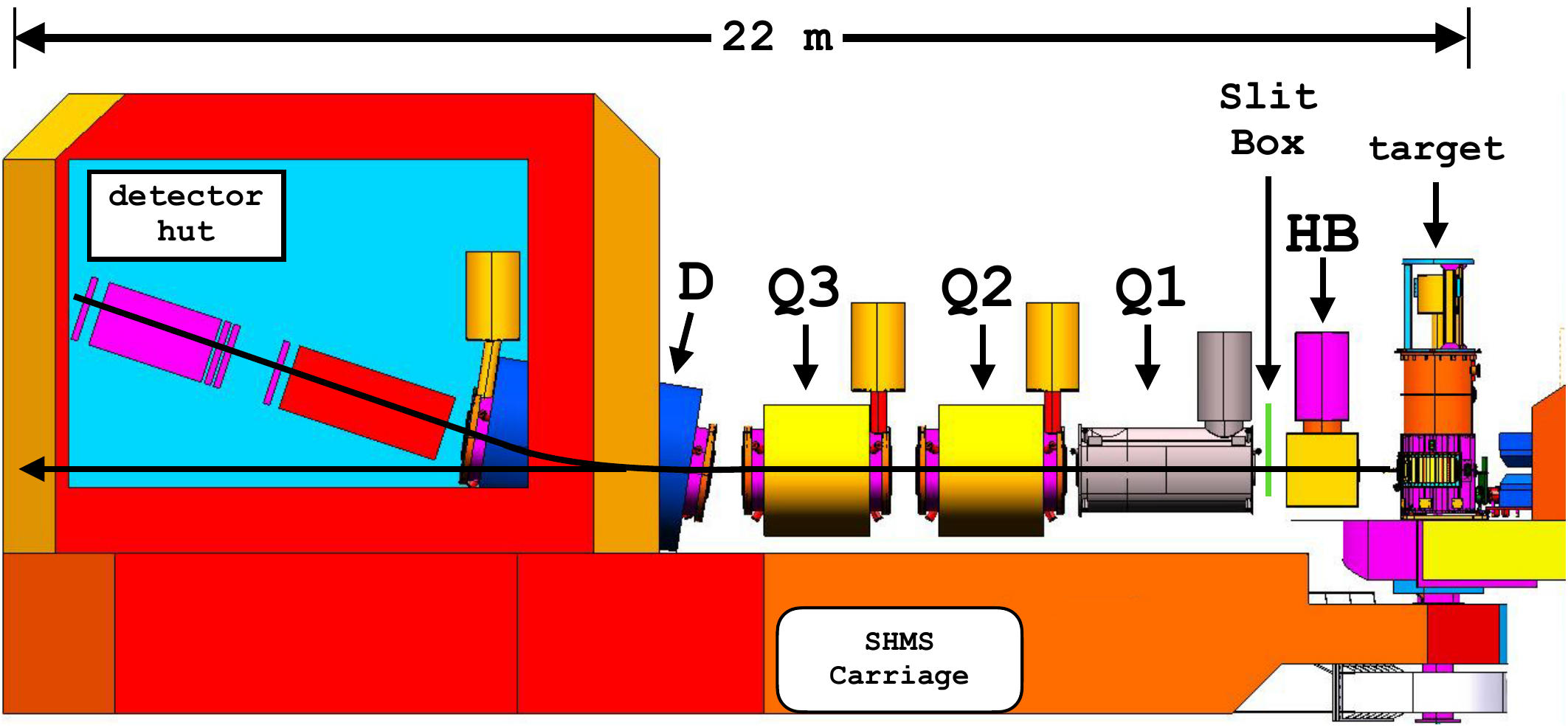}
\caption{A view of the SHMS carriage with its five magnets: a horizonal-bend (HB) magnet, a quadrupole triplet (Q1, Q2 and Q3) and the vertical-bending dipole (D). The vacuum connections between the magnets are not shown. The slit box contains the collimator and sieve slits (see Fig.~\ref{fig:SHMS_Collimators}).  }
\label{fig:sideviewshms}
\end{centering}
\end{figure}

As shown in Fig.~\ref{fig:hallcPlanView}, in order to fit within the space available in Hall~C, the SHMS must be even shorter than its lower-momentum partner, the HMS. A schematic of the SHMS carriage with the five magnets is shown in Fig.~\ref{fig:sideviewshms}. All of the SHMS magnets are superconducting so that they can provide the necessary large bending and focusing effects in the relatively short available distance. Given the 
requirement to access small scattering angles, the HB and the first two quadrupoles (Q1 and Q2) have special provisions to provide clearance for the electron beam and its vacuum pipe. HB is a ``C''-magnet so that all of the flux-return iron is on the side away from the beamline. As initially constructed, the HB leaked significant field into the beamline such that the beam would have been deflected outside of the beam dump. Simulations were done to determine the optimal shielding design to reduce the field in the beamline region for all combinations of SHMS angle and momentum and these mitigations were implemented \cite{Moore_2014}. The front of the HB cryostat, between the beamline and the magnet bore, is made very narrow. Both Q1 and Q2 have notches in their cryostats and iron yokes so that they, too, can clear the beamline when the spectrometer is configured at small scattering angles. Yoke steel for Q1 is inside the cryostat. The final quadrupole (Q3) and the dipole (D$_\mathrm{SHMS}$) have external warm yokes. Parameters of the SHMS magnets are provided in Table~\ref{tab:SHMS_Magnets}.

To minimize multiple scattering as particles pass through the SHMS, the bores of all of the magnets are evacuated. The vacuum space begins at a window on the front of HB. The entrance window into the HB is approximately 15\,cm square and is made of 0.01" thick aluminum. A vacuum connection is made between the exit of HB and Q1 entrance which is followed by the 40\,cm diameter vacuum bore in Q1. The exit of Q1 is connected to the entrance of Q2 by a vacuum pipe. The vacuum vessel bore through Q2, Q3, and D$_\mathrm{SHMS}$ is 60\,cm in diameter. The location of the end of the vacuum after the exit of D$_\mathrm{SHMS}$ depends on the needs of the experiment. If the experiment needs the Noble Gas Cherenkov (NGC) detector (described in Sec.~\ref{sec:ngcerenk}), then a window is placed at the exit of D$_\mathrm{SHMS}$ with the NGC detector placed between the exit window and the drift chambers. Otherwise, a Vacuum Extension Tank (VET) is attached to the exit of the D$_\mathrm{SHMS}$ that puts the exit window at 30\,cm from the first drift chamber in the detector stack. In both cases, the dipole exit window is made of 0.020" thick aluminum. The arrangement of the detectors and the distances between the detectors does not change when the NGC is replaced by the VET. All detectors have a fixed location in the SHMS hut on a carriage. Only the aerogel detector (see Fig.~\ref{fig:aerogelPhotoInStack}) is on rails so that it can be easily removed without changing the location of the other detectors.

\begin{table*}[bt]
\begin{center}
\begin{tabular}{|p{1.3in}|p{0.7in}|p{0.5in}|p{0.5in}|p{0.5in}|p{0.5in}|}
\hline
{\textbf{\textit{Parameter}}} & \makecell[c]{\textbf{\textit{HB}}}        
                                     & \makecell[c]{\textbf{\textit{Q1}}}         & \makecell[c]{\textbf{\textit{Q2}}}         
                                     & \makecell[c]{\textbf{\textit{Q3}}}         & \makecell[c]{\textbf{\textit{D$_{SHMS}$}}}         \\
\hline
{\raggedright Max Field or Gradient } & {\raggedright 2.6\,T  }     & {\raggedright 7.9\,T/m} & {\raggedright 11.8\,T/m} & {\raggedright 7.9\,T/m} & {\raggedright 3.9\,T  } \\
{\raggedright Effective Field Length} & {\raggedright 0.80\,m }     & {\raggedright 1.9\,m  } & {\raggedright 1.6\,m   } & {\raggedright 1.6\,m  } & {\raggedright 2.9\,m  } \\
{\raggedright Current at 11 GeV/c   } & {\raggedright 3923\,A }     & {\raggedright 2322\,A } & {\raggedright 3880\,A  } & {\raggedright 2553\,A } & {\raggedright 3510\,A } \\
{\raggedright Aperture              } & {\raggedright 14.5x18\,cm$^2$ } & {\raggedright 40\,cm  } & {\raggedright 60\,cm   } & {\raggedright 60\,cm  } & {\raggedright 60\,cm  } \\
\hline
\end{tabular}
\caption{Parameters of the SHMS Magnets 
    \label{tab:SHMS_Magnets}}
\end{center}
\end{table*}

\subsubsection{Charged Particle Transport Models}
A magnetic transport code, SNAKE~\cite{snake_code}, was used to model the transport of charged particles in the SHMS. The SNAKE model of the SHMS incorporated the mechanical sizes of the magnets, while the
magnetic fields were generated by the static field analysis code TOSCA~\cite{TOSCA} and validated with field measurements. The relative strengths of the field integrals of the
magnets were selected to maximize the acceptance while simultaneously providing the desired momentum and scattering angle resolutions. For a charged particle with relative momentum, $\delta = \frac{p-p_c}{p_c}$, where $p$ is the momentum of the particle and $p_c$ is the central momentum of the spectrometer, the transport from the target to the focal plane located
midway between the two drift chambers can be expressed in terms of a matrix representation of the solutions of the equation of motion of charged particles in magnetic fields~\cite{Penner61}.  The first-order transport matrix for the SHMS is given by:

\begin{equation}
\begin{pmatrix}
x_\mathit{fp} \\
x'_\mathit{fp} \\
y_\mathit{fp} \\
y'_\mathit{fp} \\
\end{pmatrix} =
\begin{pmatrix}
-1.5  & 0.0   & 0.0     & 0.0   & 1.65 \\
-0.5  &-0.7   & 0.0     & 0.0   & 3.2 \\
0.0   & 0.0   & -1.9    & -0.2  & -0.1 \\
0.0   & 0.0   & -3.0    & -0.8  & 0.1 
\end{pmatrix}
\begin{pmatrix}
x_\mathit{tar} \\
x'_\mathit{tar} \\
y_\mathit{tar} \\
y'_\mathit{tar} \\
\delta
\end{pmatrix}
\end{equation}

where $x_{tar}$ and $y_{tar}$ are the vertical and horizontal positions while $x^{'}_{tar} = \frac{\Delta x_{tar}}{\Delta z_{tar}}$ and $y^{'}_{tar} = \frac{\Delta y_{tar}}{\Delta z_{tar}}$
are the angles in the $z_{tar}=0$ plane, all measured relative to the central ray of the spectrometer. $x_{f}$, $y_{fp}$ and $x^{'}_{fp}$, $y^{'}_{fp}$ are the positions and angles of the
particle when transported to the focal plane. The positions, angles, and $\delta$ are in centimeters, milliradians, and \%, respectively.
 
\begin{figure*}[htbp]
	\begin{centering}
		\includegraphics[width=1\linewidth]{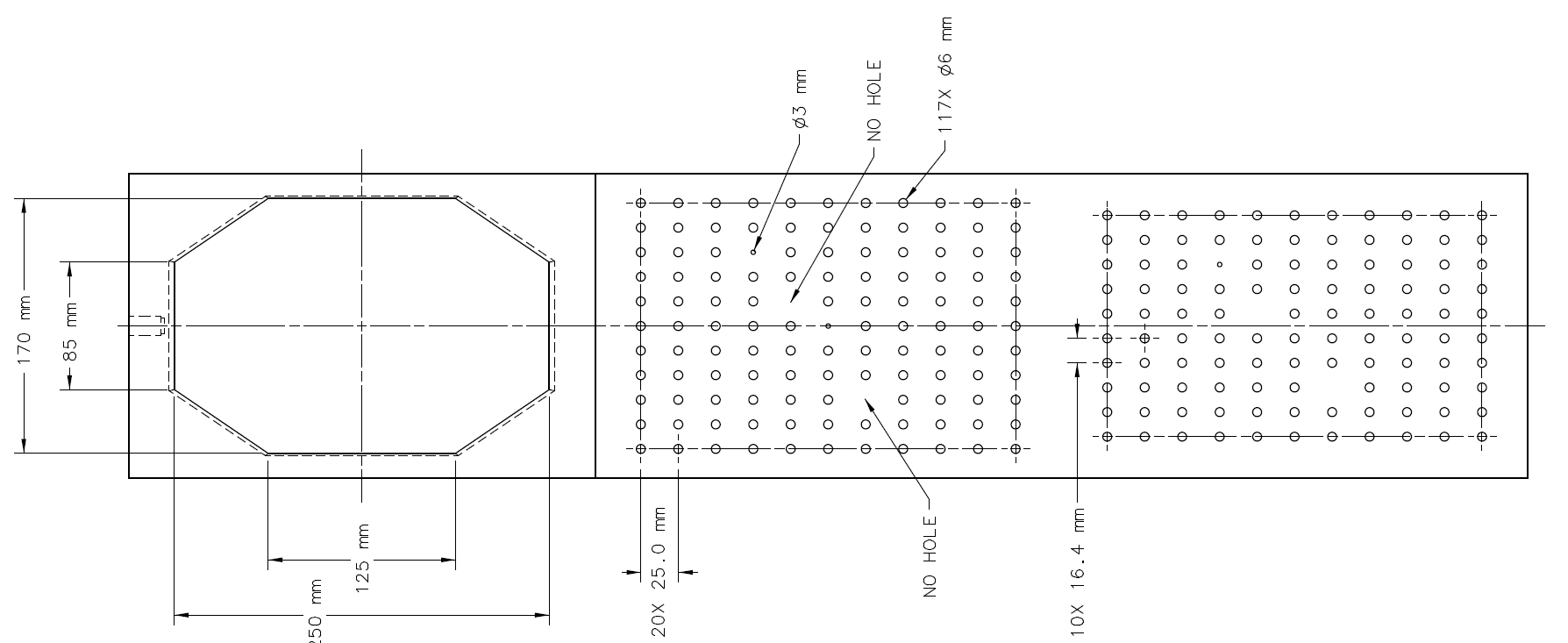}
		\caption{Schematic diagram of the SHMS collimator.\label{fig:SHMS_Collimators}}
	\end{centering}
\end{figure*}
The acceptance of the spectrometer is mainly determined by the collimator placed within a remotely operated collimator box that is installed between the HB magnet and the first quadrupole magnet.
The collimator ladder assembly within this box may be positioned at one of three settings. The top position (accessed when the assembly is at its lowest position) is a
stretched octagon with an opening height 9.843~in and width 6.693~in on the upstream side. It is 2.5~in thick. The lower two positions contain sieve slits with holes
in a rectangular pattern separated by 0.6457~in
horizontally and 0.9843~in vertically. The sieve pattern at
the middle ladder position has 11 columns of holes with
the sixth column centered horizontally. The holes on the bottom sieve are in ten columns and are offset by one half column gap from those in the middle sieve.
The sieve collimators are 1.25~in thick. The geometry is illustrated in Fig.~\ref{fig:SHMS_Collimators}. Both sieves and the octagonal collimator
are made of Mi-TechTM Tungsten HD-17 (Density 17~g/cm$^3$, and composition 90\% W, 6\% Ni, 4\% Cu)~\cite{Mitech}.

\begin{figure}[htbp]
	\begin{centering}
		\includegraphics[width=0.9\linewidth]{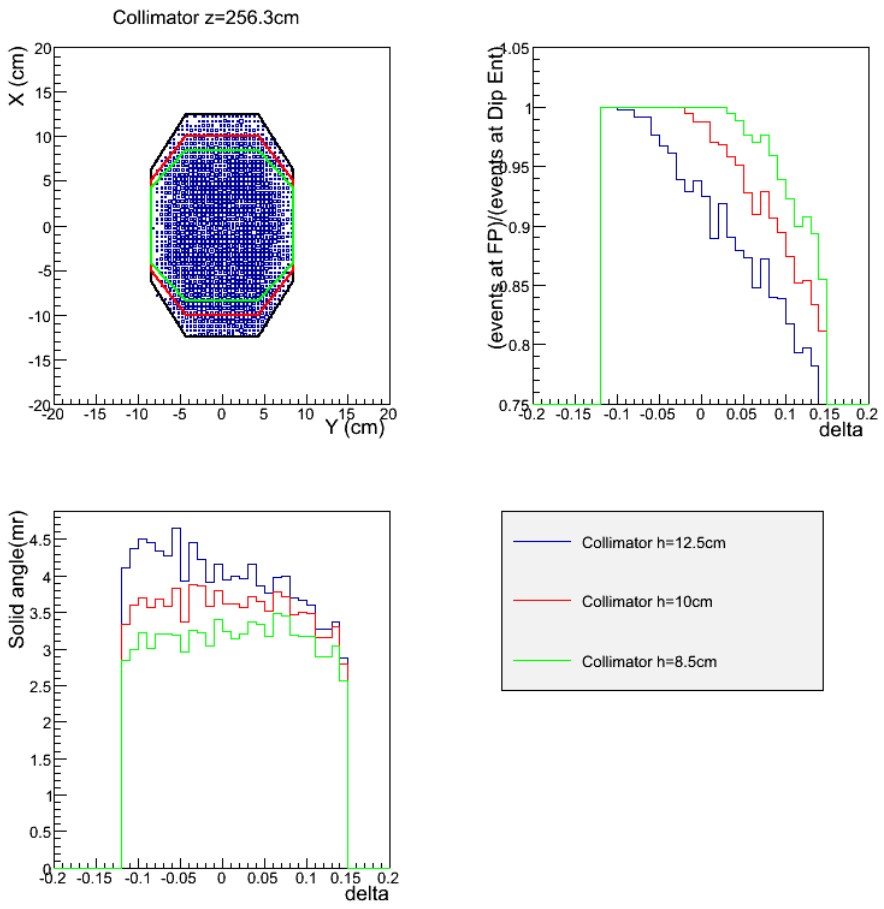}
		\caption{The upper left figure is distribution of events at the location of the collimator with three different vertical size collimators. The lower left figure is the acceptance as a function of $\delta$ for each of the collimators. The upper right figure is the fraction of events lost in the dipole bore after the dipole entrance. \label{fig:Study_coll_acceptance}}
	\end{centering}
\end{figure}

To determine the vertical size of the collimator, studies were conducted with the SNAKE model of the SHMS. Without the collimator, the vertical acceptance is mainly
determined by the mechanical exit of the HB magnet. The vertical size of $\pm$ 12.5 cm was chosen to match this vertical cut-off, to maximize the acceptance.
Two alternative vertical sizes of $\pm$8~cm and $\pm$10.5~cm for the collimators were studied. A plot of the acceptance each collimator
versus $\delta$ is shown in Fig.\,\ref{fig:Study_coll_acceptance}. The acceptance drops from an average of 4~msr for the vertical size of $\pm$12.5~cm to an average
of 3~msr for $\pm$8~cm. Another consideration was to minimize the loss of events in the bore of the vertical dipole. 
Another plot in Fig.\,\ref{fig:Study_coll_acceptance} shows the fraction of events that make it to the focal plane. The number of events lost in the dipole bore
as a function of $\delta$ is reduced by decreasing the vertical height of the collimator. With the $\pm$12.5~cm collimator vertical size, the fraction of events
reaching the focal plane drops to 75\% at $\delta$ = 0.15. In the final design, the $\pm$12.5~cm vertical opening was used to maximize the solid
angle acceptance of the SHMS at the expense of increased reliance on the modelling of the losses in the SHMS dipole bore.

\begin{figure}[htbp]
	\begin{centering}
		\includegraphics[width=.5\textwidth]{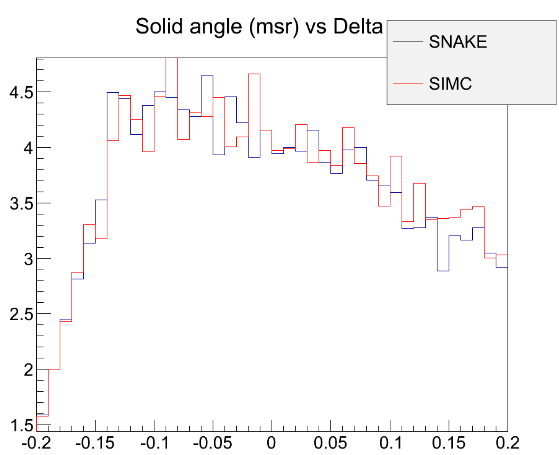}
		\caption{Comparison of predicted SHMS acceptance using the Hall C Monte Carlo (SIMC) and the magnetic transport code SNAKE. \label{fig:Snake_simc_delta}}
	\end{centering}
\end{figure}

The SNAKE-based transport model of the SHMS was also used to study the spectrometer acceptance. The acceptance of the SHMS as a function of $\delta$, as determined by the SNAKE based model
is plotted in Fig.~\ref{fig:Snake_simc_delta}. Alternatively, the spectrometer acceptance is also modeled using the Hall C Monte Carlo (SIMC) with the magnetic transport matrix obtained from the COSY INFINITY~\cite{COSY} program. The acceptance of the SHMS as a function of $\delta$ as  determined by SIMC is also plotted in Fig.~\ref{fig:Snake_simc_delta}.
The two acceptance models agree within statistics. 

The reconstruction of a charged particle’s relative momentum, horizontal target position and vertical and horizontal angles can be obtained from a polynomial expansion of the focal
plane positions and angles given by Eq.~\ref{eq:matrixeqn}.

\begin{align}
  x^{\prime}_\mathit{tar} =& \sum\limits_{ijklm} X^{\prime}_{ijklm}x^i_\mathit{fp}x^{\prime j}_\mathit{fp}y^k_\mathit{fp}y^{\prime l}_\mathit{fp}x^m_\mathit{tar}\nonumber \\
  y_\mathit{tar} =&  \sum\limits_{ijklm} Y_{ijklm}x^i_\mathit{fp}x^{\prime j}_\mathit{fp}y^k_\mathit{fp}y^{\prime l}_\mathit{fp}x^m_{tar}\nonumber\\
  y^{\prime}_\mathit{tar} =&  \sum\limits_{ijklm} Y^{\prime}_{ijklm}x^i_\mathit{fp}x^{\prime j}_\mathit{fp}y^k_\mathit{fp}y^{\prime l}_{fp}x^m_\mathit{tar}\nonumber\\
  \delta =&  \sum\limits_{ijklm} D_{ijklm}x^i_\mathit{fp}x^{\prime j}_\mathit{fp}y^k_\mathit{fp}y^{\prime l}_\mathit{fp}x^m_\mathit{tar}
\label{eq:matrixeqn}
\end{align}

The powers of each focal plane variable are given by $ijkl$ and $m$ is the power of the vertical position at the target which cannot be directly reconstructed. The transfer
coefficients for each power of the focal plane variables are given by $X^{'}, Y, Y^{'}$, and $D$ and they can be represented in a similar matrix formalism (reconstruction matrix).
The target offsets, beam offsets, and spectrometer mis-pointings are accounted for separately during the event reconstruction. The reconstruction of the 5 target variables is
under-determined, as seen in Eq.~\ref{eq:matrixeqn}, since there are only 4 measured focal plane variables. Therefore the $x_{tar}$ cannot be directly reconstructed and has to be estimated using
the beam position and the other reconstructed target position and angles. The $x_{tar}$ dependent  matrix elements ($m \neq 0$ terms in Eq.~\ref{eq:matrixeqn}) are determined using
an iterative process where the
initial values are obtained from the COSY model and the vertical beam position is assigned as the initial value of $x_{tar}$. The remaining 4 target variables are reconstructed using
these initial values and $x_{tar}$ is calculated. Using this new value of $x_{tar}$ the reconstruction matrix elements are recalculated, and the process is repeated no more than five times or until the change in the vertical angle, $x^{'}_{tar}$, between two iterations is less than 2~mrad.  

The $x_{tar}$ independent matrix elements ($m = 0$ terms in Eq.~\ref{eq:matrixeqn}) are determined using calibration data collected with a multi-foil carbon target and the sieve slit placed downstream of
the target. The sieve slit provides the ``true'' positions and angles for every particle originating from the target that passes through a particular sieve hole. These can be
determined from the beam position at the target, the location of the target foil, and the location of the sieve hole. The reconstructed $y_{tar}$ is approximately $z_{react}\sin{\theta}$,
where $\theta$ is the central angle of the spectrometer, and $z_{react}$ is the target foil position in the hall beam line coordinate system.
This information can be used to optimize the reconstruction matrix elements and thereby improve the reconstruction of the target variables. 

\begin{figure}[htbp]
  \centering
    \includegraphics[width=0.5\textwidth]{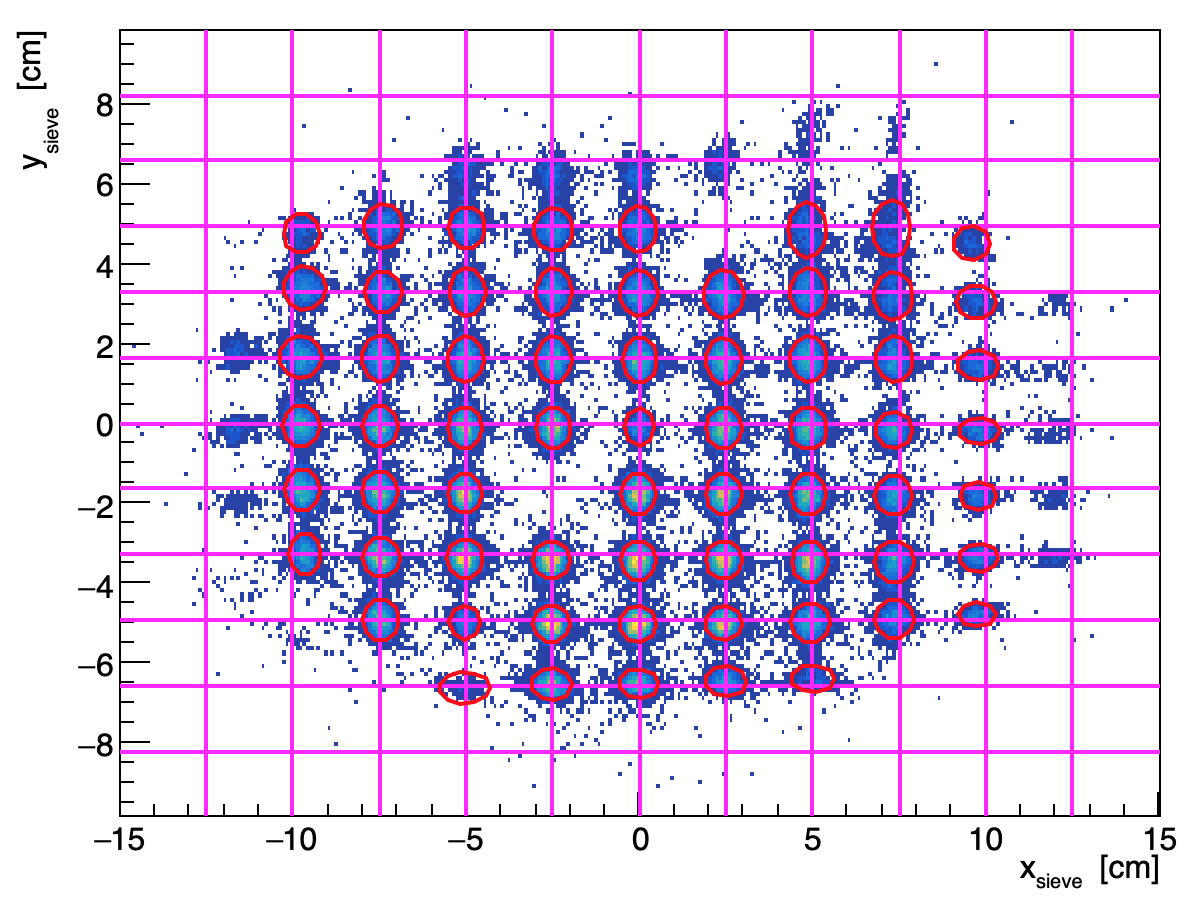}
 	 \caption[Reconstructed sieve pattern]{The sieve pattern is reconstructed here where the true sieve hole centers are indicated by the magenta cross lines and the reconstructed holes are outlined in red. The reconstructed hole positions at the edges of the sieve are somewhat shifted from the true desired values.}
  \label{fig:sievePlot}
 \end{figure}

The optimization procedure for the target
position and angles started with an initial set of matrix elements
generated by the COSY model. The data from the multi-foil target and sieve slit were used to calculate the difference between the ``true'' position and angles for each sieve slit hole
and the ``reconstructed'' position and angles. This difference is then minimized by solving a Singular Value Decomposition (SVD) to calculate the optimized/improved reconstruction matrix
elements.  To optimize over the full range of possible $y_{tar}$ values, calibration data were collected over a range of spectrometer central angles. Furthermore, data were collected with
two different sieve slits which had identical hole patterns, but one with the central hole centered on the spectrometer axis and the other with the central hole shifted by
half of the inter-hole distance. A reconstructed sieve pattern using a single carbon foil is shown in Fig.~\ref{fig:sievePlot}. In practice, the SHMS has achieved angular resolutions of $\sim$0.9\,~mrad in the horizontal direction and $\sim$1.1\,~mrad in the vertical direction.

\begin{figure}[htbp]
  \centering
    	  \includegraphics[width=0.45\textwidth]{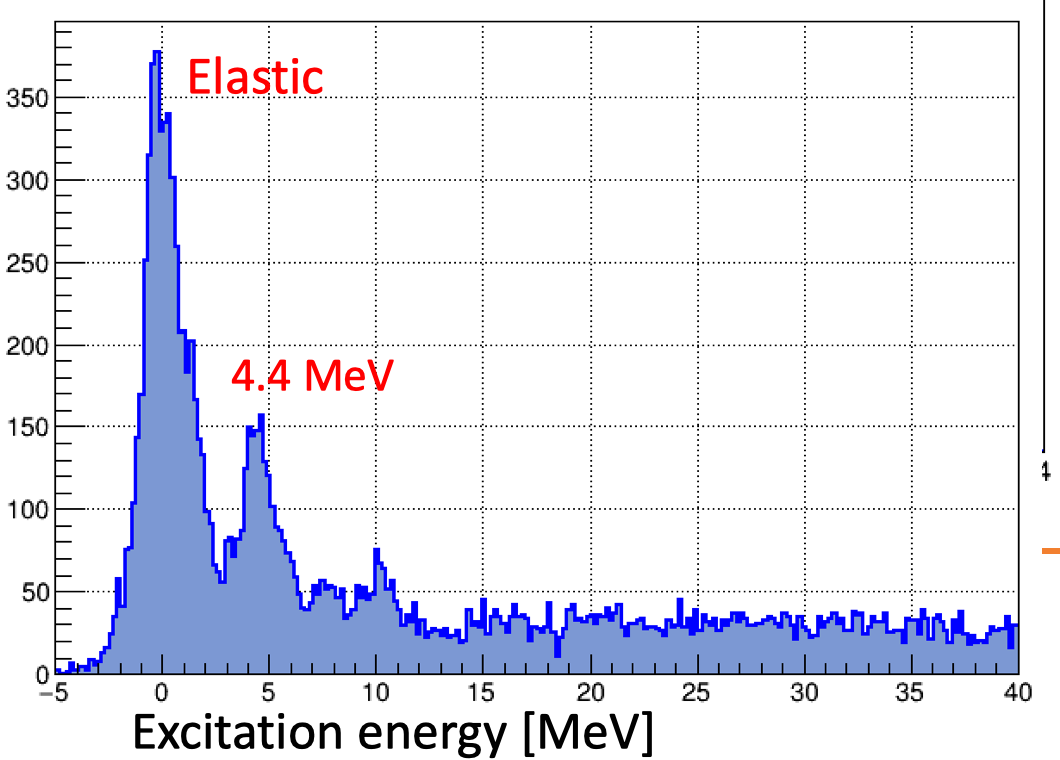}
 	 \caption[Carbon elastic energy spectrum]{The carbon scattered electron energy spectrum for events for a single sieve hole, as calculated in terms of delta from the first order optics, clearly shows the carbon elastic peak and the 4.4\,MeV excited state.}
  \label{fig:elastic}
 \end{figure}

The optimization of the $\delta$ reconstruction matrix elements was performed using carbon elastic data. Using the first-order matrix elements from the 
COSY model and selecting events originating from a carbon target which pass through a single hole in the sieve, the carbon elastic peak and the 4.4 MeV excitation state are 
identified as shown in Fig.~\ref{fig:elastic}. Additional carbon excited states are observed as the smaller peaks to the right of the 4.4 MeV peak. The $\delta$ matrix elements were
optimized by taking a series of calibration runs where the carbon elastic peak was scanned across the focal plane by varying the spectrometer's central momentum.

\subsection{Shield House Layout, Shielding Design}
\label{sec:shielding}

The radiation environment is an important consideration for the design of the SHMS shield house, in particular the effect of radiation-induced effects on the performance and reliability of detectors and electronics. It has been shown that many new commercial off the shelf components are sensitive to radiation damage and single event upsets, requiring a careful evaluation of the impact of the radiation-induced effects on their performance and reliability~\cite{griff97,under98}. A specialized SHMS shield house design was thus developed. Shielding thicknesses were optimized using a Monte Carlo simulation and benchmarked against the HMS shielding house, which has been proven to provide the necessary detector shielding over more than a decade of experiments at the 6\,GeV JLab. A full description of the shielding optimization can be found in Ref.~\cite{horn08}.

The primary particle radiation is created when the CEBAF electron beam strikes the experimental target. The main components are scattered electrons, neutral particles (photons and neutrons), and charged hadrons. The energy spectrum of this radiation depends on the incident beam energy and decreases generally as 1/$E$. It has been shown that the most efficient way to protect the experimental equipment from radiation damage is to build an enclosure around it using certain key materials. The type and thickness of the shield house walls depends on the energy and particle one needs to shield against. However, one may qualitatively expect that the largest amount of shielding material is needed on the side facing the primary source, which in the case of the Hall C focusing spectrometers is the front wall facing the target. Additional sources of radiation are the beampipe, which extends from the experimental target to the beam dump, and the beam dump area itself. Thus, the faces of the spectrometer exposed to direct sources of radiation are the front, beam side, and the back walls.

Primary and scattered electrons lose a significant amount of energy as they traverse a material by producing a large number of lower energy photons through bremsstrahlung~\cite{leo87}. It is thus important to consider shielding materials that efficiently stop the latter as well.

Neutral particles have a higher penetration power than charged particles. They are attenuated in intensity as they traverse matter, but do not continuously lose energy. Photons interact in materials almost exclusively either with electrons surrounding the atom or by pair production in the field of the nucleus. The probability for an interaction depends on the atomic number of the material. Neutrons interact with atomic nuclei in a more complicated way.

An additional source of radiation is due to charged hadrons (e.g. protons, pions). However, the probability for producing hadron radiation is relatively low, and thus will be neglected here. The shielding is, nevertheless, effective for charged hadrons. The front wall will for example stop 1 GeV protons.

\begin{figure}[htbp]
\begin{centering}
\includegraphics[width=.5\textwidth]{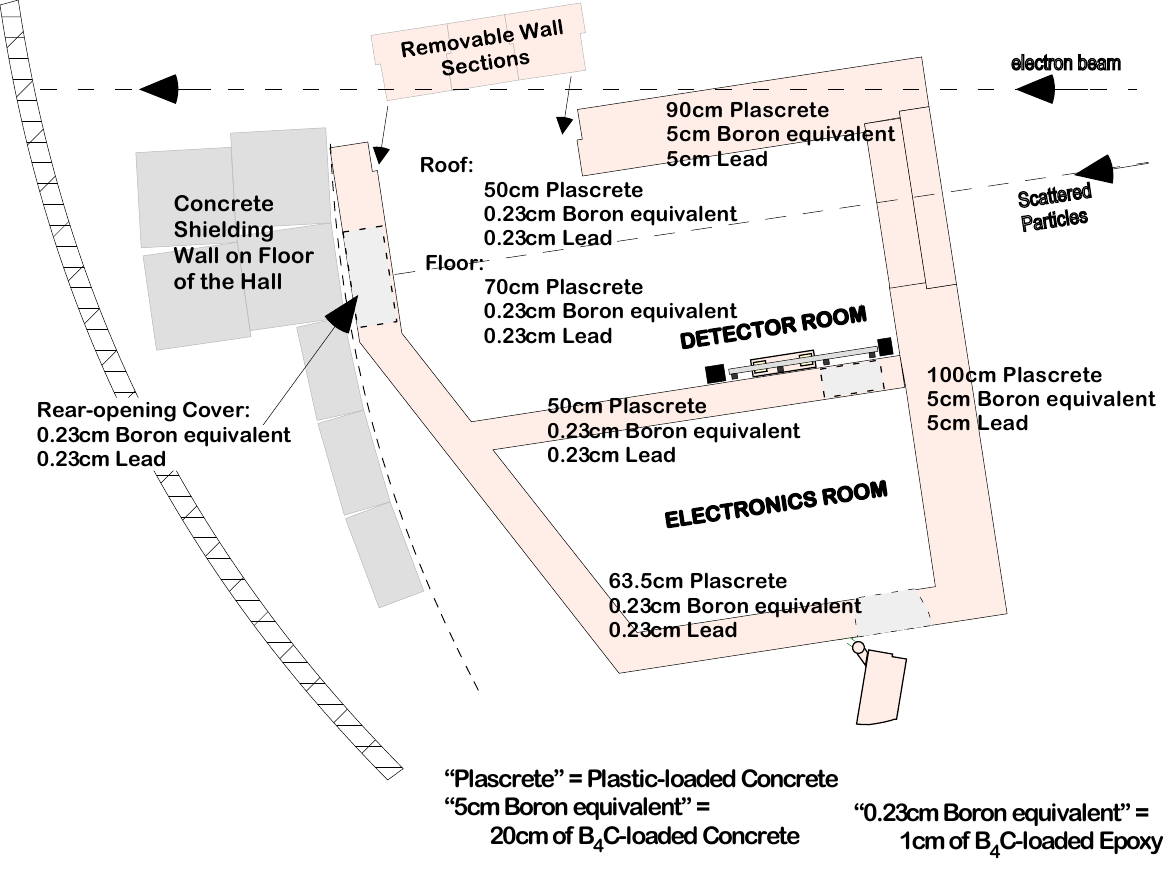}
\caption{Plan View of the SHMS Shield House showing the layout, thickness, and composition of the walls with the SHMS at 5.5$^{\circ}$. The bottom of the beam side shield wall is above the beam line, so the beam does not goes through the shield wall.\label{fig:shieldingPlanView}}
\end{centering}
\end{figure}

Fig.~\ref{fig:shieldingPlanView} shows a schematic of the SHMS shielding plan. The SHMS shield house is similar to the HMS design, but has several new features due to additional requirements. For example, the space between the beam side shield wall and the beam pipe is limited at very forward angles, and in addition, the length of the SHMS detector stack and minimum distance between the back of the detector house to the hall wall requires a reduction in thickness of the concrete shield wall.


Typical beam-target geometries were simulated using Monte Carlo techniques. Simulations were performed using the GEANT MCWORKS distribution, which includes detailed physical and geometric descriptions of the experimental hall and simulates the physics processes using standard GEANT3 \cite{GEANT3}  together with the DINREG nuclear fragmentation package. Hadronic interactions are treated using the DINREG package, which calculates the probability of such interactions using a database of photonuclear cross sections. For electron-nucleus interactions, an ``equivalent photon" representation of the electron (or positron) is used.

In this simulation, the CEBAF beam electrons start 1\,m upstream of the target, strike it head-on along the cylindrical symmetry axis, and have no momentum component transverse to the beamline. The simulation also includes the beam pipe, target entrance and exit windows, and the entire geometry of Hall C, including all elements of the beam dump. The transmission of particles through the shielding materials was calculated as a function of the material thickness and the angle relative to the beam direction.

A limitation of the radiation studies is the lack of cross section data for low-energy neutrons. The accuracy of the GEANT simulations was tested by benchmark calculations using the MCNP code~\cite{bries93} with an isotropic neutron point source of 1\,MeV located 1 m from the shield wall. The MCNP calculations suggest that 50\,cm of concrete thermalizes most of the fast neutrons, and after 1 m practically no epithermal neutrons remain. The thermalized neutrons can be captured by a 1\,cm Boron layer. In reality, however, the neutron spectrum also includes higher energy neutrons, for instance produced by electrons interacting in the concrete, and thus the actual amount of material for the walls exposed to the primary sources of radiation has to be thicker. A simple transmission calculation using GEANT4 \cite{geant4} for incident neutron beams of energies between 1 and 10\,MeV suggests that a thickness 150\,cm of concrete is sufficient to stop the majority of low-energy neutrons.

\begin{figure}[htbp]
\centering
  \includegraphics[width=.5\textwidth]{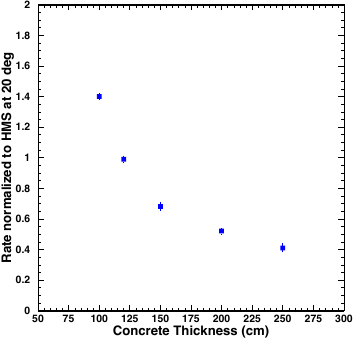} 
\caption{\label{fig-det-front-thickness} The normalized background rate vs. front wall thickness based on simulations described in the text. The rates are normalized to those found in the HMS at 20$^\circ$.}
\end{figure}

The SHMS shielding model is composed of standard concrete ($\rho$ = 2.4\,g/cm$^{-1}$). The thickness of the wall in front of the detector and electronics rooms is 100~cm, to shield from the primary radiation source around the target. Figure~\ref{fig-det-front-thickness} shows the surviving background flux for varying front wall concrete thicknesses. The results are normalized to the background flux in the HMS at 20$^\circ$. This angle was chosen as experiments in Hall C have shown that electronics problems are more frequent at lower angles. The simulation results show that 100~cm of concrete reduces the total flux to same as the flux in HMS at 20$^\circ$.

\begin{figure}[htbp]
\centering
\includegraphics[width=.5\textwidth]{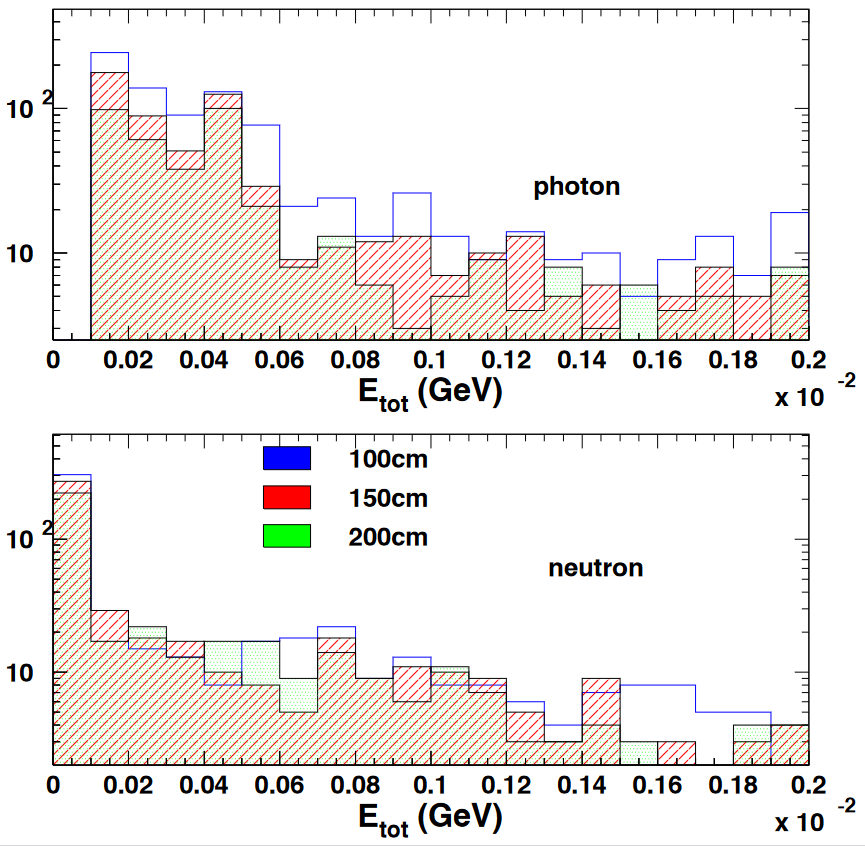}
\caption{\label{fig-det-front-etot} The outgoing, soft particle spectrum ($<$ 10\,MeV).}
\end{figure}

Figure~\ref{fig-det-front-etot} shows the energy spectra for surviving photons and neutrons with varying front wall thickness. In order to optimize the shielding, these secondary particles have to be absorbed as well. Our assumption on radiation damage is that photons below 100\,keV will not be a significant source of dislocations in the lattice of the electronics components, while neutrons will cause radiation damage down to thermal energies. Adding lead to the concrete wall reduces the photon flux significantly, but it does not help for neutrons. On the other hand, the boron reduces the flux of very low energy neutrons. Assuming that low energy photons and neutrons cause a significant fraction of the radiation damage, then adding the relevant material would be important.

The thickness of the beam-side wall (shielding from an extended source, the beamline) is constrained by the clearance with the detector stack inside the enclosure and the beamline at small angles. Conservatively assuming a clearance of 5\,cm between detector stack and the shield wall, the total concrete wall thickness is limited to 105\,cm. A 90\,cm concrete wall combined with a 5\,cm boron and 5\,cm lead layer provides the optimal shielding configuration. Adding boron is not much different from adding (or replacing) concrete, but in addition it captures thermal neutrons. 

\begin{figure}[htbp]
\begin{centering}
\includegraphics[width=.5\textwidth]{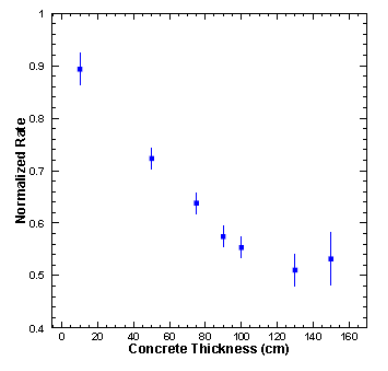}
\caption{The normalized rate versus the intermediate concrete wall thickness. \label{fig:shieldingRateVsConcThickness}}
\end{centering}
\end{figure}

The majority of charged particles are stopped by the outer walls of the spectrometer shield house. An additional source of radiation may be created from particles entering the enclosure through the magnets. To protect the electronics further, an intermediate wall was installed between the detector and electronics rooms. Figure~\ref{fig:shieldingRateVsConcThickness} shows the normalized rate as the thickness of this intermediate wall is varied. This suggests that the optimal configuration is provided by a concrete thickness of 80--100\,cm . Due to space constraints the  minimum wall thickness of 50\,cm needed to provide support for the roof of the shield house was used. Further details on shielding configurations investigated and their optimization can be found in Ref.~\cite{horn08}. 

The hydrogen-rich concrete walls function as a shield, an absorber, and a neutron moderator, and are thus placed on the outside of all faces of the shield house. On the other hand, the ordering of lead and boron to shield against the photon and neutron flux may, at first glance, not be obvious, and is discussed in detail below.

The incoming photon flux has two components: externally produced photons and bremsstrahlung photons produced by electrons in the twenty radiation lengths of concrete. The simulations have shown that the outgoing photon spectrum is soft ($<$10\,MeV). Placing a lead layer after the concrete is essential to suppress this low energy photon flux. The ($\gamma,n$) reaction in lead is not a problem. The threshold for the reaction is given by the neutron binding energy ($\sim$ 8\,MeV). At higher energies, the cross sections are in the mbarn range~\cite{biren95}. Even disregarding the low cross section, however, it is not clear that this reaction contributes to the irradiation of the electronics, because a high energy photon is replaced by a low
energy (but not thermal) neutron.

The incoming neutron flux also has two components. Neutrons from excited nuclei will typically not exceed 10\,MeV. The other neutrons are produced through direct interactions with only one nucleon in the nucleus. These will have high energies, but the flux is low. As shown by the MCNP calculation, which has reliable low energy neutron cross sections, 0.5\,m of concrete almost fully thermalizes 1\,MeV neutrons. Thus, 2\,m of concrete should be sufficient to thermalize the first component. Some of these will be captured in the concrete, but to eliminate the surviving thermal neutrons a layer of boron is needed. There are two relevant reaction channels: ($n,\gamma$) and ($n,\alpha \gamma$). The former produces high energy photons, but the cross section is relatively small. The latter produces a 0.48\,MeV photon for every captured neutron. The thermal cross section is about 10\,kbarn, and even at 1\,MeV it is still in the barn range. The majority of neutrons can thus be expected to be captured in a sufficiently thick boron layer. An optimal shielding configuration would also stop these photons produced in the capture. At 0.48\,MeV, the photoelectric effect and Compton scattering contribute about equally to the attenuation in lead. Photons from the latter will also need to be absorbed.

Thus, placing the lead in front of the boron layer has limited benefit. It will not affect the neutron flux, but will create an additional source of photons. The more lead one places after the boron, the more efficiently these photons will be suppressed. From the point of view of stopping bremsstrahlung photons, the order of boron and lead layers does not matter. Thus, all lead should be placed after the boron.

\begin{figure}[htbp]
\begin{centering}
\includegraphics[width=.4\textwidth]{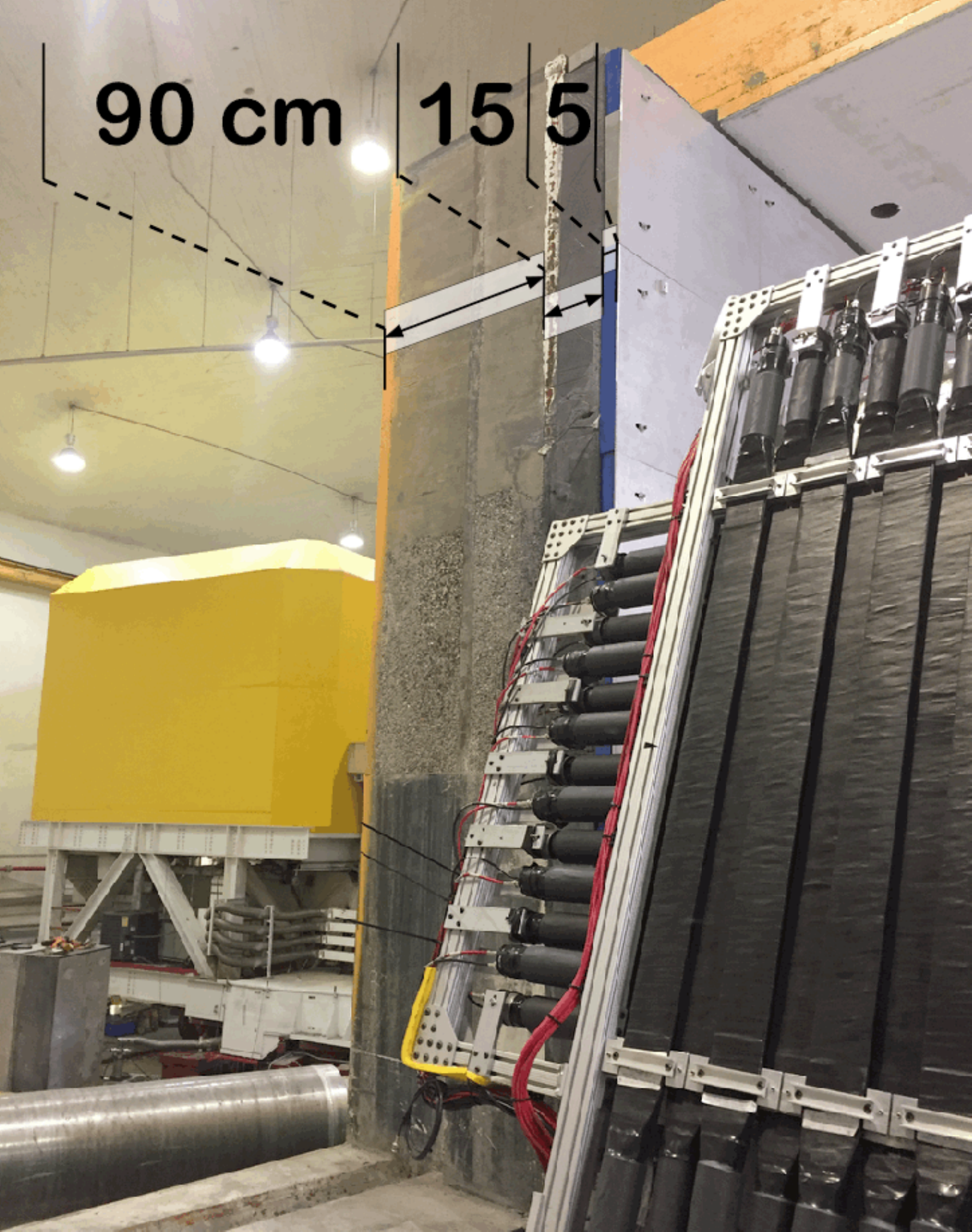}
\caption{Photograph of the SHMS beam-side shield wall in cross-section view, showing the 
layers of different materials making up the wall.\label{fig:shieldingWallXSPhoto}}
\end{centering}
\end{figure}
 
 Fig.~\ref{fig:shieldingWallXSPhoto} is a photograph showing the resulting multi-layered shielding in one of the SHMS shield house walls. The ceiling, floor, and other walls have similar compositions but varying dimensions presented earlier in Fig.~\ref{fig:shieldingPlanView}. Details about the development of custom concrete material containing boron can be found in Ref.~\cite{concrete}.

In summary, the SHMS shielding consists of concrete walls to moderate and attenuate particles. Low energy (thermal) neutrons are absorbed in a boron layer inside the concrete. Low energy and 0.5\,MeV capture photons are absorbed in lead. With this design, the rates at forward angles of 5.5$^\circ$ are estimated to be less than 70\% of the design goal (HMS at 20$^\circ$) in the detector room and below 70\% in the electronics room.


\section{Design, Construction and Calibration of the SHMS Detectors}
\label{sec:detdev}
\begin{figure}[htbp]
\includegraphics[width=\linewidth]{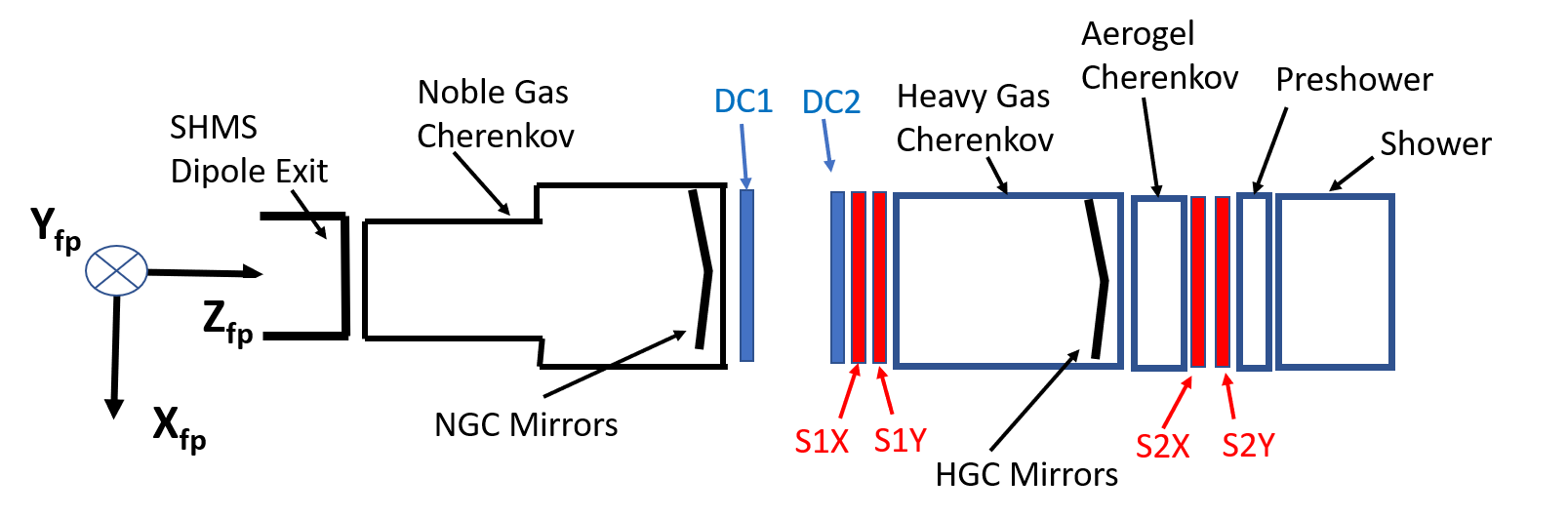}
\caption{ A drawing of the SHMS detector stack in which the detectors are designated by simple boxes. The drawing is not to scale. The labeling of the detectors is described in the text.  
\label{fig:detstack}}
\end{figure}

The layout of the SHMS detectors in the SHMS detector hut is shown in Fig.~\ref{fig:detstack} and each detector is described in detail in the following subsections. A CAD drawing of the SHMS detector stack is shown in Fig.~\ref{fig:DAQ-SHMSlayout}. After traveling from the target through the SHMS 
magnets, the scattered particles exit the SHMS dipole vacuum pipe. The first detector that is encountered is the Noble Gas Cherenkov detector (NGC) which is described in Sec.~\ref{sec:ngcerenk}. The NGC can be replaced by a vacuum tank which is designed to occupy the same space as the NGC. The tracking of the particles is done by a pair of drift chambers (DC1 and DC2) and each of drift chambers  consists of six wire planes. Details on DC1 and DC2 are described in Sec.~\ref{sec:drifts}. 
Basic trigger information comes from four planes of scintillator or fused silica hodoscopes. This first pair of hodoscope planes (S1X and S1Y) are located directly after drift chambers. The Heavy Gas Cherenkov (HGC) detector is placed after S1Y and the HGC is described in Sec.~\ref{sec:hgcerenk}. The aerogel cherenkov (AER) detector is located after the HGC and is described in Sec.~\ref{sec:aerogel}. The AER can be removed from the detector stack if it is desired without changing the positions of the other detectors. The second  pair of hodoscope planes (S2X and S2Y) are located after the aerogel cherenkov detector. The S1X, S1Y and S2X hodoscope planes are standard scintillator paddles and are described in Sec.~\ref{sec:scint}. The S2Y hodoscope plane is made of fused silica ("quartz") and is described in Sec.~\ref{sec:quartz}. The last detectors in the stack are the preshower and shower detectors which are described in Sec.~\ref{sec:shower}.

\subsection{Scintillator Trigger Hodoscope}
\label{sec:scint}

The SHMS hodoscope system provides a clean trigger and trigger time information as well as the definition of the detector package fiducial area, required for physics cross section measurements. The system is composed of four separate planes of detector paddles: S1X and S1Y located immediately after the second drift chamber, and S2X and S2Y approximately 2.6~m away along the z direction. Teh S1X plane is closest to the drift chamber and the S2Y plane is 9.6~cm away from the S1X plane along the z-direction. The S1X, S1Y, and S2X planes were built using thin plastic scintillator paddles while S2Y uses fused silica bars.
\begin{figure}[htbp]
\begin{center}
\includegraphics[width=0.4\textwidth]{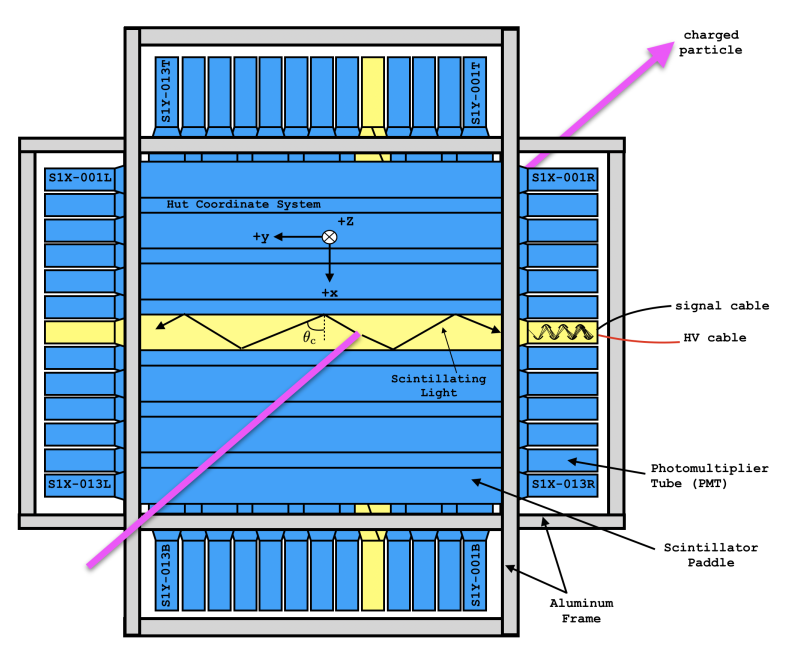}
\end{center}
\caption{ A drawing of the S1X and S1Y planes of the SHMS hodoscope. The S2X and S2Y planes are arranged in a similar manner with the S2X having 14 paddles and the S2Y having 16 paddles.
\label{fig:hodoscope}}
\end{figure}


\subsubsection {Design and Construction}

The overall dimensions and granularity of the three scintillator planes were driven by the Monte Carlo simulations of the SHMS acceptance. The S1X and S1Y planes cover a 100$\times$98\,$\rm{cm}^2$ area while the S2X plane covers 110$\times$133.5\,$\rm{cm}^2$.  Further design constraints for this detector include high ($\geq 99$\%) detection efficiency with little position dependence along the scintillator paddle; good time resolution ($\sim$100~ps) and high rate capability ($\sim$1\,MHz/cm). As the detector's lifetime is assumed to be roughly a decade, the design made use of cost effective, readily available materials and readout chain.

To meet the requirements listed above, the SHMS Hodoscope was built as a series of arrays (planes) of plastic scintillator paddles. The S1X and S1Y planes have 13 100$\times$8\,cm paddles each, while the S2X plane has 14 110$\times$10\,cm paddles. A sketch of the S1X and S2Y planes is shown in Fig.~\ref{fig:hodoscope}. For each of the three scintillator planes, the paddles were staggered in the beam direction by 0.7\,cm and overlapped transversely by 0.5\,cm. To minimize the impact of the scintillators on downstream detectors and also to ensure good timing resolution, the thickness of paddles was 0.5\,cm.

The scintillator material used was Rexon {\tt RP-408}\,\cite{Rexon_RP-408}. The paddles were wrapped by the manufacturer with millipore paper, aluminum foil, and 2" wide electrical tape. The transition between the thin scintillator material and the photomultiplier (PMT) tubes was done using a Lucite\texttrademark{} fishtail-shaped light guide. As the glued joint between the scintillator paddle and the light guide is rather fragile (0.5$\times$8.0 and 0.5$\times$10.0~cm joints, respectively) aluminum \lq\lq{}splints\rq\rq{} were used to reinforce it. The PMT to fishtail joint was originally wrapped with 2" tape as well and light-leak tested; subsequently this wrapping was reinforced with TEFLON\texttrademark{} tape and a 3" heat-shrink sleeve.

Each scintillator paddle has a PMT at each end glued to the fishtail using optical glue (BC-600\,\cite{BC-600}) matching the index of refraction of the Lucite\texttrademark{}. A combination of Photonis XP 2262\cite{Photonis_XP2262} and ET 9214B \cite{ET_9214B} 2" tubes were used. Both models have 12-stage amplification and their maximum photocathode sensitivity is in the blue--green range. The typical gain is $3 \times 10^7$. Gains were measured as a function of high voltage during the construction and the whole hodoscope was gain matched {\it in situ} once installed in SHMS.

\begin{figure}[htbp]
\begin{center}
\includegraphics[width=.4\textwidth]{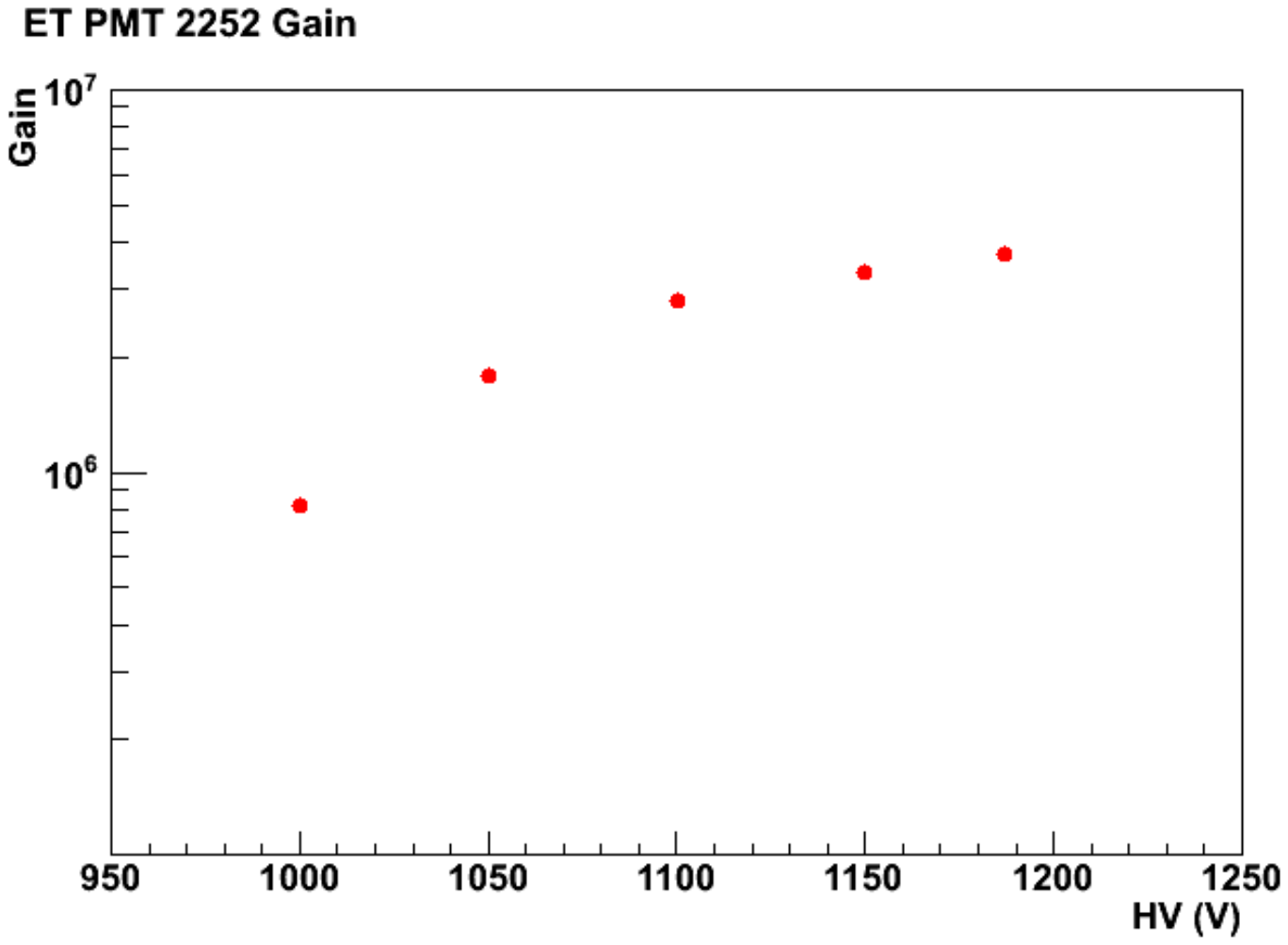}
\end{center}
\caption{Gain versus high voltage graph for an ET tube used for the scintillator hodoscope.
\label{fig:gain_vs_hv}}
\end{figure}

All hodoscope scintillator paddles and the PMTs used on the S1X, S1Y, and S2X planes were extensively tested during assembly: the dark current and the gain as a function of the high voltage were measured for each tube; the finished paddles were light-leak tested and their detection efficiency as a function of position along the paddle was measured using cosmic rays on an automated test stand. A typical gain versus HV graph is shown in Fig.\,\ref{fig:gain_vs_hv}.

\subsection{Quartz-bar Trigger Hodoscope}
\label{sec:quartz}

The SHMS hodoscope quartz plane was designed to help with neutral background rejection in the 12\,GeV high-rate environment. (The radiator material is actually fused silica, but we will often use the term ``quartz".)  It operates on the principle of Cherenkov light production by electrically charged particles. It is one of the four hodoscope planes that form the basic 3 out of 4 trigger in the SHMS. In what follows the design and construction of this detector will be presented as well as its performance with electron beam in Hall C. 

\subsubsection{Design and Construction}

Quartz bars of 2.5$\times$5.5$\times$125\,cm$^3$ dimensions with an index of refraction of approximately 1.5 were procured from Advanced Glass Industries. The Cherenkov light produced by electrically charged particles is detected by UV-glass window PMTs (model ET9814WB) and quartz window ET9814QB\,\cite{ET_9814} photomultiplier tubes optically coupled to the quartz bars through RTV\texttrademark{}615 silicon rubber of 50\,$\mu$m thickness. 
Thicker optical couplings resulted in fewer photoelectrons, so excellent quality control of the maximum thickness was required.  The 16 bars in use in the hodoscope quartz plane are staggered in the beam direction with a transverse overlap between adjacent bars of 0.5\,cm. The quartz plane frame allows for more bars to be added.
   
\subsubsection{Calibration of Hodoscope}
\begin{figure}[htbp]
\begin{centering}
\includegraphics[width=1.0\columnwidth]{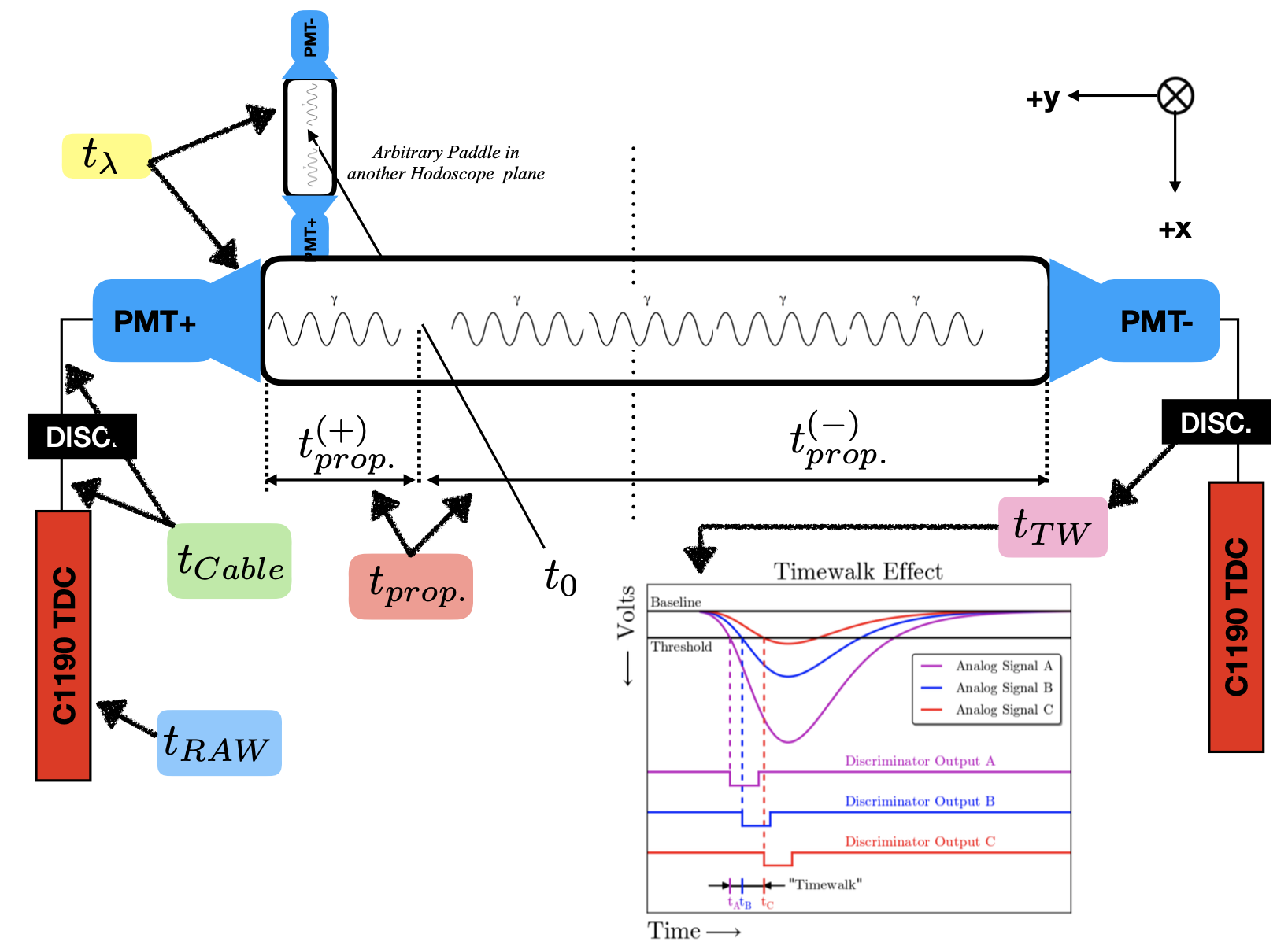}
\caption{Cartoon of individual scintillator paddles to illustrate the various timing corrections applied. \label{fig:hod_calib_diagram}}
\end{centering}
\end{figure}

The objective of the hodoscopes' calibration is to determine the the arrival time of the particles that traverse each of the scintillators (or quartz) planes relative to a reference time. When a particle traverses a plane, 
the hit paddle produces light which propagates to the PMTs at the two ends as shown schematically in Fig. \ref{fig:hod_calib_diagram}. The PMT signal is then sent to ADCs and TDCs. The raw TDC signal has multiple unwanted timing contributions which must be subtracted to obtain the true arrival time at the hodoscope plane. The corrected TDC time at each plane along with the calculated distance between each plane is then used to determine the particle velocity, $\beta = \frac{v}{c}$.
\\
The general expression for the corrected TDC time for a hodoscope PMT is
\begin{equation}
    t_{Corr} = t_{RAW} - t_{TW} - t_{Cable} - t_{\lambda} - t_{prop} \label{eq:tcorr_hod}
\end{equation}
The corrected TDC time ($t_{Corr}$) represents the particle arrival time at the scintillator paddle. The corrections from Eq.~\ref{eq:tcorr_hod} are summarized as follows:
\begin{enumerate}[itemsep=0.15cm]
    \item Time-Walk Corrections, $t_{TW}$: 
    For analog signals arriving at the Leading Edge Discriminators, larger signals fire the discriminator at earlier times. To correct for this correlation in the high resolution TDCs, the FADC Pulse Time is used as an amplitude-independent reference, with the difference of the TDC and FADC Pulse Time plotted against the FADC amplitude. A model function is fitted to this correlation, and the parameters extracted are used to correct the TDC time.
    \item Cable Time Offset Corrections, $t_{Cable}$: The time offset correction takes into account the fact that the analog signal has to propagate through a PMT and signal cables to the TDC in the Counting House. To determine this correction, a correlation between Time-Walk corrected time and hodoscope paddle track position is fitted to extract the average velocity of light propagation along the long axis of the paddle (1/slope), as well as the cable time offset (y-intercept). 
    \item Propagation Time Corrections, $t_{prop}$: This correction accounts for the light propagation time along the scintillator paddle from the particle hit location to the PMT. The correction is done in the Hall C Analyzer, hcana, using the measured average light propagation speed mentioned in the Cable Time Offset Corrections above.    
    \item Paddle Time Difference Corrections, $t_{\lambda}$:  This correction accounts for any additional time difference (other than the particle propagation time to travel across the two paddles) between each of the scintillator paddles in the different hodoscope planes. All paddle times are measured relative to paddle number 7 in the S1X plane. 
    
 \end{enumerate}   
\begin{figure}[htbp]
\begin{center}
\includegraphics[width=0.5\textwidth]{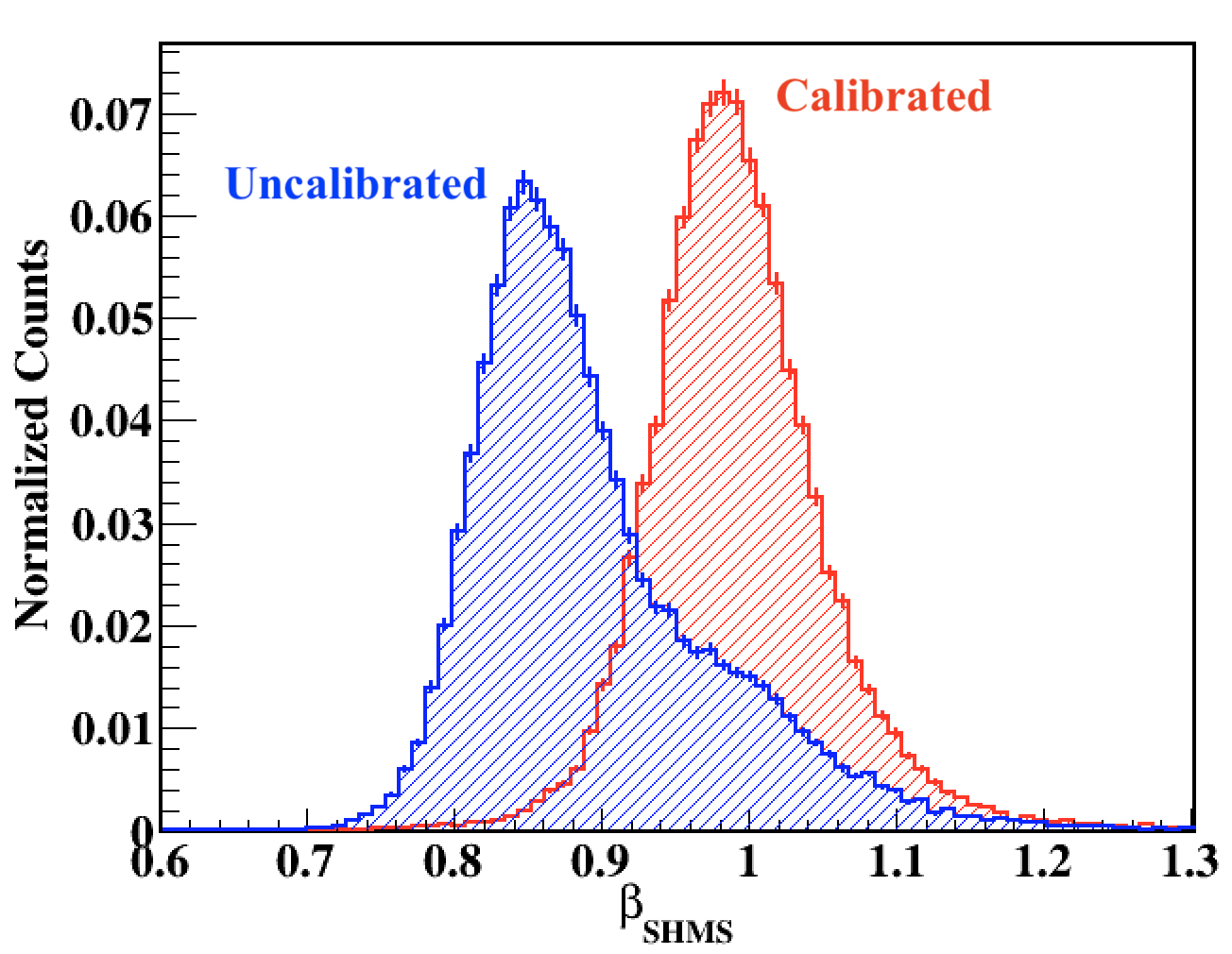}
\end{center}
\caption{SHMS Hodoscopes $\beta=v/c$ before (blue) and after (red) calibrations.\label{fig:phod_beta}}
\end{figure}

The result of a successful hodoscope timing calibration is shown in Fig. \ref{fig:phod_beta}.

\subsection{Drift Chambers}
\label{sec:drifts}

\subsubsection{Design and Construction}

%
%


The SHMS horizontal drift chambers provide information to determine the trajectory of charged particles passing through the detector stack. The drift chamber package consists of two horizontal drift chambers (DC1, DC2) separated by a distance of 1.1\,m and oriented in the detector stack such that the sense wires planes are perpendicular to the central ray. Each chamber consists of a stack of six wire planes providing information on the track position along a single dimension in the plane of the wires and perpendicular to the wire orientations to better than 250\,$\mu$m. The perpendicular distance of the track relative to the wire is determined from the time of the signal produced by the ionization electrons as they drift from their production point to the wire in an electric field of approximately 3700\,V/cm.

The design and construction technique is largely based on that of previous successful chambers built for the Hall C  6 GeV program, which have been shown to reach the resolutions and particle rate specifications of the SHMS. The open layout design consists of a stack of alternating wire and cathode foil planes; each plane consisting of 3.175~mm thick printed circuit board (PCB). These are sandwiched between a pair of aluminum plates on the outside, which provide both the overall structural support and the precise alignment of each board via dowel pins at the corners. Just inside each pair of plates is a fiberglass board with the central area cut out and covered with a vacuum-stretched film of aluminized Mylar\texttrademark{} which provides the gas window. These are sealed to prevent gas leakage via an o-ring around the gas fitting through-hole on the inside of the plate.

\begin{figure}[htbp]
    \begin{center}
        \includegraphics[width=1.0\linewidth]{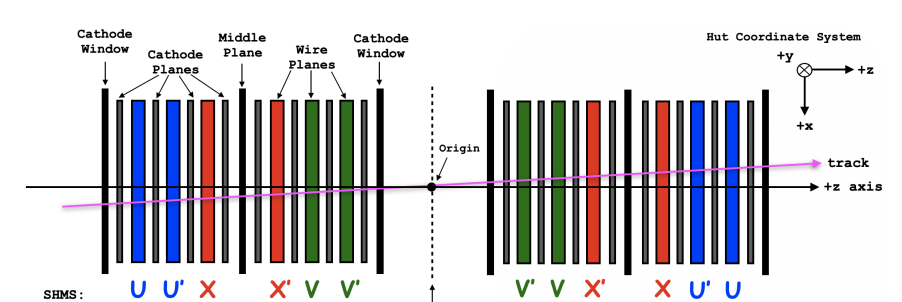}
        \caption{\label{fig:dc-sketch} Sketch of the SHMS drift chamber planes with particles tracking from left to right. A description of the labels is given in the text. The figure is from Ref.~\cite{YeroThesis}.
        }
    \end{center}
\end{figure}

Each chamber consists of two identical half chambers separated by a fiberglass mid-plane which supports the amplifier discriminator cards required for the sense wire readout. To minimize the production costs, only two unique PCB types were designed: an X-plane with wires oriented horizontally, and a U-plane with wires oriented at +60$^\circ$ relative the X-plane. All other plane orientations are generated by rotations of these two basic board types. For instance, the boards are designed such that a rotation of 180$^\circ$ in-plane about an axis through the center of the board produces boards with wires of the same orientation, but shifted by 1/2 cell width, thus allowing the resolution of left/right ambiguities. Rotation of the X-plane and the U-plane such that the top becomes the bottom produces the X' and U' orientations. The V and V' boards, with wire orientation of -60$^\circ$ relative to the X-plane, are produced by a rotation of the U and U' boards of 180$^\circ$ into the page about a vertical axis though the center of the board. Each half chamber has three planes with the first half consisting of (U, U', X) and the second half consisting of (X', V', V). The first chamber, DC1, is oriented in the SHMS frame such that the board ordering as seen by particle traversing the spectrometer is (U, U', X, X', V', V), while for the second chamber, DC2,  the ordering is reversed (V, V', X', X, U', U) as is shown in Figure~\ref{fig:dc-sketch}.

The drift gas (50/50 mixture of Ethane/Argon in production mode) flows across each board through holes in the cathode planes (k-planes) alternating from top to bottom. 
The overall dimensions of the wire chambers are driven by the desired active area for particles at the focal plane of the SHMS which was 80\,cm x 80\,cm. The active area of each wire plane consists of alternating 20\,$\mu$m diameter gold tungsten sense wires and 80\,$\mu$m diameter copper plated beryllium field wires separated by 0.5\,cm. Each wire plane is sandwiched between a pair of cathode planes with the cathode surfaces consisting of 5~mil thick stretched foils of copper plated Kapton\texttrademark{}.

\subsubsection{Calibration}
\label{Sec:DC_Cal}

As charged particles traverse the drift chambers and ionize the gas, free electrons from the ionized gas drift towards the sense wires in the chamber. This process produces measurable current pulses in the sense wires which are pre-amplified before the 16-channel input discriminators. The discriminators produce logic signals that are sent to the TDC which registers the relative time at which this signal arrives. This signal is utilised to determine the drift time, the time taken for the free electrons to drift to the sense wire, via 
\begin{equation}
    t_{D} =  t_{meas}  - \left(t_{wire}+t_{cable}\right).
    \label{eqn:Drift_Time}
\end{equation}

In Eqn.~\ref{eqn:Drift_Time}, $t_{meas}$ is the time recorded by the TDC and the term $t_{wire}+t_{cable}$ is the time it takes the signal to propagate across the sense wire, through the cable and into the TDC if the track were to pass directly through the sense wire.  When combined with information about the position of wires in each chamber, this quantity can provide coarse track information. However, this can be further refined by converting the drift time to a drift distance which is accomplished by utilising time-to-distance maps. The purpose of the drift chamber calibration procedure is to produce these per-plane look-up tables.

In the approximation that the incident particle distribution is uniform, or has a constant slope, a single cell\footnote{A cell is one sense wire surrounded by field wires such that the sense wire is at the center and the field wires are at the corners.} on average sees a flat distribution of events. The measured drift time distribution can therefore be mapped to drift distance by areal scaling. This distribution can be determined for the individual wire or averaged over an entire group (up to 16 wires per discriminator card). Associated with each drift time distribution is a time, $t_{0}$, which corresponds to the time at which ionized particles come into contact with the wire. If this value is non-zero, this is the value by which all drift times must be shifted in order to assure that $t_{0} = 0~ns$. All subsequent times in each spectrum are measured relative to this time. 
\begin{figure}[htbp]
    \begin{center}
        \includegraphics[width=1.0\columnwidth]{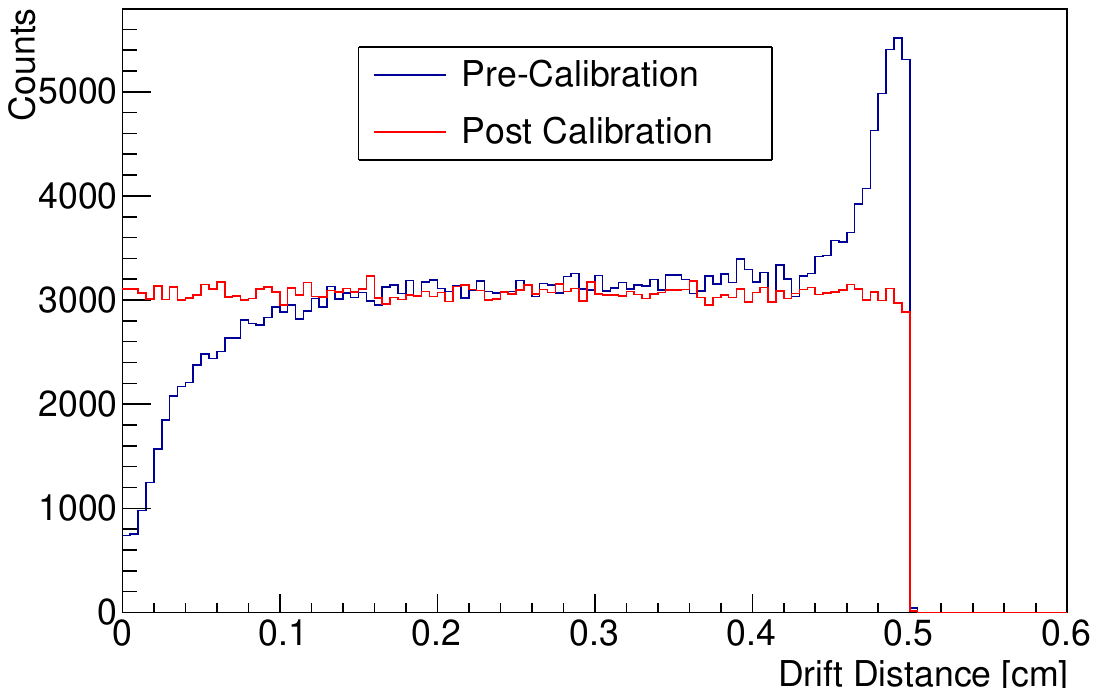}
        \caption{\label{fig:dcs-calib-DD} Example drift distance distributions for the SHMS drift chambers before (blue) and after (red) a successful calibration.}
    \end{center}
\end{figure}

From the drift time spectrum, $N(t)$, the drift distance $D(t)$ can be determined via 
\begin{equation}
    D(t) = D_{Max} \frac{\int^{t}_{t_{0}}N(t)dt}{\int^{t_{Max}}_{t_0}N(t)dt}
    \label{eqn:DriftDistInt}
\end{equation}
where $D_{Max}$ is the maximum possible drift distance, 
$t_{Max}$ is the maximum drift time and $t$ is the measured drift time. Note that $D(t_{0}) = 0~\mathrm{cm}$ and $D(t_{Max}) = 0.5~\mathrm{cm}$. Due to the finite resolution of the TDC, the integrals in Eqn.~\ref{eqn:DriftDistInt} become sums over finite bin widths and the actual expression used is 
\begin{equation}
 D(t_i) = D_{Max} 
    \frac{1}{N_{Tot}}\sum\limits_{\mathrm{i = bin}(t_{0})}^{\mathrm{bin}(t_{0}+T)} N(t_i),
    \label{eqn:DriftDistSum}
\end{equation}
which is simply a ratio of the sum of bin contents up to some maximum drift time, $T$, over all bin contents up to a maximum $t_{Max}$, called $N_{Tot}$. The results of the calibration are per-plane look up tables which utilise this ratio to map any given drift time to a drift distance for that plane. When properly calibrated, this should result in a flat distribution of drift distances for each chamber. An example drift distance spectra, showing the pre and post calibration distributions can be seen in Fig.~\ref{fig:dcs-calib-DD}.

\subsection{Heavy-Gas Cherenkov Counter}
\label{sec:hgcerenk}

\subsubsection{Design and Construction}
\label{subsec:HGC_Design}

The SHMS Heavy-Gas Cherenkov detector (HGC) is a threshold-type Cherenkov detector, designed to separate charged $\pi$ and $K$ over most of the SHMS operating momentum range, 3--11\,GeV/c. The radiator gas C$_4$F$_{10}$ at 1\,atm, with an index of refraction of $n$=1.00143 at standard temperature \cite{btev}, allows $\pi^{\pm}$ to produce abundant Cherenkov light above 3\,GeV/c momentum, while $K^{\pm}$ remain below Cherenkov threshold until about 7~GeV/c. Optimal $\pi/K$ separation at higher momenta requires a reduction in the gas pressure, down to 0.3\,atm at 11\,GeV/c.

\begin{figure}[htbp]
  \centering
  \includegraphics[scale=.3]{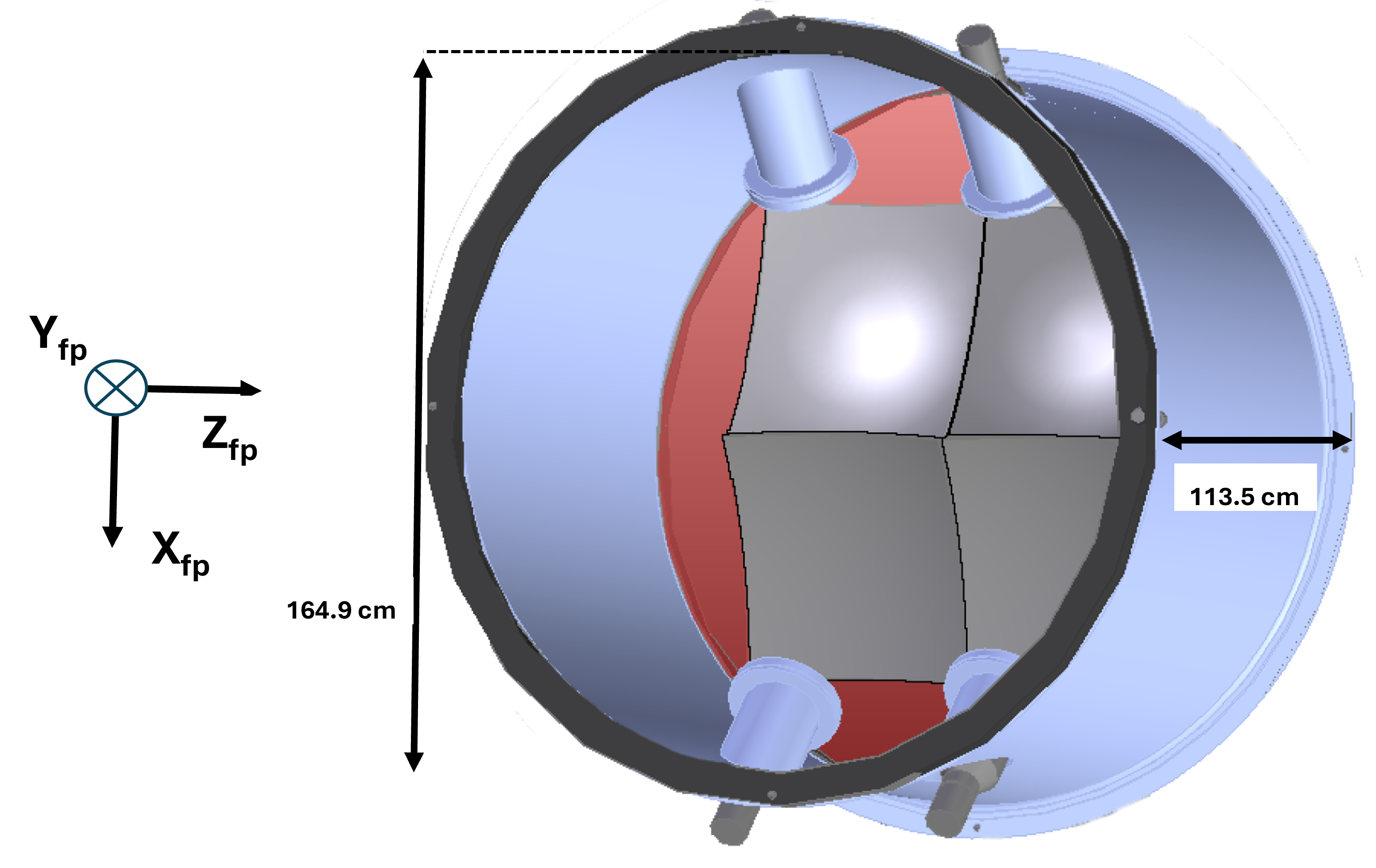}
  \caption{3D-CAD rendering of the Heavy Gas Cherenkov Detector.
\label{fig:hgc_layout}}
  \label{fig:HGC}
\end{figure}

A schematic view of the detector is shown in Fig.\,\ref{fig:hgc_layout}. The SHMS focal plane is subtended by four 55$\times$60\,cm 0.3\,cm thick glass mirrors, which reflect the Cherenkov radiation to four Hamamatsu R1584 12.5\,cm diameter photomultiplier tubes located above and below the particle envelope. The mirrors and gas are enclosed in a cylindrical aluminum tank of 164.9\,cm inner diameter and 113.5\,cm length, with entrance and exit windows of hydroformed 0.102\,cm thickness 2024 T-4 aluminum alloy \cite{ASM}. The vessel is sufficiently strong to be pumped to vacuum before introducing the radiator gas, avoiding the need to purge when filling. A unique aspect of the detector is the placement of the photomultipliers outside the gas envelope, viewing the enclosure through 1.00~cm thick Corning 7980 quartz windows. This allows the gas enclosure to be smaller in diameter than would otherwise be possible, as the full length of the PMT and base no longer need to be fully within the diameter of the vessel. It also makes the PMTs available for servicing without venting the gas.

The mirrors are inexpensive, having been produced by the slumping process \cite{sinclair}. As a result, they deviate from the desired 110\,cm radius of curvature with a slightly oblate shape \cite{wli_msc}. However, the Cherenkov cone on the mirrors for 3-7\,GeV/c $\pi^{\pm}$ in C$_4$F$_{10}$ is 7-10~cm in diameter, so optical quality mirrors are not required for this application. The UV wavelength characteristics of the respective optical components are relatively well matched. C$_4$F$_{10}$ has good transmittance down to $\sim$160~nm \cite{btev}. The quartz viewing windows provide $>$88\% transmission down to 200~nm, including the $\sim$10\% loss due to surface reflection \cite{corning}, and the optical glass faced PMTs have 70\% of their peak quantum efficiency at 200\,nm (peak at 350\,nm) \cite{hamamatsu}. Accordingly, the mirror reflectivity was optimized for $>$90\% at 270\,nm, and 75\% at 200~nm \cite{eci}.

The mirrors are arranged in a 2$\times$2 array, with two mirrors directing the light to two upper PMTs and the other two directing it to lower PMTs. Because the mirrors are curved in both the horizontal and vertical directions, it is necessary to stagger the mirrors along the tank $z$-axis to avoid dead areas. The upper left and right mirrors are the most forwards, with the lower left and right being behind. The mirrors overlap slightly to give good $x-y$ coverage. The geometry near the center of the tank, where the mirrors make their closest approach, is complicated, and some shadowing for certain Cherenkov light trajectories is unavoidable. This leads to a small region of lower detection efficiency at the center of the tank. This is further discussed in Sec.\,\ref{Subsec:HGC_Perf}.

Each mirror is clamped individually along its two outer edges and is held in place by 3 flexible three-point mounts extending from the tank to the mirror clamps. This allows each mirror to be optically aligned in 3 dimensions separately from the others. The mirror positions were fine-tuned with the use of an LED-light array clamped to the front of the tank. The reflected light from each LED onto the PMT positions was compared to predictions of a Geant4 simulation and adjustments made until they came into close agreement.

\subsubsection{Calibration}
\label{Sec:HGC_Cal}

The goal of the SHMS HGC calibration procedure is to generate an accurate translation from raw FADC channels (or charge in pC) to the number of photoelectrons emitted from the cathode surface of the PMT (NPE). This is achieved by isolating the single photoelectron (SPE) peak, yielding a calibration, and then verified by examining the regular spacing of the first few photoelectron contributions in the ADC spectrum.


To isolate the SPE peak, tracking cuts are applied to the data.
As a charged particle passes through a mirror quadrant, the produced Cherenkov cone allows some light to be incident on adjacent mirrors. As each mirror is focused on a single PMT, one PMT will receive most of the produced light while the other three ``off-axis'' PMTs receive much smaller amounts. This small signal on the 3 ``off-axis'' PMTs allows the SPE peak to be measured, yielding a reliable calibration. 
To select this adjacent mirror light, cuts (based on the physical dimensions of the mirrors) are placed on the tracked coordinates of the charged particles, extrapolated to the HGC mirror plane,
\begin{eqnarray}
  \label{eq:HGCtracking}
  x_\text{HGC} = x_\text{Focal Plane} + x'_\text{Focal Plane}\cdot z_\text{HGC}\\ 
  y_\text{HGC} = y_\text{Focal Plane} + y'_\text{Focal Plane}\cdot z_\text{HGC},
  \end{eqnarray}
where $z_\text{HGC} = 156.27\,\text{cm}$ is the distance from the focal plane to the HGC mirror plane.
The coordinate axis for the HGC is the convention used in charged particle transport in dispersive magnetic systems. The $x$-axis is the direction of increasing particle momentum, the $z$-axis is the direction of particle travel through the spectrometer, and the $y$-axis is deduced from $z \times x$. 
Additionally, timing cuts are applied to the HGC data, collected using the high resolution pulse time setting in the FADC250's FPGA \cite{JLABF250}. The pulse time reconstruction in firmware corresponds to the time at which a pulse reaches half of its maximum amplitude after passing a threshold of 5\,mV. Lastly, a cut on particle velocity, $\beta$, is applied.


\begin{figure}[htbp]
  \centering
  \includegraphics[scale=0.5]{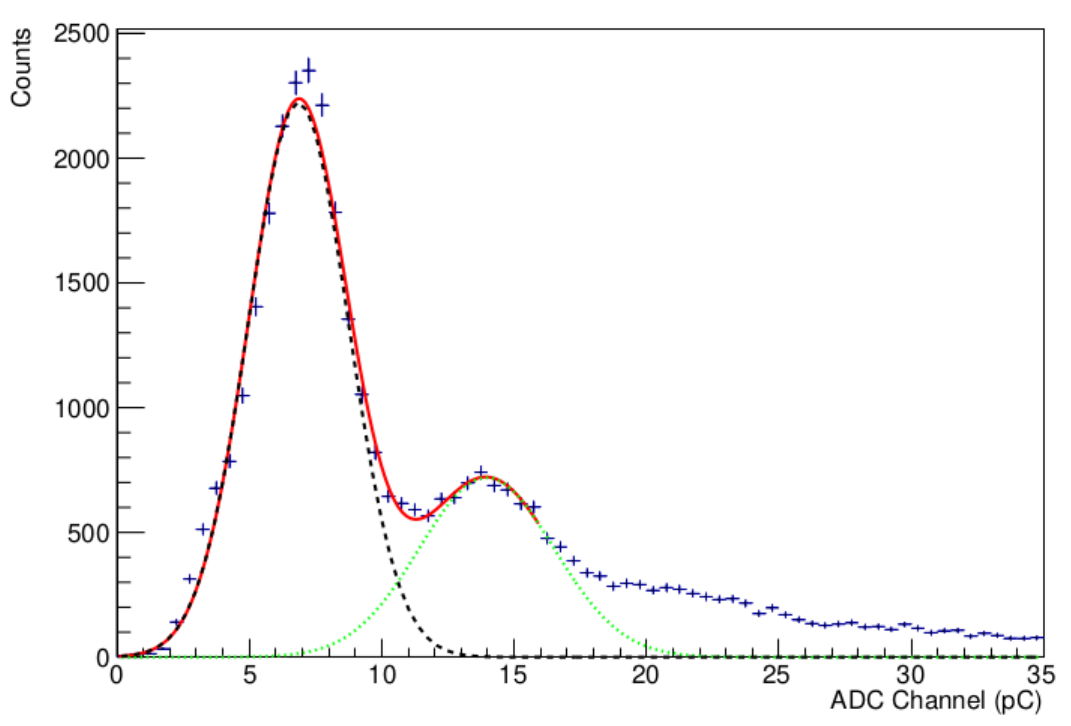}
  \caption{
  The single (dashed black) and two (dotted green) photoelectron peaks for the lower right PMT \#2, and their sum (solid red), obtained by selecting tracks in the upper right quadrant \#4. Three such adjacent mirror plots are obtained for each PMT. The light from the mirror closest to the PMT is far more intense, with too few SPE events available to yield a reliable calibration.}
  \label{fig:Calib}
\end{figure}

\begin{figure}[htbp]
  \centering
  \includegraphics[scale=0.5]{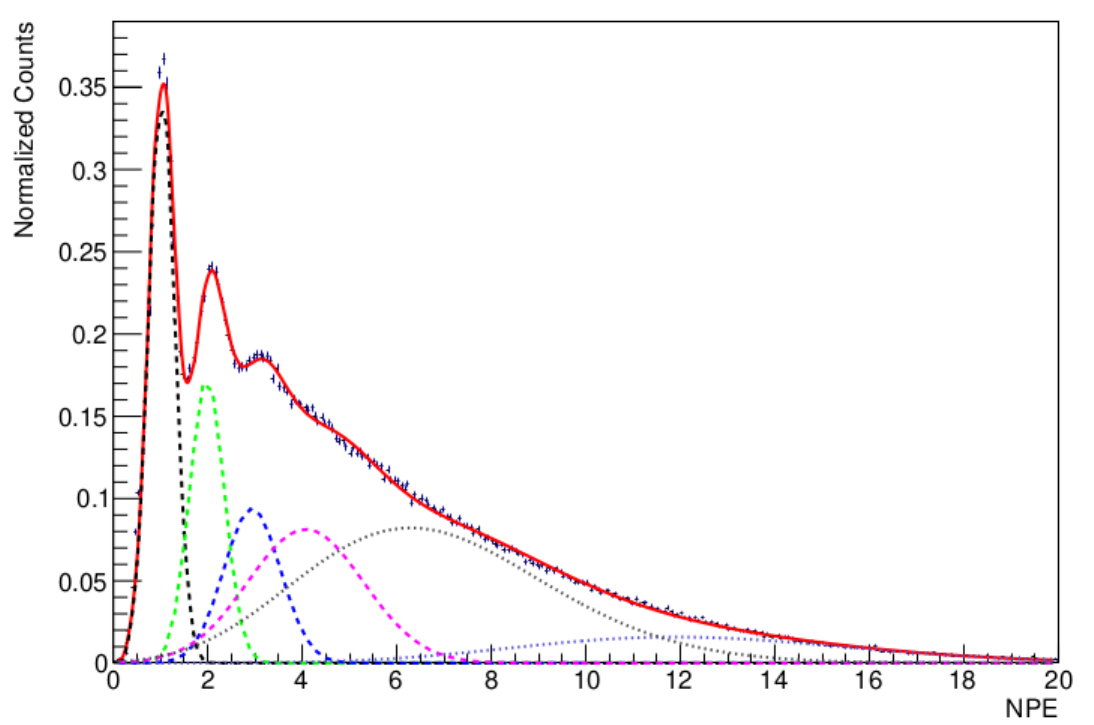}
  \caption{Results from a successful calibration of the HGC. Shown is the NPE distribution of the lower right PMT \#2 obtained from all four mirrors. The 1, 2, and 3 NPE peaks are shown, indicated by dashed Gaussian distributions. Two Poisson distributions (dotted lines) provide a good description of the nearest mirror events with large NPE, and a broad Gaussian near 4 NPE fills in the gap with the lower NPE peaks. The sum of all 6 distributions is shown as the solid red curve.}
  \label{fig:Poisson}
\end{figure}

An example of a completed calibration is shown in
Figs.~\ref{fig:Calib}, \ref{fig:Poisson}. For this run, the HGC was filled with C$_4$F$_{10}$ at 1\,atm, and the SHMS central momentum was 2.583\,GeV/c with polarity set to detect positively-charged particles. Cherenkov radiation is produced by $\pi^+$ traversing the HGC with momentum $>2.598$\,GeV/c. This can occur only for $\delta>+0.5\%$, which corresponds roughly to the bottom half of the HGC. Subthreshold $\pi^+$ with $\delta<+0.5\%$, as well as $K^+$ and $p$, may produce low-level light in the HGC via knock-on electron emission and scintillation in the radiator gas. The adjacent mirror cuts described above produce a clear SPE peak in Fig.~\ref{fig:Calib}.

A histogram of light collected in one PMT from all four mirrors is shown in Fig.~\ref{fig:Poisson}, where the average number of photoelectrons detected per event is higher due to the more intense light from the closest mirror. In this figure, the spectrum is fit with a sum of four Gaussian and two Poisson distributions, shown by the solid red line.


An inherent systematic uncertainty is present in the HGC calibration due to statistical errors in determining the location of the SPE peak in the various mirror quadrants. This uncertainty was quantified by recording the locations of the SPE across several runs,  for the different adjacent mirror combinations for each PMT, as well as by varying the contribution of the higher PE tail extending underneath the SPE peak, as in Figs.~\ref{fig:Calib}, \ref{fig:Poisson}. The systematic uncertainty in the calibration is taken to be the root mean square of this set of values, giving $\pm 1.5$\%. It should be noted this uncertainty is somewhat larger than the statistical uncertainty of the SPE peak, which is typically 0.2 to 0.6\%.

\subsubsection{Gain Matching}
\label{subsec:HGC_Gain_Matching}
\begin{figure}[htbp]
  \centering
  \includegraphics[scale=0.4]{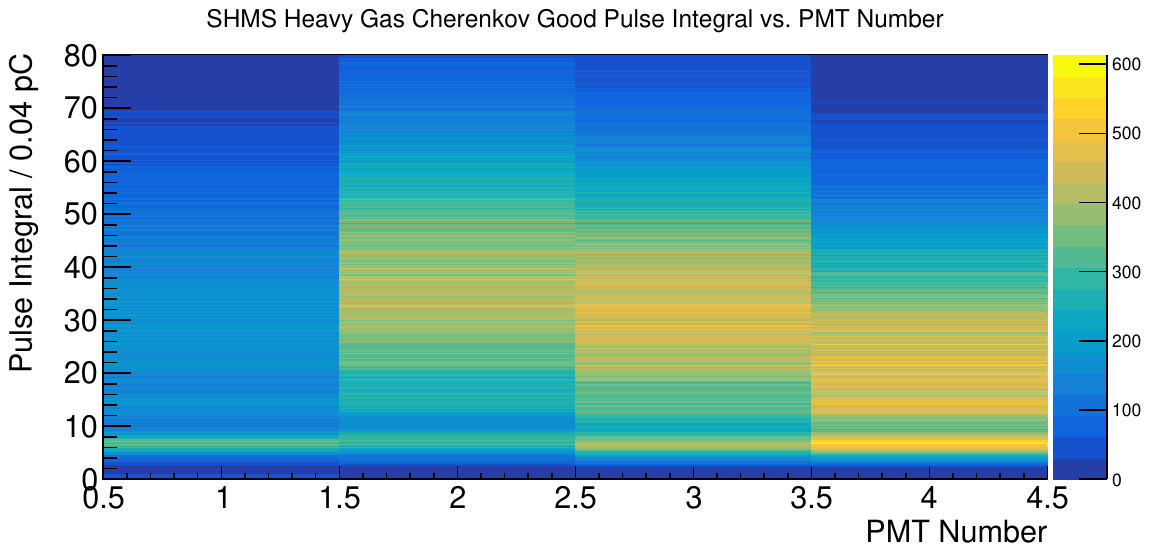}
  \caption{Demonstration of gain matching between PMTs by the
  alignment of the single photoelectron, indicated by the yellow band about 6.825\,pC. The horizontal axis refers to PMT number, the vertical axis to Pulse Integral in bins of 0.04\,pC. The color axis represents the number of events filling each bin.}
  \label{fig:gain}
\end{figure}

To ensure each PMT has a similar response to incident light, the voltages of each PMT were adjusted to obtain accurate gain matching. This can be seen in Fig.\,\ref{fig:gain} by the alignment of the SPE at approximately 6.825\,pC, represented by the common band across all four PMTs at that value. Additionally, the gain of each PMT was tested by the manufacturer, Hamamatsu, and at Jefferson Lab. The results of each test are shown in Table~\ref{table:GainCompare}.
\begin{table}[b] 
  \begin{center}
    \begin{tabular}{|c|c|c|} \hline
      PMT & JLab Gain & Hamamatsu Gain \\\hline
      PMT 1 & $(2.79\pm0.01)\times10^7$& $0.969\times10^7$\\\hline
      PMT 2 & $(6.55\pm0.04)\times10^7$& $3.60\times10^7$\\\hline
      PMT 3 & $(7.12\pm0.05)\times10^7$& $5.79\times10^7$\\\hline
      PMT 4 & $(5.35\pm0.04)\times10^7$& $3.20\times10^7$\\\hline
    \end{tabular} 
    \caption[Gain Comparison]{Gain characteristics for the PMTs in the HGC. Two measurements were performed, one at Jefferson Lab in an experimental setting, and one by the manufacturer Hamamatsu. The set voltage for the gain measurements is 2000 V for each PMT.}
    \label{table:GainCompare}
  \end{center}
\end{table}
The Hamamatsu data were taken directly at 2000\,V in a highly controlled environment, thus leading to small uncertainty in the gain which was not quoted. The Jefferson Lab measurements were also taken at 2000\,V, but in an experimental environment. This gives rise to an uncertainty in the JLab gain data on the order of 1\%, larger than the Hamamatsu data.


%

\subsection{Noble-Gas Cherenkov Counter}
\label{sec:ngcerenk}

\subsubsection{Design and Construction}

Analyzing momenta up to 11\,GeV/c at scattering angles from 5.5$^{\circ}$ to 40.0$^{\circ}$, the SHMS will encounter pion background rates which exceed the scattered electron signal rate by more than 1000:1. The suppression of these anticipated pion backgrounds while maintaining high electron identification efficiency is one of the main duties of the SHMS PID detectors, including the Noble Gas Cherenkov (NGC). 
The critical role for the NGC arises 
for momenta between 6 GeV/c and 11 GeV/c. This momentum range is challenging because it not only requires a gas with a low index of refraction so that pions will be below Cherenkov threshold, it requires that the radiator be quite long to obtain enough photoelectrons for efficient electron identification. 
Operating at 1~atm, the NGC will use a mixture of argon and neon as the radiator: pure argon with an index of refraction $n$=1.00028201 for a momentum of 6\,GeV/c, pure neon with an index of refraction $n$=1.000066102 at 11\,GeV/c, and a mixture of argon and neon at intermediate momenta. 

The SHMS NGC design was constrained by the available space and the need to have good discrimination at the highest momenta. The number of photoelectrons is maximized in this design by the use of quartz window PMTs and mirrors with excellent reflectivity well into the UV.


The NGC consists of four main elements: 1) a light tight box with thin entrance and exit windows (hence the requirement of operation at 1~atm); 2) four spherical mirrors held in a rigid frame; 3) four 5~inch quartz window photomultipliers (PMTs) and 4) the radiator gas.

\begin{figure}[tbhp] 
\centering
\includegraphics[width=0.45\textwidth]{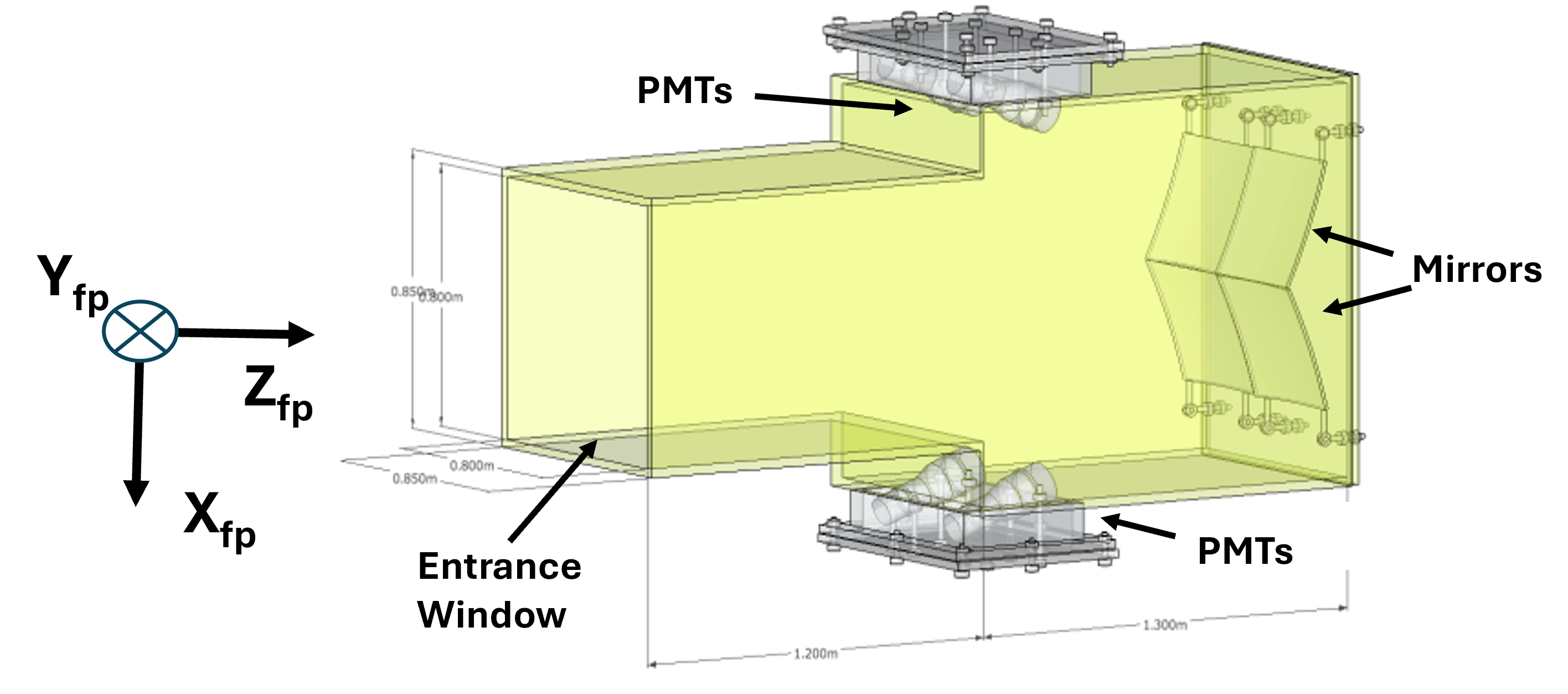} 
\caption{ Schematic of the Nobel Gas Cherenkov detector. Particles enter the
the detector through the entrance window. The mirrors focus light on the PMTs at the top and bottom.
\label{fig:NGC_tank}
}     
\end{figure}

The tank in Fig.\,\ref{fig:NGC_tank} was fabricated with an internal rigid aluminum t-slot frame and thin aluminum walls welded together. It has an active length of 2\,m along the beam direction and approximately 90\,cm perpendicular to the beam direction. The main access is provided through a large side `door', and four small panels provide modest access to the PMTs. The tank has feedthroughs for gas management as well as for HV and signal cables. The interior was coated with a black flat paint to absorb stray light from cosmic rays or hall background. Thin entrance and exit windows are made of two layers of 51~$\mu$m  Dupont Tedlar (CH$_2$CHCl)$_n$. The PMTs were positioned outside the beam envelope, achieved by a $15^\circ$ tilt of the mirrors.
 
Four spherical thin glass mirrors of radius 135\,cm, square in shape with edge lengths of 43\,cm, focus the Cherenkov light onto the PMTs. The glass blanks were manufactured by Rayotek Scientific \cite{Rayotek} from borosilicate glass of 3\,mm thickness by slumping over a polished steel mold and then cutting to dimensions. Simulations showed a reduction of collection efficiency due to incoming photon losses at the exposed edges of the mirror. As such, the edges were bevelled away from the active surface to minimize scattering from these edges.

The final batch of the glass blanks was shipped to Apex Metrology Solutions of Fort Wayne for coordinate measuring machine shape measurements. Apex's measurements were performed on a grid of 1806 points. The data were fitted with spherical, conical and elliptical functions for each mirror. Though the elliptical fit described the surface slightly better than the spherical fit, the updated simulation with the real measured parameters showed almost no difference in the collection efficiency between the two. In addition, the same fitting was performed for 5 selected locations on the mirror: entire mirror, the center, and 4 quadrants. Based on the spherical fit results, ``best'' mirrors and ``best'' corners for each mirror were identified. The 4 mirrors come together and overlap at the center of the acceptance where a majority of the scattered electrons are focused. Care was taken to locate the best corners of the best 4 glass pieces in the overlap region. The radii of the 4 best pieces of glass, from fitting, were found to never vary by more than 2\,cm from the contracted value of 135\,cm in the fit areas described above.

\begin{figure}[tbhp]
\centering
\includegraphics[width=.5\textwidth]{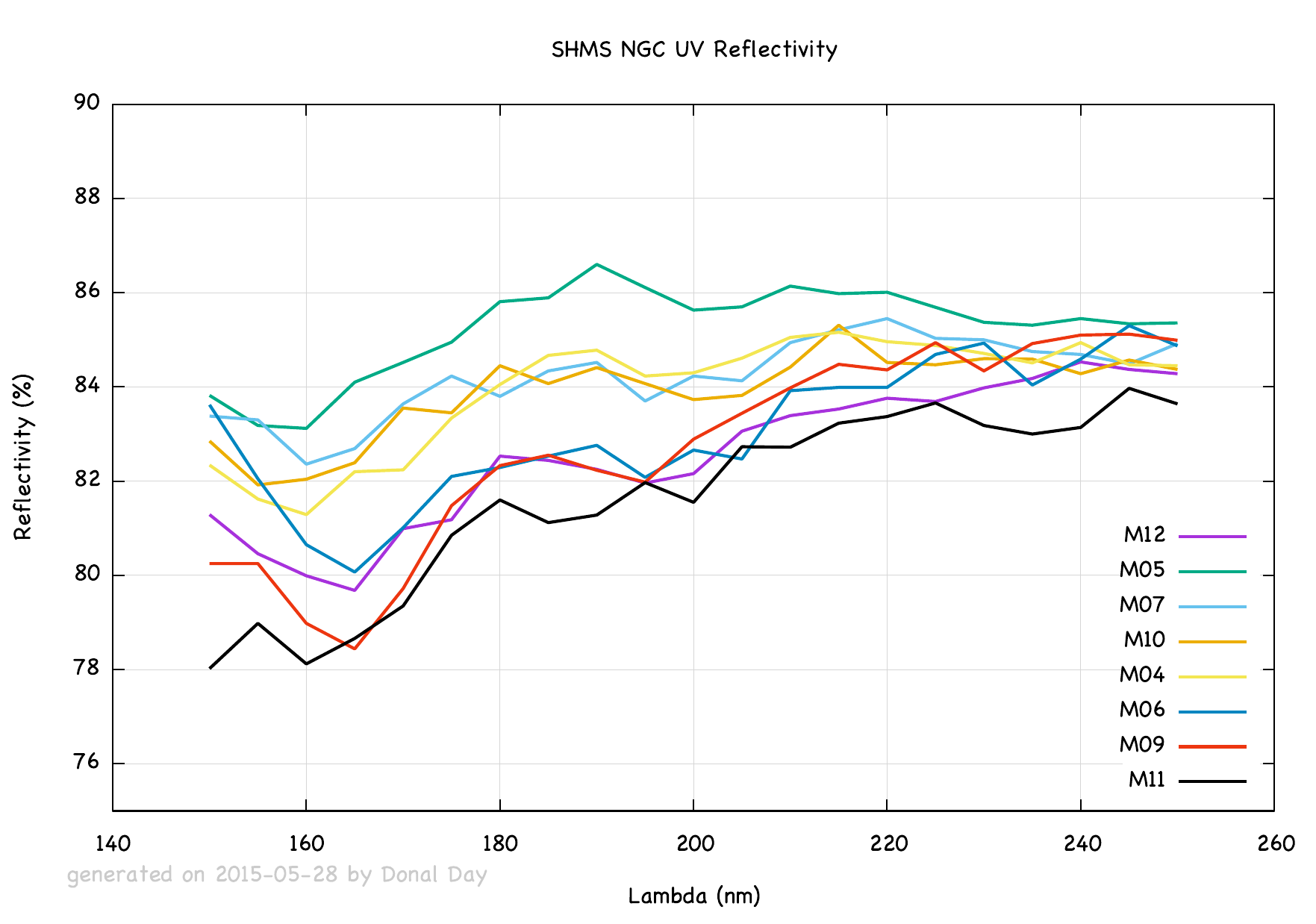} 
\caption{The UV measured reflectivity of the finished mirrors, coated at CERN which is no less than 78\% at 150\,nm. Between 250\,nm and 600\,nm the reflectivity rises to almost 90\%.\label{fig:Reflect}}
\end{figure}

The blanks were coated by the Thin Film and Glass Service of the Detector Technologies Group at CERN \cite{DTG}. The reflectivity was also measured at CERN and found to be excellent well into the UV (Fig.~\ref{fig:Reflect}).

\begin{figure}[tbhp] 
\centering
\includegraphics[width=0.35\textwidth,angle=-90]{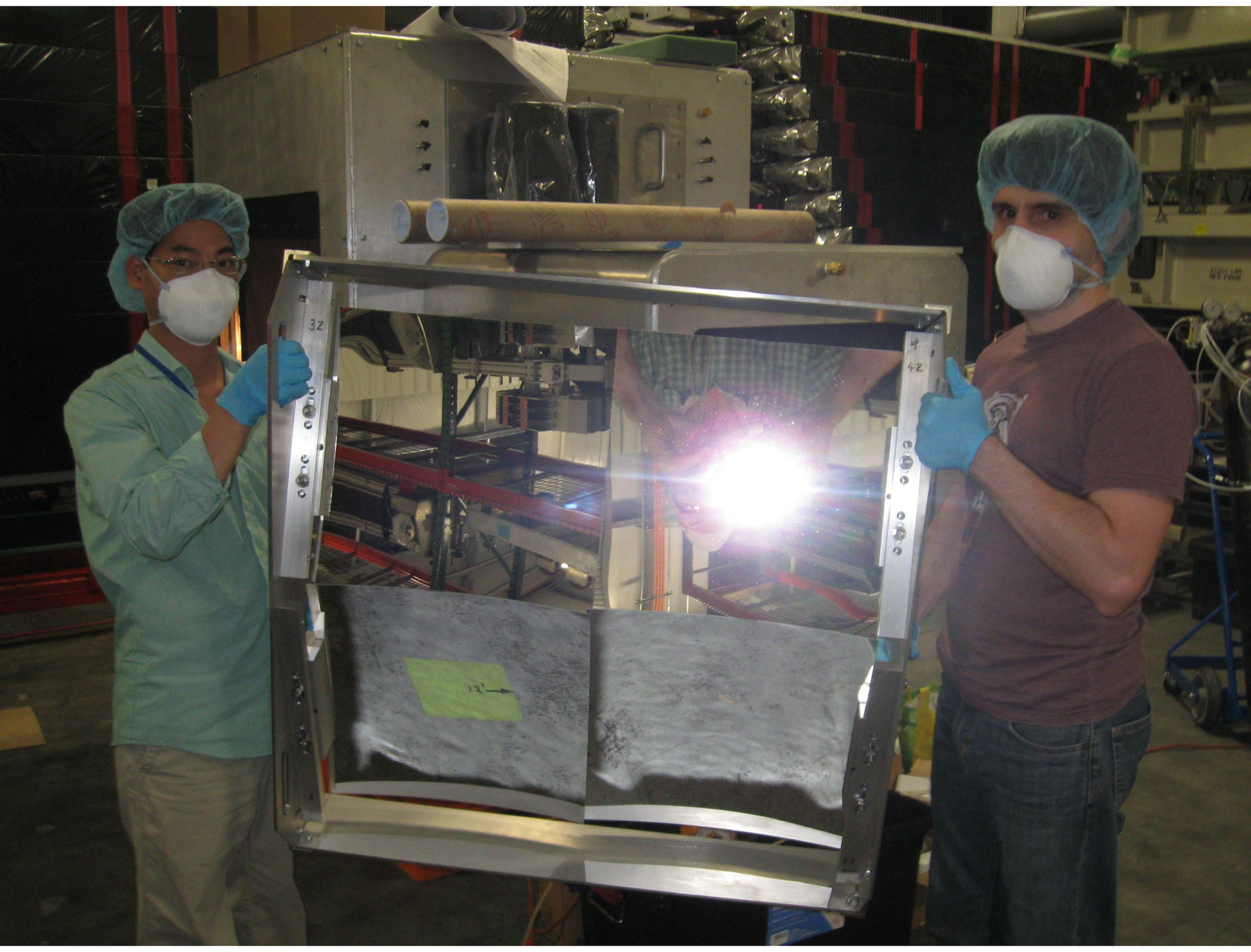} 
\caption{Frame with mirrors about to be moved into tank.\label{fig:install}}
\end{figure}

Like the HGC, the four mirrors are arranged in a 2 by 2 array with a small overlap in the center, providing full coverage over the active area. In order to accomplish this without mechanical interference, the mirrors were staggered along the tank $z$-axis which is the direction of the incoming particles.  The mirrors were mounted in a monolithic frame installed as single unit (see Figure~\ref{fig:install}).

%

The four PMTs are 14~stage 5" quartz window PMTs manufactured by Electron Tubes Enterprises \cite{ETTubes}, model 9823QKB04. The 9823QKB04 has a quantum efficiency above 5\% at 150\,nm and 30\% at 350\,nm. The tubes are surrounded by a mu-metal shield and the positive HV is distributed to the stages by a resistive divider. After evidence of corona or arcing inside one of the bases due to the relatively low dielectric breakdown strength of the gas mixture, the problem was resolved by applying silicon conformal coating to suspect areas.  
 

\subsubsection{Calibration}

As with the HGC (see Sec.~\ref{Sec:HGC_Cal}), the goal of the NGC calibration is to generate an accurate transformation from raw FADC channels to the number of photoelectrons (NPE).
The NGC calibration method can be broken down into three steps:

\begin{enumerate}
    \item Selecting an appropriate data set.
    \item Selection cuts to identify a clean electron sample for each PMT.
    \item Using the clean electron sample to fit the pulse integral distribution for each PMT. This is used to determine the calibration constants.
\end{enumerate}

\paragraph{Selecting an appropriate data set}

The NGC calibration requires electron events in the SHMS. Any data set with the SHMS running with negative polarity can in theory be utilised for calibrations. However, for best results, a data sample with an even distribution of events across all PMTs in the NGC should be utilised. Additionally, the data set should contain on the order of $\sim10^{6}$ events or more.

\paragraph{Selection Cuts}

To obtain a clean electron sample from the data, several selection cuts are applied to the data. Cuts are applied on:

\begin{itemize}
    \item $-10 \leqslant \delta \leqslant 20$, a nominal acceptance cut, removing events outside this range.
    \item $0.7 \leqslant E_{TotTrackNorm} \leqslant 2.0$, a calorimeter based PID cut using the normalized calorimeter energy to remove pion/hadron background events.
    \item NGC multiplicity and position cuts. These are used to select events where the majority of the Cherenkov light was deposited in a single PMT. 
\end{itemize}

After selection cuts, the PMTs can be calibrated.

\paragraph{Determining Calibration Constants}

\begin{figure}[thbp]
\begin{centering}
\includegraphics[width=\columnwidth]{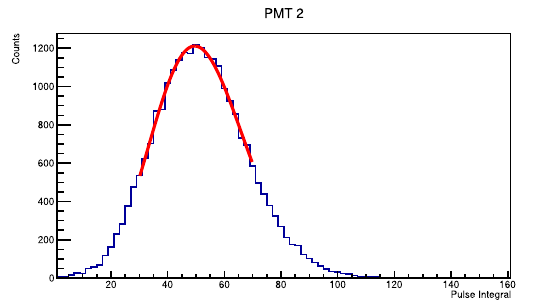}
\caption{A sample pulse integral distribution fitted with the function described by Eqn.~\ref{eqn:NGC_Cal_Fit} shown in red.\label{fig:NGC_Cal_Fit}}
\end{centering}
\end{figure}

After selection cuts, the pulse integral distributions for each of the NGC PMTs are fitted with the function 
\begin{gather}
    \label{eqn:NGC_Cal_Fit}
    f(x) = A \frac{\lambda^{\frac{x}{\mu}}e^{-\lambda}}{\Gamma\left(\frac{x}{\mu}+1\right)},
\end{gather}
where $x$ is the pulse integral in pC,  $A$ is a normalization factor to account for the number of events in the dataset being fit, $\lambda$ is the mean NPE
for an event above the Cherenkov threshold, and $\mu$ is the calibration constant in units of pC/pe.
This value is determined for each PMT. An example pulse integral distribution and the associated fit can be seen in Fig.~\ref{fig:NGC_Cal_Fit}.

The NGC PMTs were also gain matched in a similar manner to the HGC. Refer to Sec.\,\ref{subsec:HGC_Gain_Matching} for details on this procedure.


\subsection{Aerogel Cherenkov Counter}
\label{sec:aerogel}

\subsubsection{Design and Construction}

\begin{figure}[tbhp]
\begin{centering}
\includegraphics[width=0.4\textwidth]{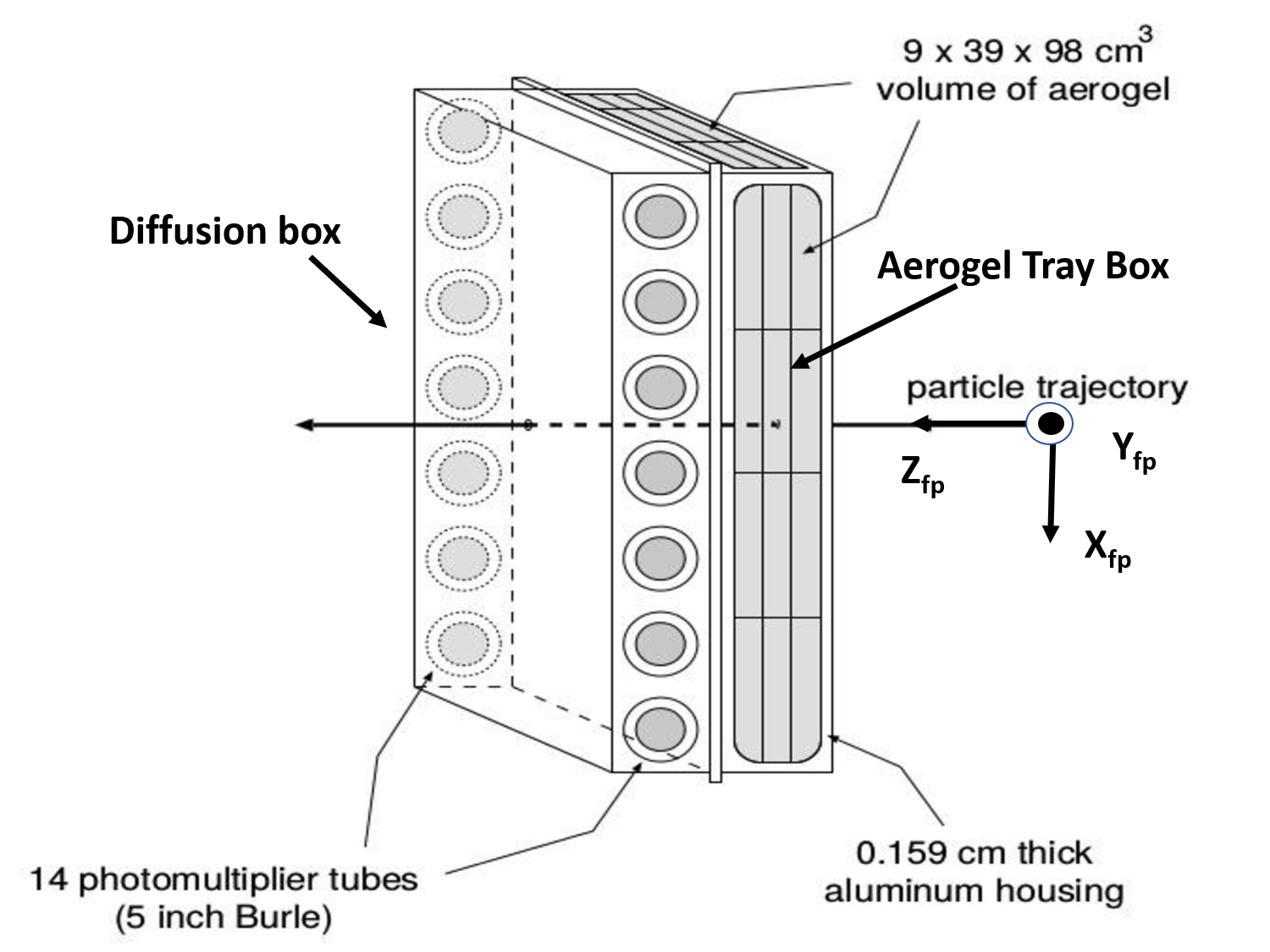}
\caption{ Schematic view of the aerogel detector with the tray which holds the aerogel tiles and the diffusion box with the PMTs. \label{fig:aerogelPhotoInStack}}
\end{centering}
\end{figure}

Fig.~\ref{fig:aerogelPhotoInStack} shows a drawing of the aerogel counter installed downstream of the cylindrical HGC in the SHMS detector stack. The detector consists of two main components: a tray which holds the aerogel material, and a light diffusion box with photomultiplier tubes (PMTs) for light readout. Four identical trays for aerogel of nominal refractive indices of 1.030, 1.020, 1.015 and 1.011 were constructed. These allow for particle identification over the wide range of momenta summarized in Table \ref{tab:aerogel-thresholds}. 
The design allows for easy detector assembly and replacement of the aerogel trays. The aerogel tray is 9,cm thickness and  the total thickness of the detector is 24.5\,cm along the optical axis of the SHMS. A detailed discussion of the detector, characterization of its components, and performance tests can be found in 
Refs.~\cite{AeroNIM17}.

\begin{table}[tbh]
\begin{center}
\caption{ Threshold momenta P$_\mathit{Th}$ in GeV/c for Cherenkov radiation for charged muons, pions, kaons, and protons in aerogel of four refractive indices ranging from $n$=1.011 to 1.030.}
\label{tab:aerogel-thresholds} \smallskip
\small
\begin{tabular}{||l|c|c|c|c|} \hline
Particle & P$_{Th}$   & P$_{Th}$    & P$_{Th}$    & P$_{Th}$    \\
         & $n$=1.030 &  $n$=1.020  & $n$=1.015  &  $n$=1.011 \\
\hline 
 $\mu$   &  0.428    &  0.526      & 0.608      &  0.711     \\
 $\pi$   &  0.565    &  0.692      & 0.803      & 0.935      \\
 $K$     &  2.000    &  2.453      & 2.840      &  3.315     \\
 $p$     &  3.802    &  4.667      & 5.379      &  6.307     \\
\hline
\hline\end{tabular}
\end{center}
\end{table}


The diffusion box is made of the aluminum alloy 6061-T6. The side panels are constructed of $\sim$2.5\,cm (1-inch) plates. The back cover is $\sim$1\,mm (1/16") thick. The inner dimensions of the box are $\sim 103\times 113\times 17.3$~cm$^3$ (40.5" $\times$ 44.5" $\times$6.82"). To optimize light collection, the inner surface of the diffusion box is lined with either 3\,mm (covering $\sim$60\% of the surface) or 1\,mm (remaining $\sim$40\% of the surface) thick GORE diffuse reflector material~\cite{Gore}. This material has a reflectivity of about 99\% over the entire spectrum.

The light collection is handled by 5" diameter photomultiplier tubes (XP4500). The 5.56" (14.1\,cm) diameter cylindrical housings holding the PMTs are mounted upon 14 waterjet cut circular openings on the left and right (long) sides of the diffusion box, with minimum spacing of 14.92\,cm (5.875") between the centers. The PMTs are sealed into their housing using a light-tight synthetic rubber material (Momentive RTV103 Black Silicone Sealant) and the whole assembly is sealed light-tight. 

The trays contain two different types of aerogel. The aerogel tiles of refractive indices $n$=1.030 and $n$=1.020 were manufactured by Matsushita Electric Works, Ltd prior to 2010. The tiles of refractive indices $n$=1.015 and $n$=1.011 were manufactured by the Japan Fine Ceramics Center between 2010 and 2013. All tiles have dimensions of approximately 11\,cm by 11\,cm by 1.1\,cm and are hydrophobic. The depth of the aerogel radiator in their trays is on average $\sim$9\,cm thick (8 layers). For the SP-30, SP-20 and SP-15 trays the aerogel covers and area of 110\,cm x 100\,cm. In the SP-11 aerogel tray the radiator covers an active area of 90\,cm x 60\,cm. To improve the reflectivity, and thus light collection, inside the trays for the lowest two refractive indices 1\,mm thick GORE diffusive reflector material (DRP-1.0-12x30-PSA) with reflectivity of about 99\% was used. The 
SP-30 and SP-20 aerogel trays were covered with 0.45\,$\mu$m thick Millipore paper membrane GSWP-0010 (Millipore). 

Based on prior experience all aerogel trays originally featured a net of stainless steel wires close to the aerogel surface to minimize damage during handling and installation. This, however, proved insufficient and dangerous to the aerogel tiles as discussed below. Since 2022 the two lowest refractive index aerogel trays feature an additional net of mylar strips holding the aerogel tiles in place and protecting them from being damaged by the wire mesh.

\subsubsection{Performance aspects}

\begin{figure}[tbhp]
\begin{centering}
\includegraphics[width=\textwidth]{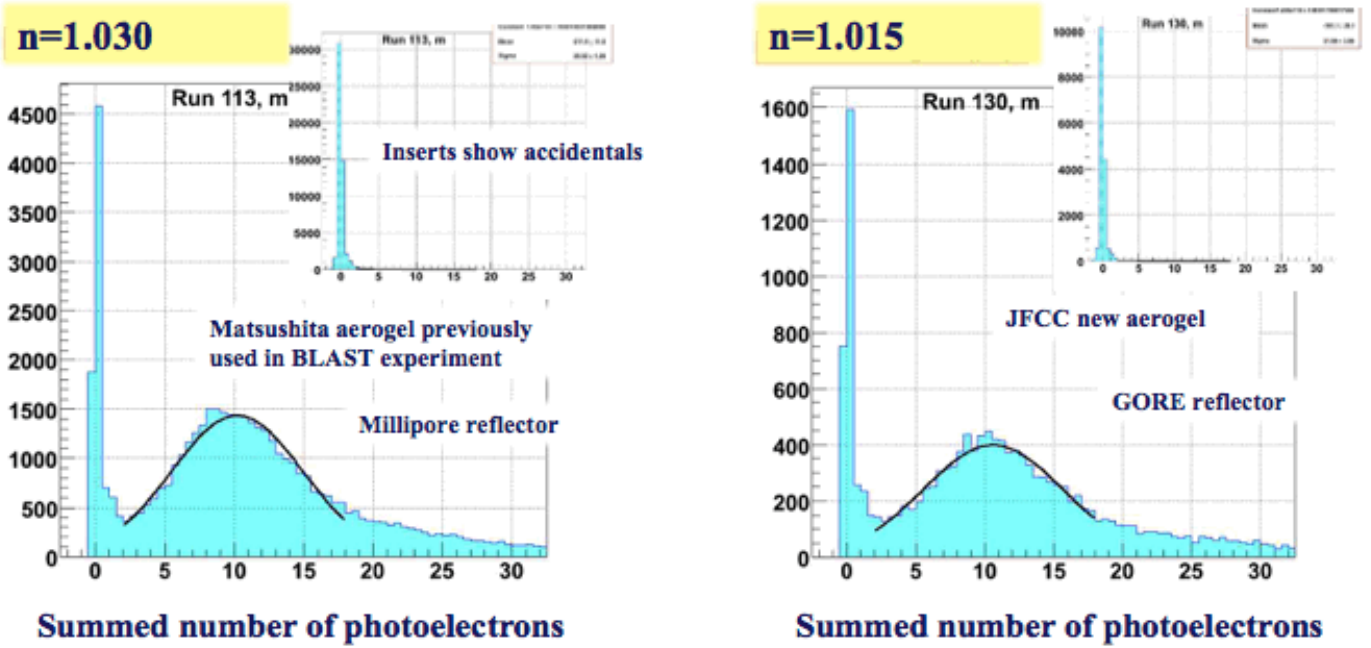}
\caption{Numbers of photoelectrons observed in the Aerogel Cherenkov. \label{fig:aerogelNPE}}
\end{centering}
\end{figure}

The light collection performance of the detector was tested with cosmic rays and electron beam. The detector signal showed good uniformity along the vertical coordinate of the detector surface, but had a significant dependence in the horizontal direction. Mitigation of this included a position-dependent threshold and an optimized selection of the PMTs installed on the right and left side of the detector. The response of the detector to particles is shown in Fig.~\ref{fig:aerogelNPE}. 

The mean number of photo-electrons in saturation for the tray filled with $n$=1.030 ($n$=1.020) refractive index aerogel is $\sim$10 ($\sim$8), close to expectation from Monte Carlo simulation. For the trays filled with $n$=1.015 and $n$=1.011 refractive index aerogel, high numbers of photoelectrons were obtained with the use of higher reflectivity GORE material to cover the tray, $\sim$10 and $\sim$5.5 respectively. This result could be fully reproduced by our Monte Carlo simulation with the assumption of an aerogel absorption length of order 220\,cm.

\subsubsection{Results from Aerogel Detector Operations}

The SHMS aerogel detector was installed in the SHMS in 2015 and since then has been used by experiments requiring pion/proton and kaon/proton separation. At particle momenta of 3.319 GeV/c (5.389 GeV/c) clear pion/proton (kaon/proton) separation was achieved with the SP-15 (SP-11) aerogel tray. Many of these experiments required on the order of tens of aerogel tray exchanges over the course of the experiment.

After the first round of experiments, an inspection of the aerogel trays revealed noticeable shifts in the aerogel tile stacks. This was traced back to the instrumentation, the overhead crane, used for the installation/exchanges of the aerogel trays. The wire mesh kept the shifted tiles from falling out of the tray, but damaged the surface of tiles that were in direct contact. To prevent further shifts of the aerogel tiles, which in the best case would result in non-uniform aerogel thickness and in the worst case total loss of the aerogel, an additional mylar strip grid was installed in the SP-15 and SP-11 trays. This solution has performed well since 2022, but it does require occasional re-tightening of the mylar strips depending on usage.


\subsection{Preshower and Shower Counters}
\label{sec:shower}

Broadly speaking, the approved experiments demand a suppression of pion background for electron/hadron separation of 1000:1, with suppression in the electromagnetic calorimeter alone on the level of 100:1. An experiment to measure the pion form factor at the highest accessible $Q^2$ at JLab with an 11\,GeV beam requires an even stronger suppression of electrons against negative pions of a few 1,000:1, with a requirement on the electromagnetic calorimeter of a 200:1 suppression.

Particle detection using electromagnetic calorimeters is based on the production of electromagnetic showers in a material. The total amount of the light radiated is proportional to the energy deposited in the medium. Electrons (as well as positrons and photons) will deposit their entire energy in the calorimeter. Thus, $E/p$ for electron showers is close to 1, with hadrons typically giving a significantly smaller value. 

Charged hadrons entering a calorimeter have a lower probability to interact and produce a shower, and may pass through without interaction. In this case, they will deposit a constant amount of energy in the calorimeter. However, they may undergo nuclear interactions in the radiator (in our case lead-glass) and produce particle showers.
Hadrons that interact near the front surface of the calorimeter and transfer a sufficiently large fraction of their energy to neutral pions will mimic electrons. The maximum attainable electron/hadron rejection factor is limited mainly by the cross section of such interactions. 
At the cost of a small loss of electron efficiency, the tendency of electrons to deposit significant energy in the initial preshower layer can be used to improve hadron rejection further.  

A nearly complete description of the design, pre-assembly component checkout, and construction of the SHMS calorimeter can be found in Ref.~\cite{mkrtchyan}. This section provides a brief summary. 


\subsubsection{Design and Construction}
\label{calo_construct}
%


As a full absorption detector, the SHMS calorimeter is situated at the very end of detector stack of the spectrometer. The relatively large beam envelope of the SHMS dictated a design of a wide acceptance coverage. The general requirements for the SHMS calorimeter were: 
\begin{itemize}
    \item Effective area: ${\sim}120 \times 140\,{\rm cm}^2$.
    \item Total thickness: $\sim$20\,rad. length.
    \item Dynamic range: 1.0 - 11.0\,GeV/c.
    \item Energy resolution: $\sim 6\%/\sqrt E $, $E$ in GeV. 
    \item Pion rejection: $\sim$100:1 at $P\gtrsim$1.5-2.0\,GeV/c.
    \item Electron detection efficiency: $>98\%$.
\end{itemize}

\begin{figure}[hbtp]
\begin{centering}
\includegraphics[width=\linewidth,height=2.5in,keepaspectratio]{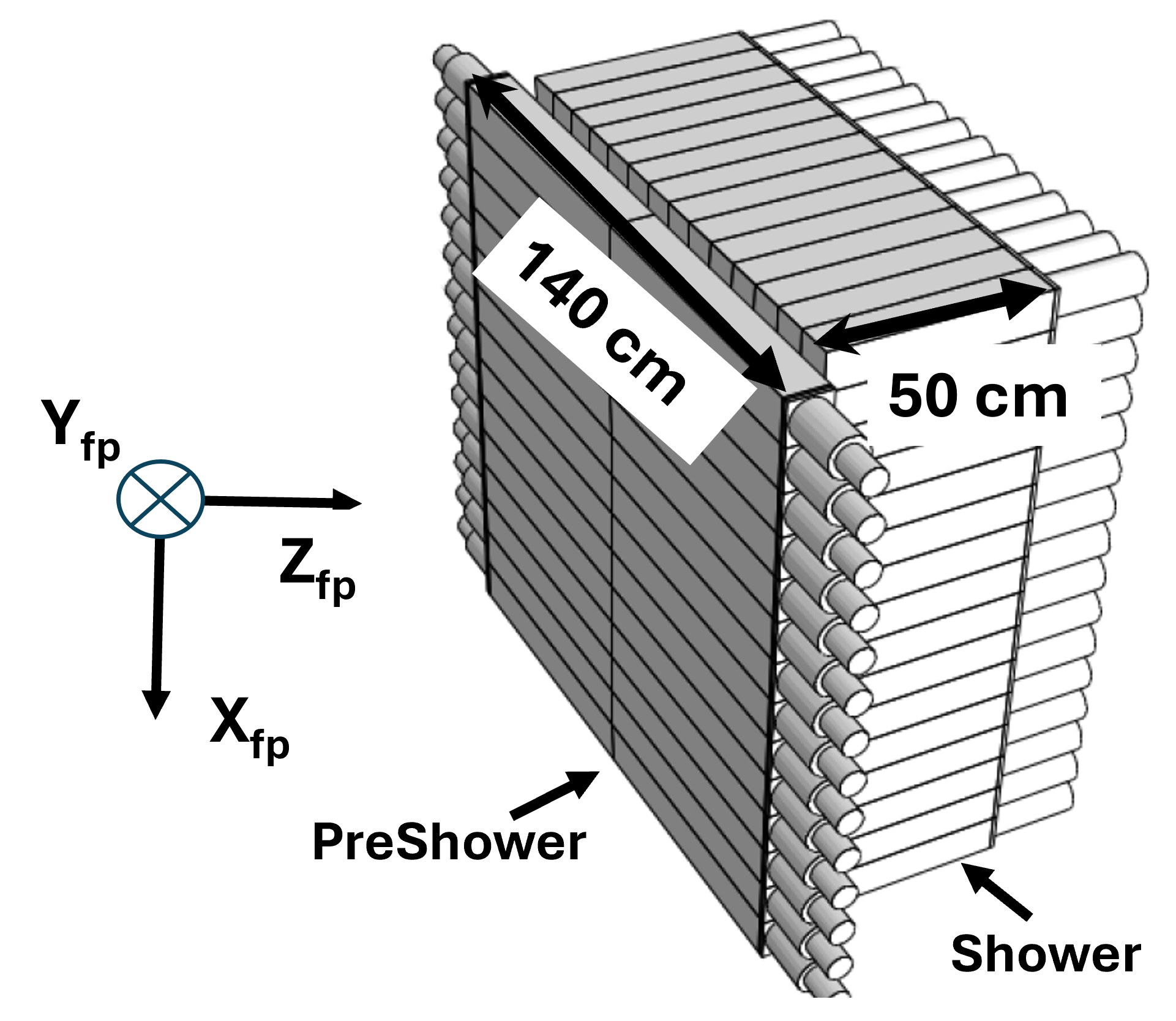}
\caption{A sketch of SHMS calorimeter. Particles enter obliquely from the left. Electron showers initiating in the transverse blocks of the Preshower are largely absorbed in the longitudinal blocks of the Shower. Support structures are omitted.}
\label{fig:shms_calo_sk}
\end{centering}
\end{figure}

The SHMS calorimeter consists of two parts (see Fig.~\ref{fig:shms_calo_sk}): a Preshower at the front of the calorimeter,
and the main part, the Shower, at the rear.
An expedient and cost-effective choice was to use modules from the decommissioned HERMES calorimeter \cite{Avakian:1998bz} for the Shower part, and modules from the decommissioned SOS calorimeter \cite{mkrtchyan} for the Preshower. 
With these choices, the Preshower is 3.6 radiation lengths thick, and the Shower is 18.2 radiation lengths deep. The combination almost entirely absorbs showers from $\sim$10\, GeV electrons.  The SHMS Preshower consists of one layer of 28 TF-1 type lead glass blocks stacked in two columns in an aluminum enclosure (not shown in Fig.~\ref{fig:shms_calo_sk}). Each Preshower block is an $10\times10\times70\,{\rm cm}^3$ block of TF-1 lead-glass with a Photonis XP3462B PMT attached.
The Shower consists of 224 TF-101 type lead glass modules stacked in a ``fly's eye'' configuration of 14 columns and 16 rows. Each Shower block is has a size of $8.9\times8.9\times50\,{\rm cm}^3$ and is optically isolated with a Photonis XP3461 PMT attached.

\subsubsection{Choice and studies of PMT bases}
\label{xp3462-xp3461}

The Preshower PMT high voltage base is optimized for good linearity (better than 1\%), high rate capability and a weak variation of PMT gain with anode current~\cite{Amatuni96}.
A design was selected which is a purely resistive, high current (2.3\,mA at 1.5\,kV), surface mounted divider ($\sim0.640\,M\Omega$), operating at negative HV. The relative applied HVs down the dynode chain (from cathode to anode) are: 3.12/1.50/1.25/1.25/1.50/1.75/2.00/2.75/2.75. The supply voltage for a gain of $10^6$ is approximately 1750\,V.

The PMT resistive base assembly is linear to within $\sim 2\%$ up to the peak anode current of 120\,$\mu$A ($\sim 5\times 10^4$ pe). The dark current is typically less than 3\,nA. The base has anode and dynode output signals.

\subsubsection{Monte Carlo simulations}
\label{calo_simulations}

Prior to construction, the calorimeter design was simulated in order to optimize the setup and get predictions for key characteristics. The simulations were based on the GEANT4 package \cite{geant4}, release 9.2. As in the simulations of the HMS calorimeter (see \cite{mkrtchyan}), the QGSP\_BERT physics list was chosen to model hadron interactions \cite{qgsp-bert}. The code closely followed the parameters of the detector components. Other features are added into the model to make it more realistic, such as:

\begin{itemize}
    \item Light attenuation length in the lead glasses and its block to block variation according to our measurements.
    \item PMT quantum efficiencies from the graphs provided by vendor.
    \item Passive material between the spectrometer focal plane and the calorimeter. 
    \item Sampling of incoming particles at the focal plane of the spectrometer.
\end{itemize}

The Cherenkov light propagation and detection were handled by a custom code, using an approximation of strict rectangular geometry of the lead glass blocks with perfectly polished surfaces. Light reflection and absorption by the Mylar wrapping was modeled via aluminum complex refractive index, with Mylar support facing the block, and a thin air gap between the wrapping and the block. Both light passage to the PMT photocathode through the optical grease and the PMT window, and reflections from the block sides, were modeled using the approximation of thin dielectric layers (\cite{BornWolf}, p. 360). The electronic effects, such as pedestal widths and channel to channel PMT gain variations, were assumed as for the HMS calorimeter before the 12\,GeV modifications.

The simulations revealed no flaws in the design construction of the SHMS calorimeter, and performance similar to other lead glass calorimeters. Studies indicated gain in pion suppression by a factor of several times after combining signals from the Preshower with the total energy deposition in the calorimeter.

\subsubsection{Calorimeter Gain Matching}
\label{calo_gain_match}

Approximate gain matching of calorimeter PMTs is important for uniformity of the pre-trigger efficiency and energy resolution over the spectrometer's acceptance. The gain matching was done in two steps.

\begin{figure}[hbtp]
\begin{centering}
\includegraphics[width=0.7\columnwidth]{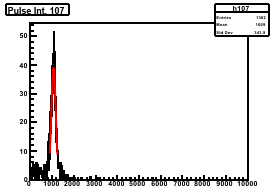}
\caption{Distribution of ADC signals of a Shower module from minimum ionizing pions. The red line is a Gaussian fit to the MIP peak.}
\label{fig:shms_calo_mip_signal}
\end{centering}
\end{figure}

In the first step, MIP signals from pions were used. MIP pion candidates for the Shower gain matching were selected by requesting 
signals from the Heavy Gas Cherenkov with fewer than 2 p.e., and with normalized energy deposited in the Preshower close to the MIP peak value, within a range from 0.02 to 0.15. Even higher MIP purity in the Shower itself was ensured by selecting single block events.
The resultant MIP peaks in the ADC signal distributions were fitted by Gaussians
as in Fig.~\ref{fig:shms_calo_mip_signal}.

\begin{figure}[hbtp]
\begin{centering}
\includegraphics[width=\columnwidth]{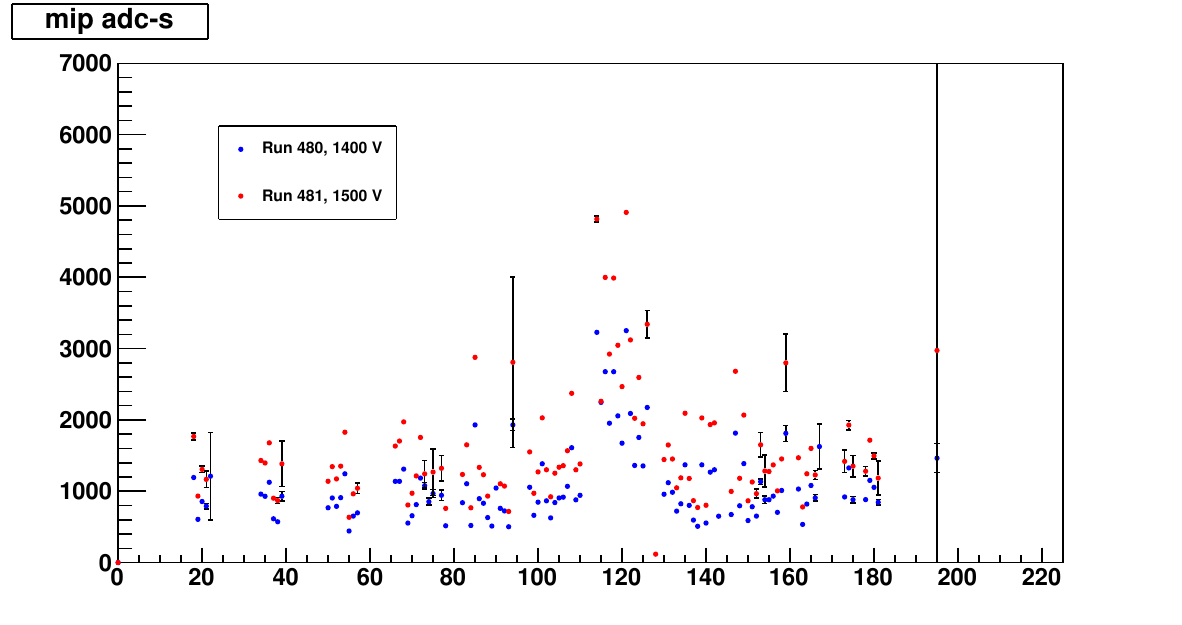}
\caption{Amplitudes of ADC signals from MIP pions in a set of Shower channels, for supply voltages of 1.4\,kV and 1.5\,kV.}
\label{fig:shms_calo_mip_positions}
\end{centering}
\end{figure}

As gain matching had to be achieved by adjustment of high voltages on the PMT bases, knowledge of gain versus supplied HV was needed. These were obtained by measuring signals from MIP pions at two supply high voltages for all the Shower channels, at 1.4\,kV and 1.5\,kV (see Fig.~\ref{fig:shms_calo_mip_positions}). By assuming gain dependence on the supplied voltage of proportional to $V^{\alpha}$ \cite{hamamatsu}, the average exponent $\alpha$ was found to be 5.70 $\pm$ 0.01 for a subset of $\sim$100 channels.

From a reference run with supply voltages $V_{REF}$ = 1.4\,kV in all the Shower channels, MIP ADC signal amplitudes $A_{REF}(i)$ were obtained.
For a desired final MIP signal amplitude of $A_{SET} = 1000$ ADC channels, the final set voltages $V_{SET}(i)$ were estimated via

\begin{equation}
\label{eq:calo-mip-vset1}
{V_{SET}(i) = V_{REF} \cdot \left( \frac{A_{SET}}{A_{REF}(i)} \right) ^{1/\alpha}} .
\end{equation}

Gain matching using MIPS alone may be biased since they emit Cherenkov light uniformly along their path, whereas electron showers deposit most of their energy in the front of the shower blocks. Furthermore, typical energies in MIP calibrations were only a few \% of full scale.  
So in a second step, a consistency check with electrons in the SHMS was done.  

The SHMS optics were set at 3\,GeV/c central momentum, but in a defocused mode, which allowed for illumination and calibration of more than 150 Shower modules. For deposited energy $E_{Dep}$ in a given module with PMT gain $g$, and  signal amplitude $g E_{pe}$, then the calibration constant $c$ is defined by $E_{dep} = c g  E_{pe}$. To the extent that all blocks have gone through rigorous quality control and have similar light transmission, and all PMTs have similar quantum efficiencies and gain vs HV performance, then $E_{pe}/E_{Dep} = (cg)^{-1}$ is approximately the same for all channels and 
the following relationship between set voltages and fitted calibration constants should roughly hold: 
\begin{equation}
\label{eq:calo-mip-vset2}
{ V_{SET}(i) = V_{REF}  \left( \frac{c_{REF}(i)}{c_{SET}}  \right) ^{1/\alpha}  } .
\end{equation}

\begin{figure}[hbtp]
\begin{centering}
\includegraphics[width=\columnwidth]{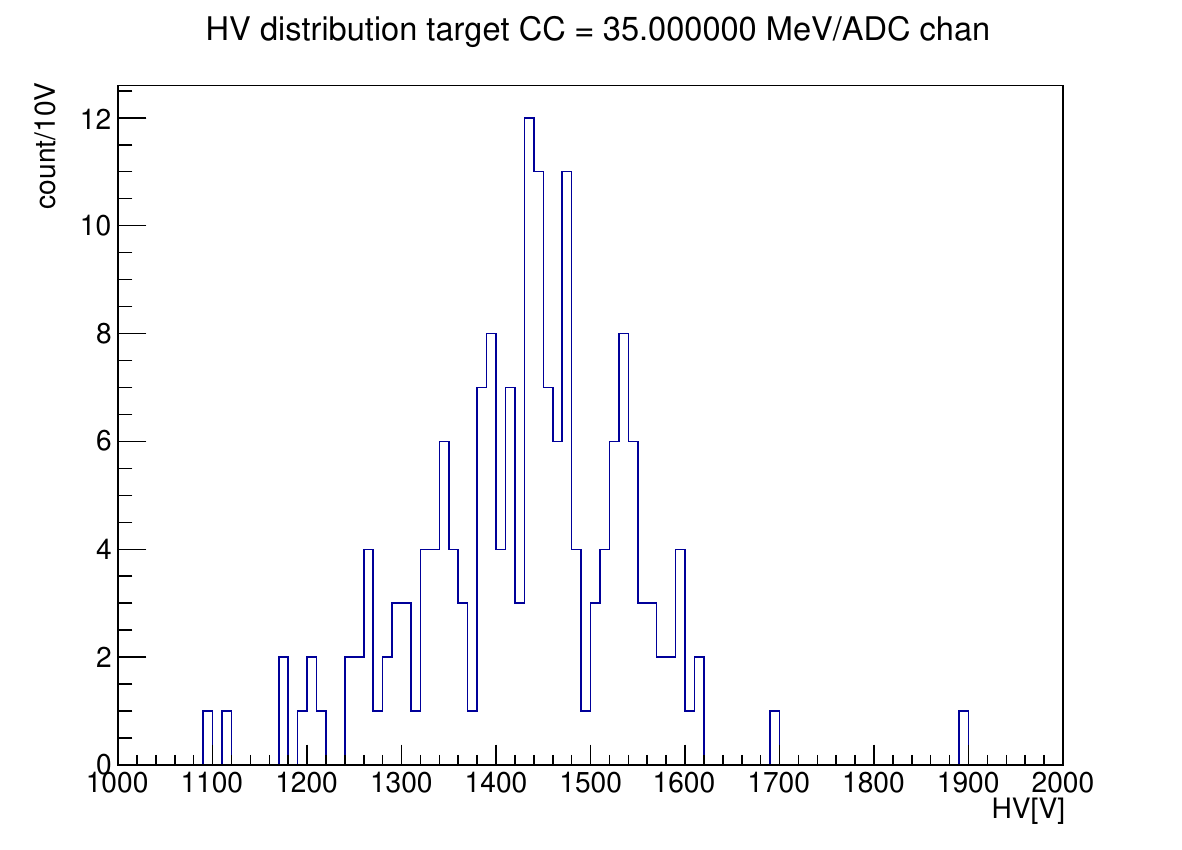}
\caption{Gain matched high voltage settings for the Shower PMTs
(see text for details).}
\label{fig:shms_calo_hv_distrib}
\end{centering}
\end{figure}

The HV settings from the second method, for $c_{SET}$ = 35 MeV/ADC~ch are within the range from 1.2\,kV to 1.6\,kV, grouped around 1.4\,kV (Fig.~\ref{fig:shms_calo_hv_distrib}). A few settings above the hard limit of 1.7\,kV were set to the limit. The HV settings from the two methods are consistent.

Note that for blocks out of the SHMS acceptance, and hence not gain matched, the HV was left at the nominal 1.4\,kV. All chosen voltages were conservative, lower than the HV settings at which modules were operated in the HERMES calorimeter.

\begin{figure}[hbtp]
\begin{centering}
\includegraphics[width=\linewidth]{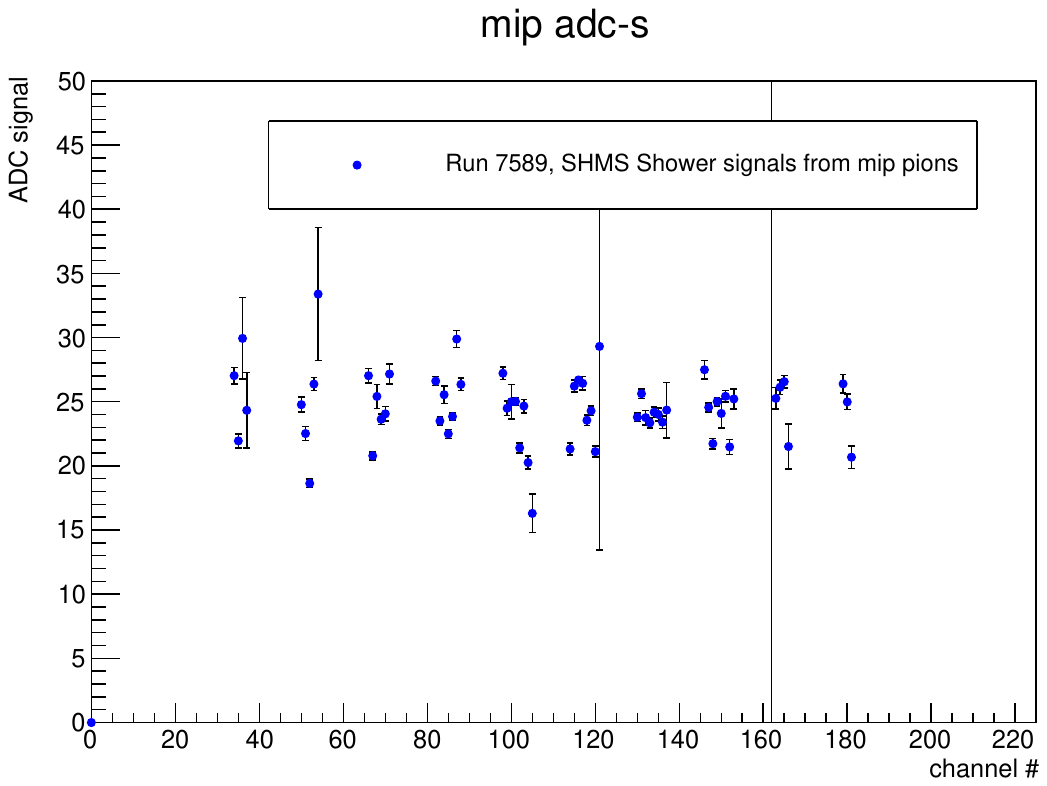}
\caption{Amplitudes of ADC signals from MIP pions in a set of Shower channels after gain matching.}
\label{fig:shms_calo_mip_amp}
\end{centering}
\end{figure}

The amplitudes of ADC signals from MIP pions after the gain matching are shown in Fig.~\ref{fig:shms_calo_mip_amp}. The majority of amplitudes are grouped between 20 and 30 ADC channels. The spread in signals among hit channels is much less than in the case of constant supply voltages shown in Fig.~\ref{fig:shms_calo_mip_positions}.

\begin{figure}[hbtp]
\begin{centering}
\includegraphics[width=\linewidth]{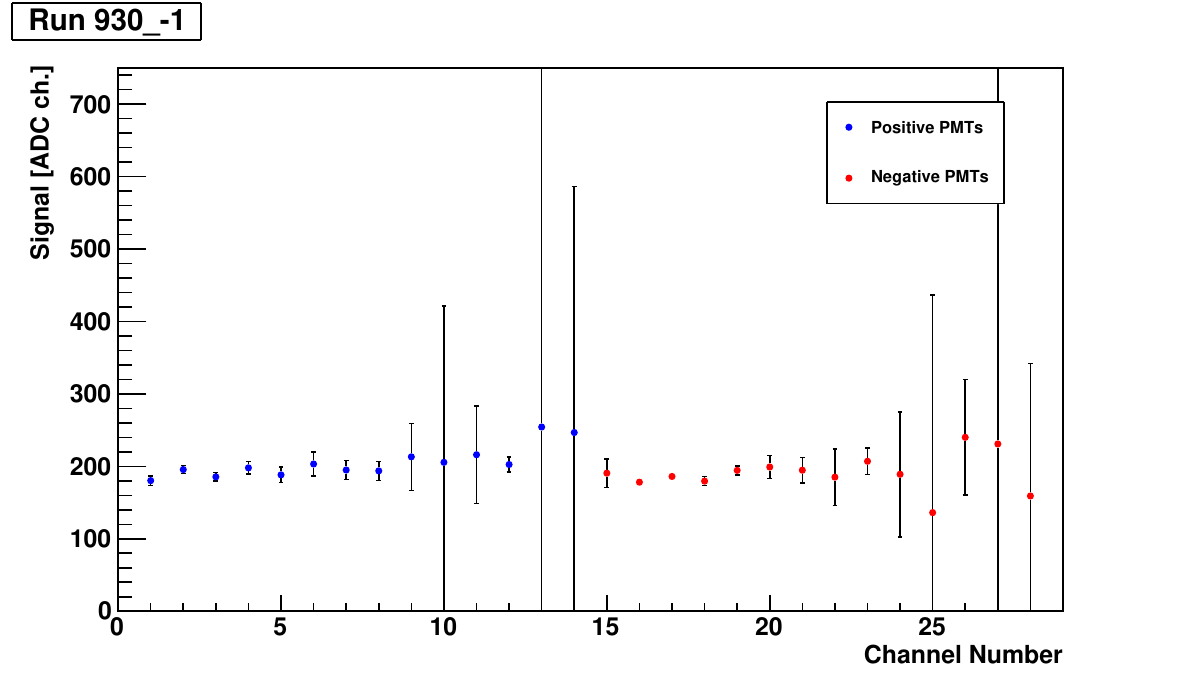}
\caption{Amplitudes of ADC signals from cosmic muons in the Preshower channels after gain matching.}
\label{fig:shms_prsh_mip_amp}
\end{centering}
\end{figure}

The Preshower detector was gain matched with cosmic rays prior to installation in the spectrometer. The coincidence of signals from scintillator counters positioned above and below the detector served as a trigger. The gain matching was adjusted after the installation, again with cosmics but this time passing through the detector stack. Muons were identified as events of a single track in the drift chambers and single hit module in the Preshower. A new set of voltages was calculated based on MIP peak positions and according to  formulae similar to Eqns \ref{eq:calo-mip-vset1}, \ref{eq:calo-mip-vset2}. The voltages span the range from 1.1\,kV to 1.7\,kV. The quality of gain matching was ensured by taking cosmic data with the new HV settings (Fig.~\ref{fig:shms_prsh_mip_amp}).

\subsubsection{Calorimeter Calibration}
\label{calo_calibr}

Particle identification in a calorimeter is based on differences in the energy deposition from different types of projectiles. The deposited energy is obtained by converting the recorded ADC channel value of each module into an equivalent energy.



The calorimeter calibration procedure corrects for the gain differences between channels. 
Good electron events are selected by utilising the gas Cherenkov detector(s). The standard calibration algorithm~\cite{amatuni} is based on minimization of the variance of the estimated energy with respect to the calibration constants, subject to the constraint that the estimate is unbiased (relative to the primary energy). 

\begin{figure}[hbtp]
\begin{centering}
\includegraphics[width=\columnwidth]{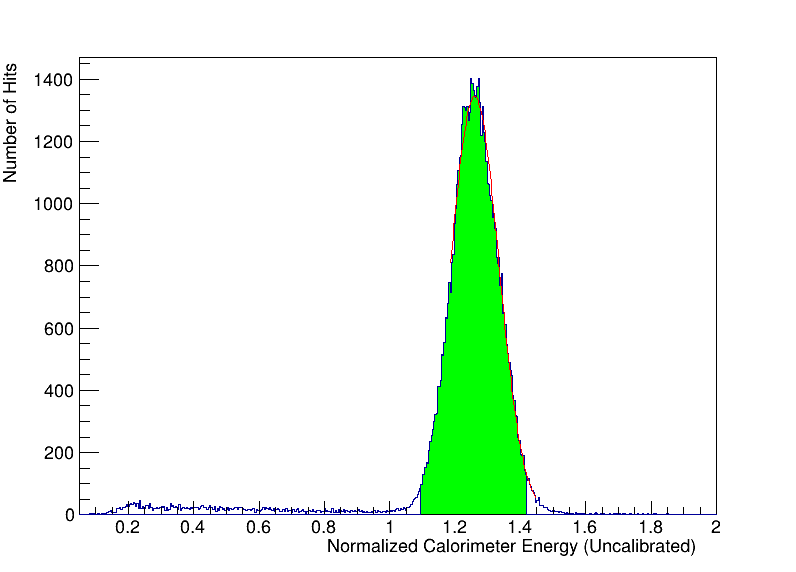}
\caption{\label{fig:calo_uncal}An electron sample (selected through Cherenkov PID) in the calorimeter before calibration. The peak of the $E_{Norm}$ distribution is clearly greater than 1 and is relatively wide.}
\end{centering}
\end{figure}

\begin{figure}[hbtp]
\begin{centering}
\includegraphics[width=\columnwidth]{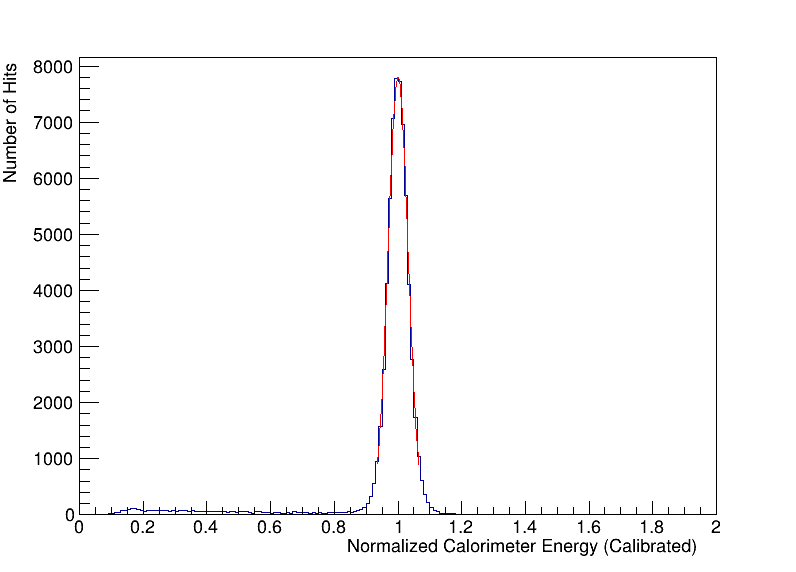}
\caption{\label{fig:calo_cal}An electron sample (selected through Cherenkov PID) in the calorimeter after calibration. The peak of the $E_{Norm}$ distribution is now much narrower and centered at 1 as expected for electrons.}
\end{centering}
\end{figure}

The deposited energy per channel is estimated by
\begin{equation}
\label{eq:calo-module-e}
{ e_i = c_i \times A_i },
\end{equation}
where $i$ is the channel number, $c_i$ is the calibration constant, $A_i$ is the FADC pulse integral signal. Note that the Preshower signals are corrected for the light attenuation dependence versus horizontal hit coordinate $y$. The calorimeter calibration can be checked by comparing the track momentum to the energy deposition in the calorimeter. The ratio 
\begin{equation}
    E_{norm}=\frac {E_{Dep}}{P_{Track}},
\end{equation}
is referred to as the \emph{normalized energy}. For electrons, $E_{Norm}$ should be close to 1.
An example of the normalized energy distribution for electron tracks can be seen before and after a successful calibration in Figs.\,\ref{fig:calo_uncal} and \ref{fig:calo_cal}.

In the calorimeter analysis code, hits on adjacent blocks in the Preshower and in the Shower are grouped into clusters. For each cluster, the deposited energy and center of gravity are calculated. These clusters are matched with tracks from the upstream detectors if the distance from the track to cluster is less than a predefined ``slop'' parameter (usually 7.5\,cm). For the Preshower, the distance is calculated only in the vertical direction.

\section{Trigger and Data Acquisition}
\label{sec:trigdaq}

%
%


The Hall C data acquisition (DAQ) system is designed to meet the needs of a high luminosity, dual spectrometer (SHMS + HMS) configuration, with the capability of extracting polarization-dependent absolute cross sections with precision at the 1\% level or better. JLab's CODA data acquisition software \cite{CODA} provides a framework that ties together a distributed network of read-out controllers (ROCs) controlling multiple crates of digitization hardware, event builders to serialize the data, and event recorder processes to write the data to disk. It also provides a graphical control interface for the users.

\begin{figure}[htbp]
  \begin{centering}
    \includegraphics[width=\columnwidth]{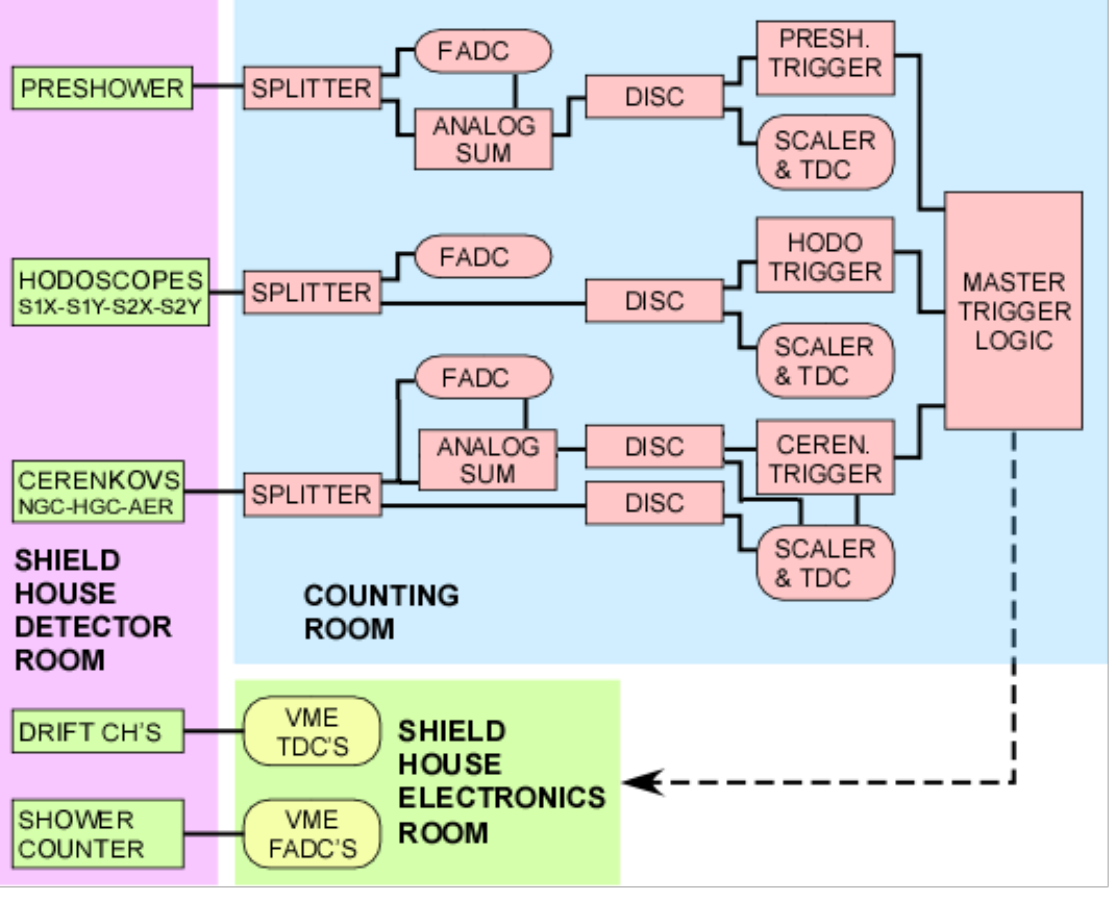}
    \caption{Block diagram of high-level trigger formation for SHMS.
      See Sec.\,\ref{sec:DAQtrigs} for detailed description.}
    \label{fig:BlockTrigger}
  \end{centering}
\end{figure}

The Hall C DAQ system can run in dual-arm trigger mode that requires a coincidence between both spectrometers, or each arm's DAQ may be run entirely independently. Incorporating additional detector systems into the standard two-arm design is also straightforward. A high-level block diagram of trigger formation and readout for SHMS is depicted in Fig.\,\ref{fig:BlockTrigger}.

The hardware DAQ and trigger designs were strongly influenced by the preceding 6\,GeV HMS and SOS configurations. This choice was made to provide a careful and systematic migration from the very well understood systematics of the 6\,GeV system while incorporating and characterizing a new generation of FPGA-based logic and readout electronics. To this end, the present system relies on a combination of legacy NIM and CAMAC discriminators and logic modules to form readout triggers, but utilizes a full set of modern high speed payload and front-end modules to allow a transition to a firmware based trigger and fully pipelined readout in the future.

In the present configuration, the DAQ has a nominal maximum trigger acceptance rate of 4\,kHz with a deadtime of $\approx$20\%. Dead times are measured using the Electronic Dead Time Measurement (EDTM) system outlined in Sec.\,\ref{sec:DAQ-EDTM}. The underlying hardware supports running in a fully pipelined mode, and should be capable of running at trigger rates exceeding 20\,kHz with minimal deadtime using firmware based triggers similar to those employed in Halls B and D. This capability was not part of the initial 12\,GeV upgrade plan for Hall C, but may be pursued in the future (see Sec.\,\ref{sec:DAQ-future}).

Signals from the scintillator planes, Cherenkov detectors, and calorimeter detectors in the SHMS and HMS detector stacks are processed to form \emph{pre-triggers}. Those pre-triggers can serve as \emph{event triggers} themselves (that initiate a recorded event), or be combined to bias data collection towards particular particle types (\emph{i.e.} electrons \emph{vs.} pion) and suppress backgrounds. Each running DAQ can be fed up to six independent triggers simultaneously and the experimenter can control what fraction of each is recorded to disk run-by-run through an integrated pre-scale feature.

\ltwo{Standard Triggers}\label{sec:DAQtrigs}

All trigger-related PMT signals from both the SHMS and HMS are routed out of the experimental Hall to a dedicated electronics room on the main level of the Hall C Counting House using low-loss RG-8 air-core signal cables. Those signals are then split with one output running into a JLab F250 flash analog to digital converter (FADC)\cite{JLABF250}, while the second output is processed and discriminated. All discriminated pulses are delivered to scalers for rate information, TDCs for precision timing measurement, and to form pre-triggers as described below. This design allows direct access to all raw signals that may participate in a trigger during beam operations and has proven invaluable during the debugging and commissioning phases of Hall operations.

Non-trigger related signals include wire-chamber readouts and the Shower (but not Preshower) layer of the SHMS calorimeter. The readout electronics for those sub-detectors remain inside their respective detector huts within the experimental Hall. All SHMS calorimeter PMT signals are fed into F250 FADCs configured to provide timing, integrated energy, pulse amplitude, and (optionally) pulse profile data as desired. The wire-chamber timing signals are digitized using multi-hit CAEN v1190 modules \cite{CAEN1190}.

The CAEN v1190 payload modules provide 128 independent multi-hit/multi-event TDC channels with a user configurable resolution ranging from 52\,$\mu$s to 100\,ps per bin. They provide a 32 kilo-word deep output buffer and can be read out asynchronously with respect to the event triggers. Typical Hall C operation has all units configured for 100\,ps/bin.

\lthree{JLab F250 Flash ADCs}
The JLab F250 flash ADC modules are an FPGA-based design developed by the Jefferson Lab Fast Electronics group \cite{JLABF250} and are used lab wide. Each F250 module provides 16 independent 50\,$\Omega$ input channels. The voltage at each input channel is continuously digitized into an 8\,$\mu$s ring buffer at 250\,MHz, with a resolution of 12\,bits, and a hardware adjustable full-scale range. When a module receives a readout trigger, digitized sample data stored in the ring buffer are processed in parallel without incurring front-end deadtime. In typical operation each `hit' over a pre-programmed threshold is assigned an interpolated leading-edge threshold time ($<$1\,ns resolution), integrated energy (analogous to a charge-integrating ADC value), a peak amplitude, and a measurement of the DC offset (pedestal) present on the channel prior to the detected pulse. Full pulse-profile data for each hit may also be stored if desired. However, that mode increases the data rate by several orders of magnitude, and is generally used only for debugging or limited duration pulse characterization runs.

\begin{figure}[htbp]
  \begin{centering}
    \includegraphics[width=\columnwidth]{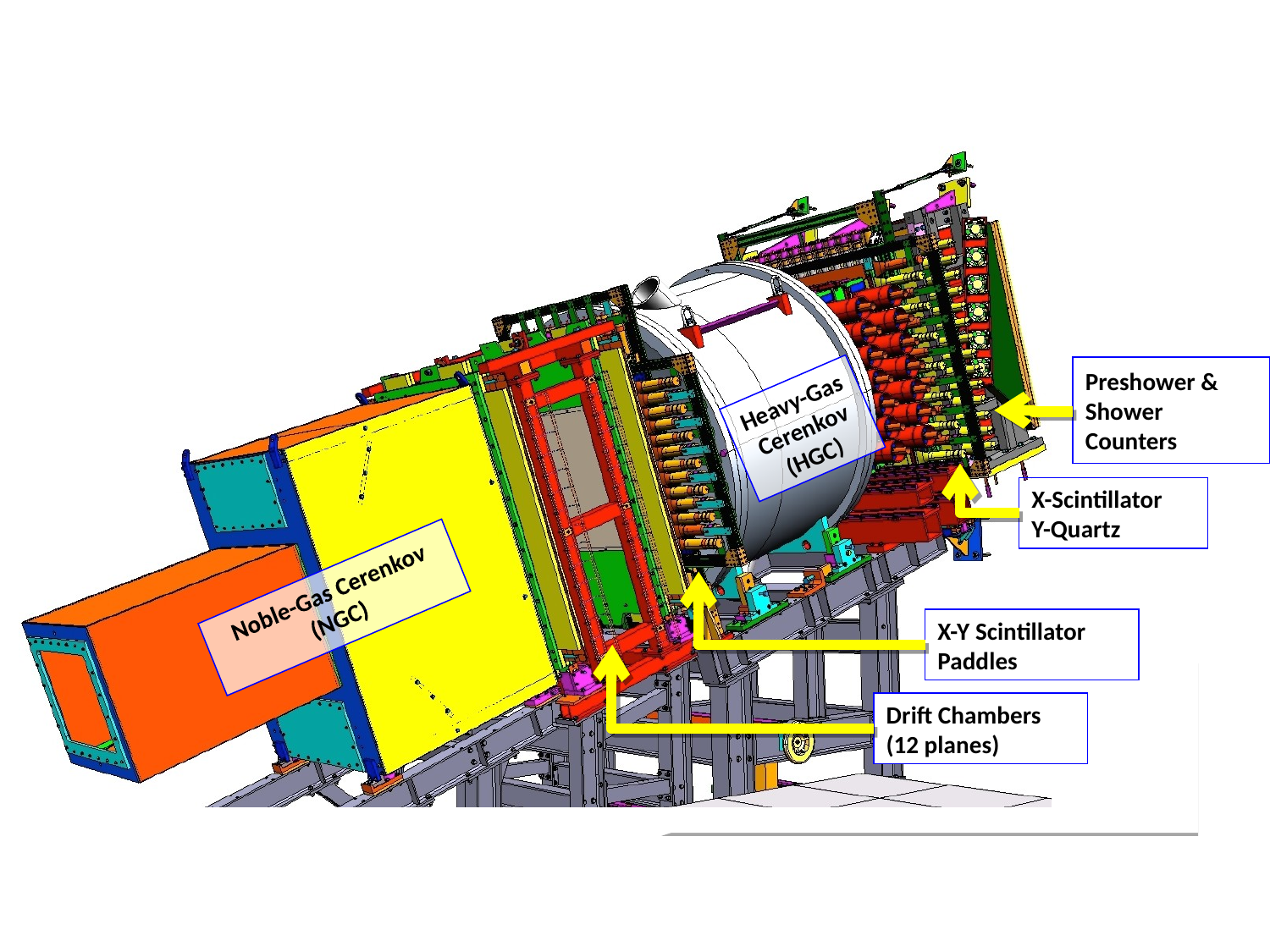}
    \caption{Typical detector layout for the SHMS.}
    \label{fig:DAQ-SHMSlayout}
  \end{centering}
\end{figure}

\lthree{SHMS Triggers}\label{sec:SHMS-trigs}
The SHMS detector stack layout is described in Sec.\,\ref{sec:shielding}. A representative detector layout is presented in Fig.\,\ref{fig:DAQ-SHMSlayout}.

\begin{figure}[htbp]
  \begin{centering}
    \includegraphics[width=0.95 \columnwidth]{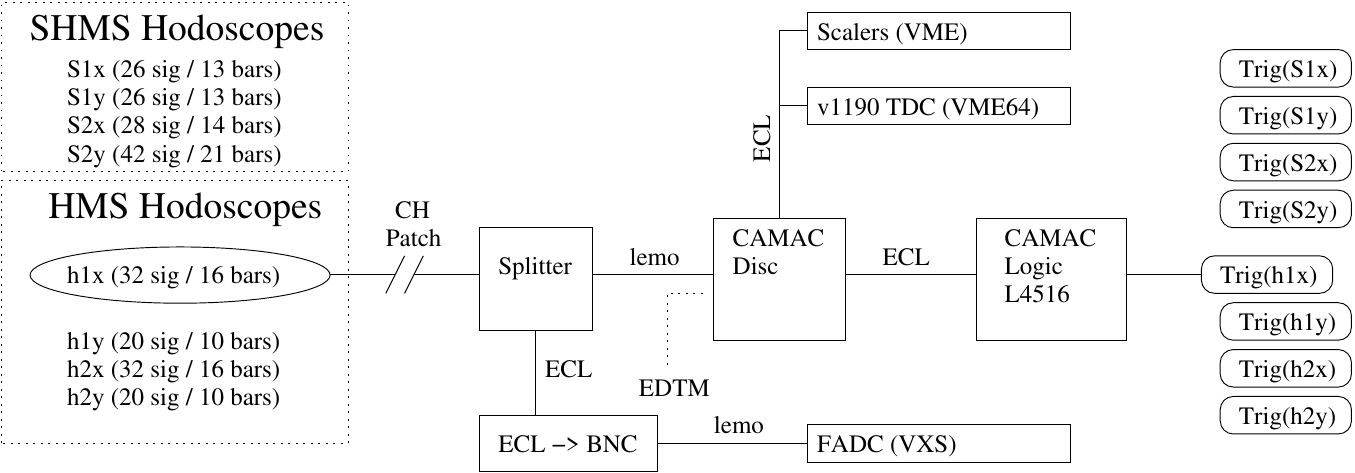}
    \caption{Block diagram for SHMS and HMS hodoscope pre-trigger formation.}
    \label{fig:DAQ-hodo-pretrig}
  \end{centering}
\end{figure}

Each hodoscope plane, described in Secs.\,\ref{sec:scint} and \ref{sec:quartz}, is constructed from an array of horizontal (or vertical) scintillator bars with a PMT on each end. Signals from those PMTs are split and one analog output is delivered to F250 FADCs. The second analog output is discriminated and sent to CAEN 1190 TDCs for precision timing information, to scalers for raw rate information, and to logic modules to provide the hodoscope pre-triggers plane by plane. A pre-trigger for each plane is generated by OR'ing the discriminated signals from each side of a hodoscope plane together, then AND'ing the resulting two signals together. The pre-triggers are designated S1X, S1Y and S2X, S2Y; where 1(2) denote the up(down)stream plane, and X(Y) denote the horizontal(vertical) scintillator bar orientation (see Fig.\,\ref{fig:DAQ-hodo-pretrig}).

It should be noted an optimal design would generate an AND between the PMTs on each side of every bar first, and OR the resulting per-bar coincidences to form a pre-trigger for the plane. The compromise above was driven by constraints of the legacy LeCroy 4564 CAMAC logic units held over from the 6\,GeV era.

The SHMS detector stack includes a permanent Heavy Gas Cherenkov (HGC) (see Sec.\,\ref{sec:hgcerenk}), but also includes space for a \emph{Noble Gas} Cherenkov (NGC) (see Sec.\,\ref{sec:ngcerenk}). Each SHMS gas Cherenkov detector incorporates four PMTs, with each PMT detecting light from one of four mirrors inside their respective gas volumes. Analog signals from the PMTs are split (50:50) with one path plugged into an FADC. The second copies from each PMT are summed, and the summed output is discriminated to form a Cherenkov pre-trigger for that Cherenkov detector (HGC and NGC). The pre-triggers are also routed to scaler channels and a v1190 TDC.

\begin{figure}[htbp]
  \begin{centering}
    \includegraphics[width=0.95 \columnwidth]{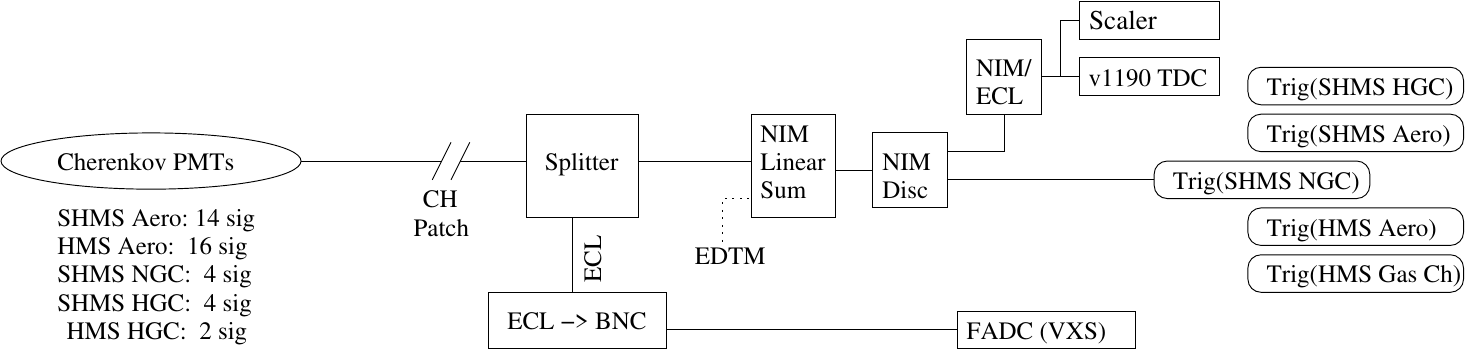}
    \caption{Block diagram for SHMS and HMS Cherenkov pre-trigger formation.}
    \label{fig:DAQ-cer-pretrig}
  \end{centering}
\end{figure}

An optional SHMS aerogel Cherenkov detector (AER), as detailed in Sec.\,\ref{sec:aerogel}, may also be installed. It employs seven PMTs on each side of its diffusion box. The signals from all 14 PMTs are handled analogously to the gas Cherenkov detectors, with each analog signal being split and read out by an individual FADC channel, with the second outputs being summed and discriminated to form an associated aerogel pre-trigger. The pre-trigger is routed to a scaler and v1190 TDC as well.
A block diagram for the Cherenkov pre-triggers is presented in Fig.\,\ref{fig:DAQ-cer-pretrig}.

\begin{figure}[htbp]
  \begin{centering}
    \includegraphics[width=0.95 \columnwidth]{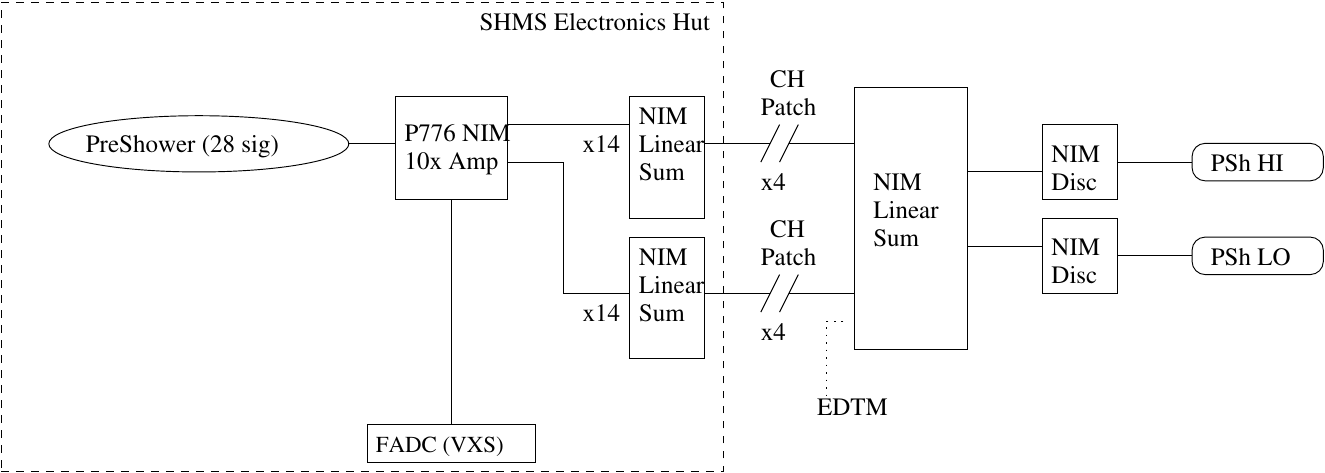}
    \caption{Block diagram for SHMS Preshower summing trigger.}
    \label{fig:DAQ-SHMS-PSh-pretrig}
  \end{centering}
\end{figure}

For the SHMS Preshower, described in Sec.\,\ref{sec:shower}, a  pre-trigger is formed using the 28 analog signals from PMTs which are are split and summed in 3 groups of 4 rows, and 1 group of 2 rows. Each of the 4 group sums is read out by an FADC channel for cross checks. The 4 group sums are summed in turn to provide a total Preshower sum which is then discriminated and provides the SHMS \emph{PSh} pre-trigger. Provision is made to generate independent pre-triggers for both low and high energy depositions in the Preshower layer (\emph{PSh\_Lo} and \emph{PSh\_Hi}, respectively) as seen in Fig.\,\ref{fig:DAQ-SHMS-PSh-pretrig}.

The aforementioned pre-triggers are then combined to form a set of triggers capable of initiating a DAQ event. These combinations are often adjusted or optimized to serve the needs of particular experiments but a set of commonly available event triggers is outlined in Sec.\,\ref{sec:EventTrigs}.

\lthree{HMS Triggers}\label{sec:HMS-trigs}

\begin{figure}[htbp]
  \begin{centering}
    \includegraphics[width=1.0\columnwidth]{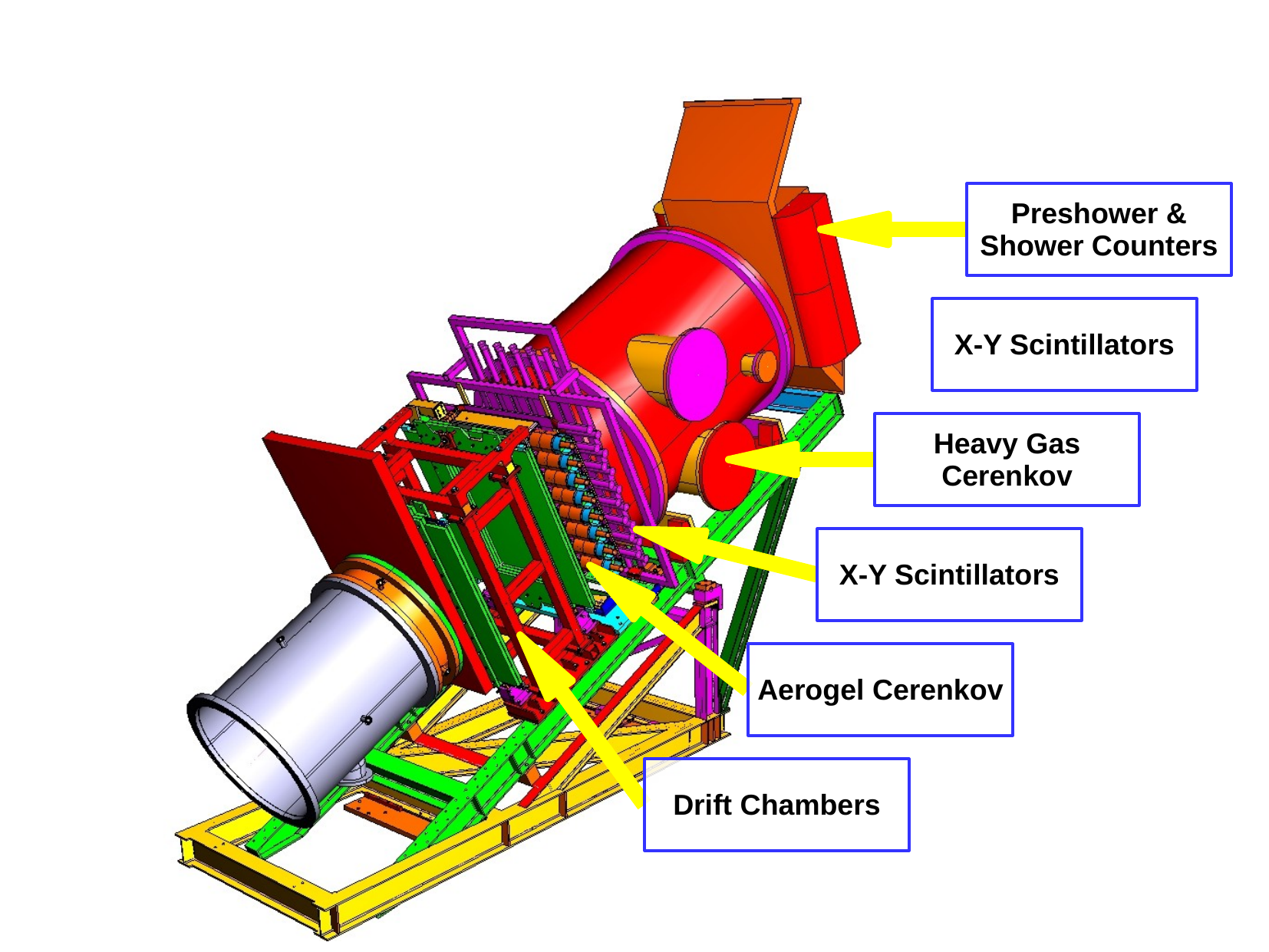}
    \caption{Typical detector layout for the HMS.}
    \label{fig:DAQ-HMSlayout}
  \end{centering}
\end{figure}

The standard HMS detector stack is the predecessor of the SHMS system and shares a nearly identical design as seen in Fig.\,\ref{fig:DAQ-HMSlayout}. It consists of a pair of scintillator-based hodoscope planes in an X+Y configuration, a gas Cherenkov detector, a second pair of X+Y hodoscopes, and a Preshower + Shower Calorimeter. Provision is also made for an optional Aerogel Cherenkov to be inserted into the detector stack just downstream of the drift chambers for supplemental particle identification (PID).

The trigger and readouts designs follow the patterns described in Sec.\,\ref{sec:SHMS-trigs}, with a modest difference associated with the HMS Calorimeter.

Signals from the four HMS hodoscope planes, denoted h1x, h1y, h2x, h2y, are split, discriminated, and recombined to form a \emph{Scin} trigger following the same logic as the SHMS hodoscopes described previously.

The HMS gas Cherenkov detector incorporates two PMTs detecting light from two mirrors inside the HMS Cherenkov tank. Analog signals from the PMTs are split (50:50), with one path plugged into an FADC. The second copies from each PMT are summed, and the summed output is discriminated to form the Cherenkov pre-trigger. That pre-trigger is also routed to a scaler and v1190 TDC.

The HMS Aerogel employs eight PMTs on each side of its diffusion box. The signals from all 16 PMTs are split and read out by an individual FADC channel, with the second copies being summed and discriminated to form the associated aerogel pre-trigger. The pre-trigger is routed to a scaler and v1190 TDC as well.

The HMS calorimeter is composed of four layers of lead glass blocks. Each layer has 13 lead-glass blocks arranged horizontally, and the layers are denoted A, B, C and D as seen by a particle passing through the detector stack. Layers A and B have PMTs bonded to each end of their blocks, while Layers C and D have a single PMT on one side only. Analog signals from the PMTs are split 50:50 with one copy being delivered to an FADC. The copies are formed into an analog sum for each side of each layer, denoted hA+, hA-, hB+, hB-, hC, and hD. Layer sums hA and HB are formed by summing hA+ and hA-, and hB+ and hB-, respectively (hC and hD are already layer sums).

One copy of each layer sum is sent to an FADC for monitoring and cross checks. A Preshower pre-trigger is formed by summing and discriminating Layers A + B, and a \emph{Shower Low} pre-trigger is formed by summing and discriminating Layers A+B+C+D. Copies of the Preshower and Shower sums are sent to FADCs and copies of the discriminated pre-trigger signals are sent to scalers and 1190 TDCs.

\begin{figure}[htbp]
    \includegraphics[width=0.95 \columnwidth]{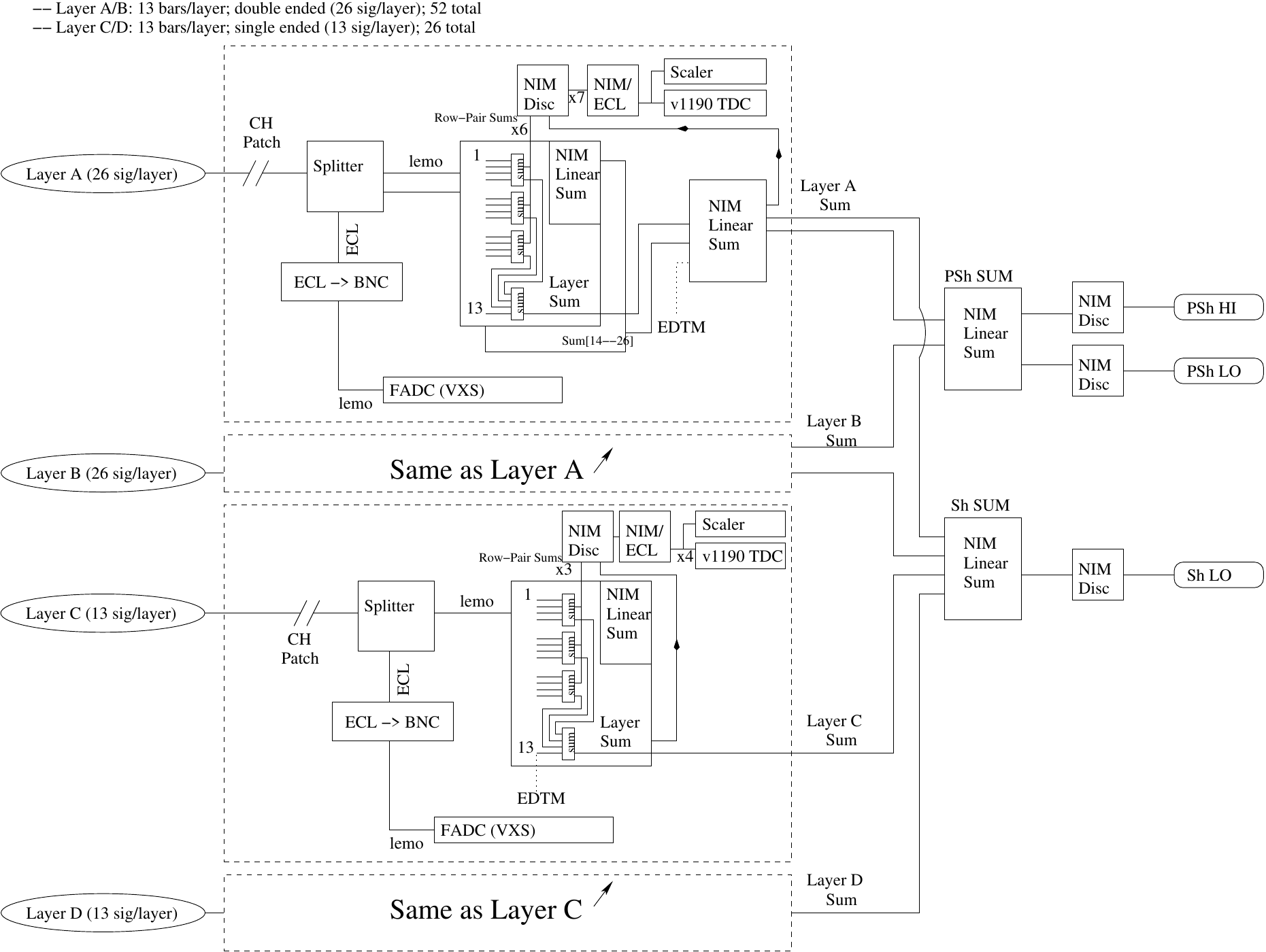}
    \caption{Block diagram for HMS Shower and Preshower summing triggers.}
    \label{fig:DAQ-HMS-shower-pretrig}
\end{figure}

Fig.\,\ref{fig:DAQ-HMS-shower-pretrig} depicts a block diagram of the HMS Calorimeter pre-triggers.

\lthree{Event Triggers}\label{sec:EventTrigs}

The aforementioned pre-triggers are then combined to form a set of triggers capable of initiating a DAQ event. The `default' single-arm trigger is formed by 3 out of 4 hodoscope planes firing in coincidence. Often referred to as the \emph{3 of 4} or \emph{Scin} trigger, it provides a high-efficiency ($>99$\%) general-purpose charged particle trigger.

A second standard trigger is referred to as \emph{El\_Clean}. It implements particle discrimination at the trigger level by forming a coincidence between the \emph{Scin} pre-trigger, one (or more) Cherenkov pre-triggers, and (optionally) the pre-shower (\emph{PSh}) and/or calorimeter total-sum (\emph{ShTot} pre-triggers).

\ltwo{Electronic Dead Time Measurement System (EDTM)}\label{sec:DAQ-EDTM}

The DAQ and trigger system for each spectrometer also includes an Electronic Dead Time Measurement (EDTM) system. This is implemented by replicating a pulse from a pulse-generator circuit and feeding into every pre-trigger leg as close to the analog signals as possible. The timing of those duplicated pulses is adjusted to match those generated by a real particle passing through the detector stack. A copy of each synthetic EDTM trigger is counted in a deadtime free scaler and sent to a dedicated TDC channel in each arm. The presence of an appropriately timed hit in that TDC channel tags an event as having been generated by an EDTM trigger.

During beam operations, this allows a direct measurement of the fraction of triggers that are lost due to some component of the DAQ being busy. This is known as the system \emph{deadtime}. By inducing synthetic signals as early in the trigger electronics as possible, this system is sensitive to high-rate signal pile-up in the full front-end trigger logic chain (loosely referred to as ``electronic deadtime"), as well as digitization and read out related deadtimes implicit in the non-pipelined DAQ operation presently in use in Hall C.

In addition to the above function, the system has proved useful for pre-beam trigger verification and end to end checkout of the DAQ system.
\begin{itemize}
  \item It allows rough timing on all trigger legs to be verified without beam.
  \item It allows coincidence timing between the SHMS and HMS arms to be roughed in and tested without beam.
  \item It allows the entire DAQ system to be stress tested under controlled conditions without beam.
\end{itemize}


%
%
%
%
%
%

\ltwo{Auxiliary Data Collection}\label{sec:DAQ-aux}

The standard method for slow controls data logging is through the Experimental Physics and Industrial Control System (EPICS) \cite{EPICS}. EPICS is a system of open source software tools and applications used to provide control user interfaces and data logging for systems such as high- and low-voltage detector power supplies, target systems, spectrometer magnets, vacuum, and cryogenic systems, etc.

Long-term, persistent storage of EPICS based slow controls data is provided through an independent archiving system managed by the Accelerator Division's MYA archiving system. An experimentally relevant subset of EPICS data (beam and target characteristics; magnet, spectrometer and detector settings, etc.) are also stored in the experimental data files at regular intervals whenever the DAQ is running.


\ltwo{Online Hall C Computing Environment}\label{sec:DAQ-hcdaq}

Hall C employs a dedicated stand-alone computing cluster with redundant multi-core servers focused on prompt online analysis, high volume local data storage, and 1--10\,Gb ethernet interconnects. There are dedicated hosts for each independent DAQ system (\emph{ex.} SHMS and HMS), and auxiliary machines for polarimetry, target controls, spectrometer slow controls, etc.

Experimental control and operational feedback is provided to users in the Hall C Counting house through a collection of multi-screen computer workstations and a set of large wall-mounted displays for critical data.

All systems have direct access to the JLab centrally managed Scientific Computing resources. This includes multi-petabyte tape storage and online disk facilities, as well as a several thousand core compute farm for simulation and offline data analysis \cite{JLABFARM}.

\ltwo{Future Plans / Pipeline trigger}\label{sec:DAQ-future}

During the early stages of the 12\,GeV Hall C upgrade plan, it was concluded that the risks of moving to a fully pipelined DAQ system with a firmware driven trigger were not justified by the needs of the initial experimental program. In general, those experiments did not impose a too heavy burden on the DAQ, and the more conventional trigger design with its well understood characteristics
was preferred.

However, provision was made to design and build the low-level DAQ system with an upgrade path in mind. To that end, a full complement of trigger and payload modules compatible with the pipelined systems being implemented for Halls B and D was selected.

A phased transition from the NIM/CAMAC trigger system to a fully pipelined approach would involve implementing the present trigger logic within the existing JLab FADC and VXS Trigger Processor (VTP) boards, and a thorough validation of the firmware based trigger decisions against the well understood conventional trigger. Once the firmware is fully debugged/characterized, the DAQ could transition to pipelined mode and take advantage of significant boost in trigger accept rates into the 10's of kHz range with minimal deadtime. At that point, the next DAQ bottleneck would likely be rate limitations in the detector systems themselves (signal pile-up in the front-end, track reconstruction limitations, etc.)

\section{Software}
\label{sec:software}

Hall C Data is analyzed by the Hall C analysis package \texttt{hcana}.
This package does full event reconstruction for the SHMS used alone or in coincidence with other detectors. \texttt{hcana} is based on the modular Hall A \texttt{analyzer} \cite{HallAAnalyzer} ROOT \cite{ROOT} based C++ analysis framework. This framework provides for run time user configuration of histograms, ROOT tree contents, cuts, parameters and detector layout.

\texttt{hcana} includes C++ classes for detectors, spectrometers, and physics analyses. Instantiation of these classes as objects is configured at run-time through a ROOT script which also sets up the configuration of analysis replay. Due to the similarity of the SHMS and HMS spectrometers and their detector packages, the same spectrometer and detector classes are used for both spectrometers. For example, the drift chamber package class is instantiated for both spectrometers with each object configured by its specific parameters and geometry. Additional modules such as new front end decoders, detectors, or physics analysis modules can easily be added to \texttt{hcana}. These modules can either be compiled into the analyzer or be compiled separately and dynamically loaded at run time.

Event analysis is segmented into 3 steps of spectrometer and detector specific analysis.
\begin{enumerate}
    \item Decoding: Detector requests from the low level decoder produce a list of hits sorted by detector plane and counter number. A minimal amount of processing is done to make data available for low level histograms.
    \item Coarse Processing: Tracks are found in the drift chambers. Hits and clusters in the hodoscope, shower counter and other detectors are matched to the tracks to determine time-of flight. The various detectors provide information for particle identification.
    \item Fine processing: Particle identification information is refined, tracks in the focal plane are traced back to the target coordinate system, and particle momentum is determined.
\end{enumerate}

Each of these steps is completed for all detectors before proceeding to the next step. Some limited information is passed between detectors at each step. For example, timing information from the hodoscopes is used to obtain the start time for the the drift chambers in the coarse processing step, and tracks obtained from the drift chambers are associated with shower counter hit clusters in the fine processing step.

\subsection{Online Monitoring}

After each data taking run is started, a subset of the data is analyzed with \texttt{hcana}. An easily configurable histogram display GUI is used to view diagnostic histograms and compare them to reference histograms.
The EPICS \cite{EPICS} control system alarm handler is used to monitor experiment settings and beam conditions. This includes spectrometer magnet settings, detector high voltages, drift chamber gas, cryogenic systems and spectrometer vacuum.

\section{SHMS Performance: Operating Experience and Commissioning Results}
\label{sec:performance}



\subsection {Acceptance}

\begin{figure}[htbp]
	\includegraphics[width=\linewidth]{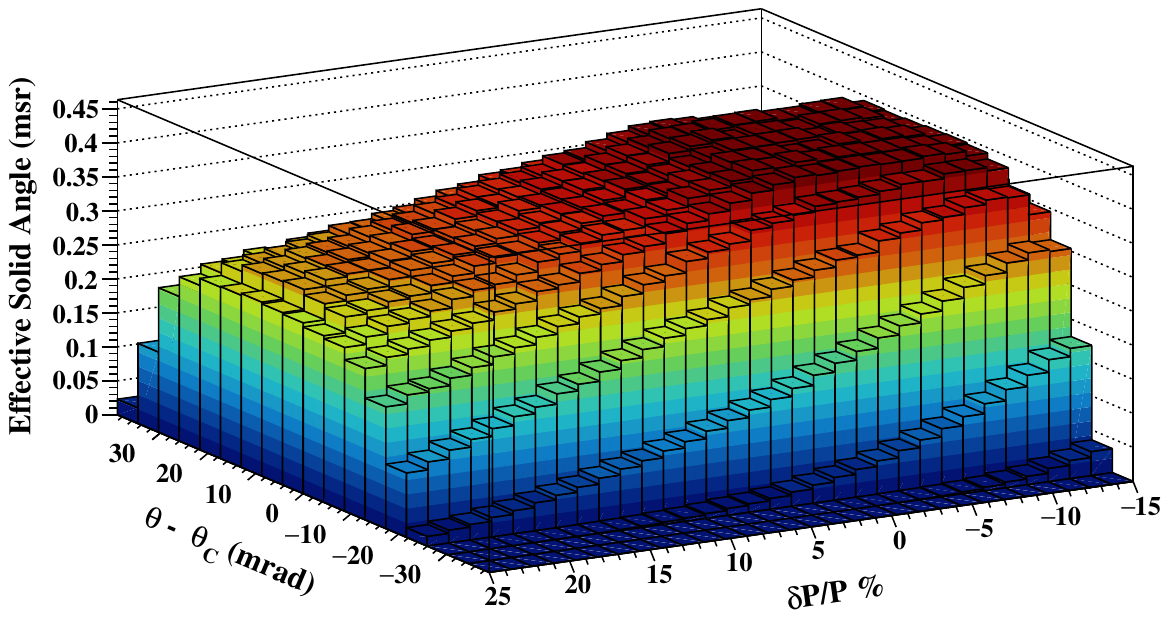}
	\caption{SHMS effective solid angle as a function of $\delta P / P$ and $\theta$. SHMS $\theta_{central} = 21^{\circ}$ and $P_{central} = 3.3\,$GeV/c.}
	\label{fig:2dacceptance}
\end{figure}

The acceptance of the SHMS can be determined from simulation and defined as $A(\delta,\theta)=N_{sus}(\delta,\theta)/N_{gen}(\delta,\theta)$, where $N_{gen}$ is the number of events generated into a particular $\delta,\theta$ bin and $N_{sus}$ is the number of events that successfully reached the detector stack. Since $A(\delta,\theta)$ depends on the generation limits of the simulation, a more useful quantity is the effective solid angle, $\Delta\Omega_{eff}=A(\delta,\theta)*\Delta\Omega_{gen}$, where $\Delta\Omega_{gen}$ is the solid angle generated into for each bin. Fig.\,\ref{fig:2dacceptance} shows the effective solid angle of the SHMS at a central angle of $21^{\circ}$ and central momentum of $3.3$\,GeV/c for a $10$ \,cm liquid hydrogen target.
 
\begin{figure}[htbp]
\centering
		\includegraphics[width=0.8\linewidth]{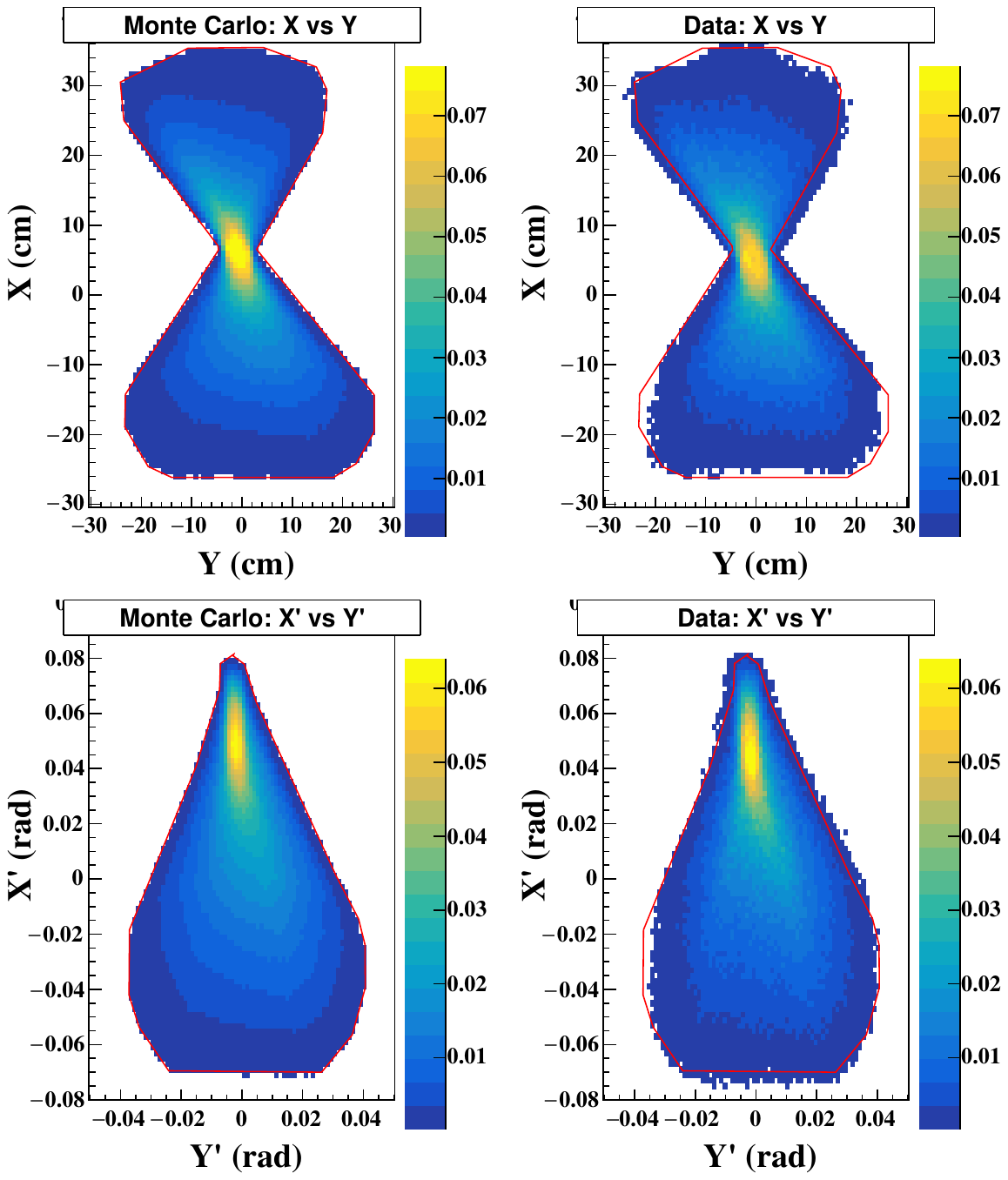}
		\caption{Comparison of SHMS focal plane quantities, simulation is on the left and data is on the right. The top plots are the position at the focal plane and the bottom is the angles at the focal plane determined from tracks formed by the drift chamber planes. The red outline represents the expected shape determined from simulation.}
		\label{fig:focalPlane2}
\end{figure}

Fig.\,\ref{fig:focalPlane2} shows the position and angular distribution of tracks formed from the drift chambers at the focal plane. A good agreement between simulation and data reflects our understanding of both the magnetic forward transport and physical locations of the apertures which determine the acceptance. 

\begin{figure}[htbp]
  \includegraphics[width=0.8\linewidth]{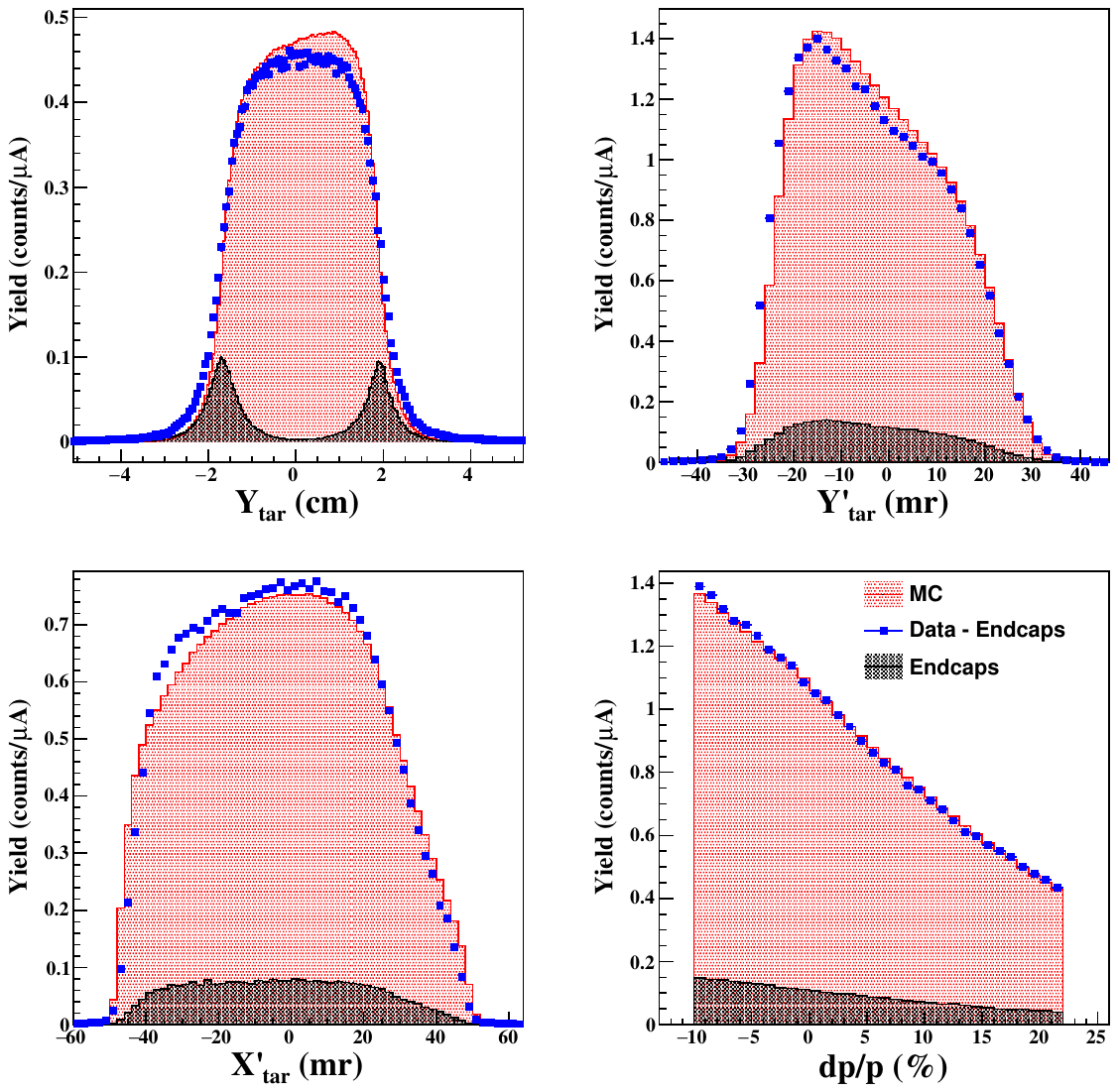}
  \caption{Target variable comparison of data versus Monte Carlo simulation from \cite{E12-10-002}. After subtracting the aluminum cell walls (black histogram) of the hydrogen target using dummy foil data, the agreement between data (blue histogram) and Monte Carlo (red histogram) is reasonable.}
  \label{fig:acceptance3.3gev}
\end{figure}

Fig.\,\ref{fig:acceptance3.3gev} demonstrates the agreement between simulation (after subtracting the cell walls) of the target variables 
$x_{tar}, y'_{tar}, x'_{tar}, $ and $\delta$ that were described in Sec.\,\ref{sec:optics}.

\begin{figure}[htbp]
  \begin{centering}
  \includegraphics[width=0.5\linewidth]{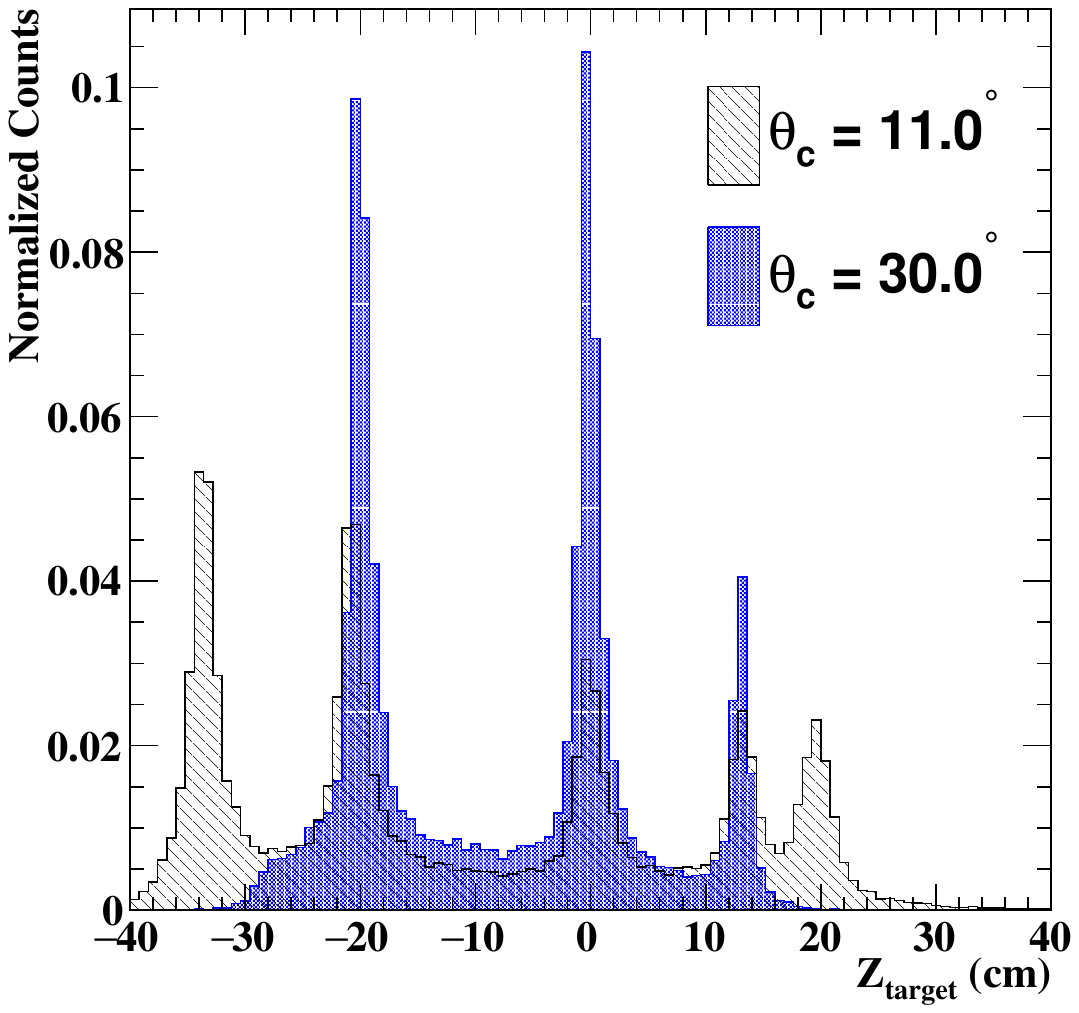}
  \caption{Reconstructed $z_{target}$ for a carbon foil optics target at SHMS central angles of $11^{\circ}$ and $30^{\circ}$. Carbon foils were located at approximately -20, 0, 13.3 and 20.0\,cm. The peak located at -35\,cm is from the beam pipe exit window. The target region was not under vacuum and therefore a background from air is present in the data and not subtracted here.}
  \label{fig:ztar2}
  \end{centering}
\end{figure}

To demonstrate how large the SHMS acceptance is in $y_{tar}$, optics data were taken during the $A_1^n$ experiment. Fig.\,\ref{fig:ztar2} plots the reconstructed position along the beam line, $z_{target}$. 


\subsection{Rates and Live time}

\begin{figure}[p]
\centering
\includegraphics[width=0.55\linewidth]{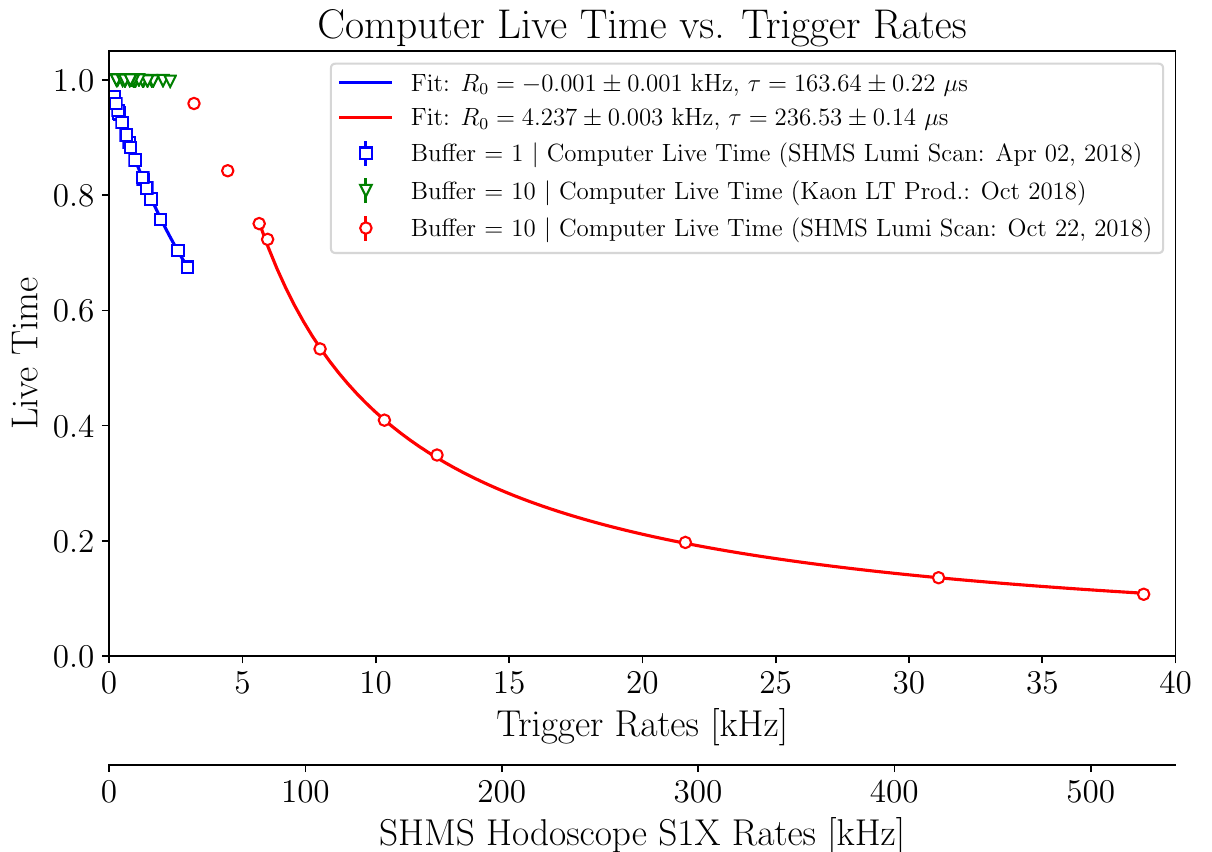}
\caption{Computer live time vs. trigger rates (top x-axis) and SHMS hodoscope S1X plane rates (bottom x-axis) for DAQ buffer levels 1 and 10.}
\label{fig:cpuLT}
\end{figure}

\subsubsection{Computer Live Time}

The computer live time efficiency of the DAQ is defined as
\begin{equation}
\epsilon_{\mathrm{CLT}} = \frac{N_{\mathrm{(phy+edtm), TDC}} -  N_{\mathrm{(edtm), TDC}}}{N_{\mathrm{(phy+edtm), SCL}} - N_{\mathrm{(edtm), SCL}}},
\end{equation}
where  the numerator is the total number of EDTM-subtracted TDC counts (total accepted physics triggers) and the denominator is the total number of EDTM-subtracted scaler counts (total physics pre-triggers). The EDTM events must be subtracted because they are generated by a fixed rate clock as described in section \ref{sec:DAQ-EDTM}, hence would bias the calculation.
A beam current cut was also applied so that the live time calculation matched the period of good physics production. 

The computer live time data shown in Fig.\,\ref{fig:cpuLT} is plotted against the un-prescaled input trigger rates (top x-axis) and the total rate in the first plane (S1X) of the SHMS Hodoscopes (bottom-axis). The data were obtained from the SHMS luminosity scans and the Kaon LT experimental data taken in Fall 2018. The Spring 2018 scans (blue squares) were taken with the DAQ in buffer level 1 (unbuffered mode) and the Kaon LT data (green triangles) and Fall 2018 scans (red circles) were with the DAQ in buffer level 10 (buffered mode). The advantage of buffered mode is that the DAQ is capable of accepting higher trigger rates while keeping the computer live time efficiency $\sim100\%$. Both buffered and unbuffered modes exhibit a characteristic fall-off of the live time as a function of the trigger rate which has been modeled using the fit function,
\begin{equation}
    f_{\epsilon_{\mathrm{CLT}}} (R) \equiv \frac{1}{1 + (R-R_{0})\tau},
\end{equation}
where $R$ is the input trigger rate, $R_{0}$ describes a horizontal offset between the unbuffered and buffered modes and $\tau$ represents the averaged data readout time (deadtime) before the DAQ is ready to accept another pre-trigger. The fit function, however, is unable to describe the ``flat'' region where the live time is nearly 100 $\%$. From the fit parameters, the fall-off behavior of buffered mode starts at a trigger rate of $R\sim 1/\tau$ which corresponds to $\sim4.2$\,kHz.

Since fall 2018, the DAQ has been operated in buffered mode which has proven to be advantageous for high-rate experiments at Hall C.


\subsection{Subsystem Performance}


\subsubsection{Hodoscope Performance}

\begin{figure}[htbp]
\includegraphics[width=\linewidth]{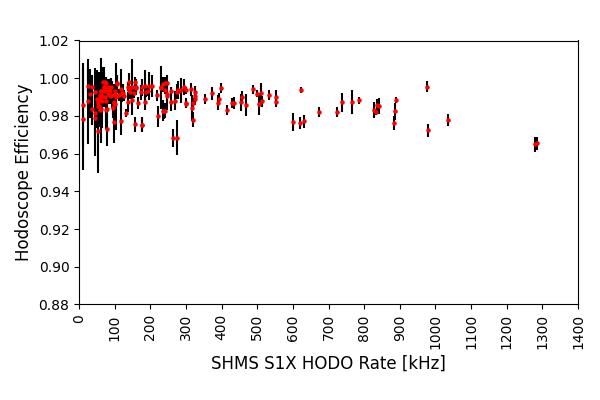}
\caption{Hodoscope efficiency as a function of rate in the first hodoscope plane, S1X.
\label{fig:Hodo_Eff_Lumi}}
\end{figure}

Once installed in the SHMS detector hut, all paddles were retested and gain matched. During the Hall C commissioning experiments, carried out during spring 2018, the scintillators performed as expected with no major problems. The hodoscope efficiency as a function of S1X rate (first hodoscope plane) can be seen in Fig.\,\ref{fig:Hodo_Eff_Lumi}. 

\begin{figure}[p]
\includegraphics[width=\linewidth]{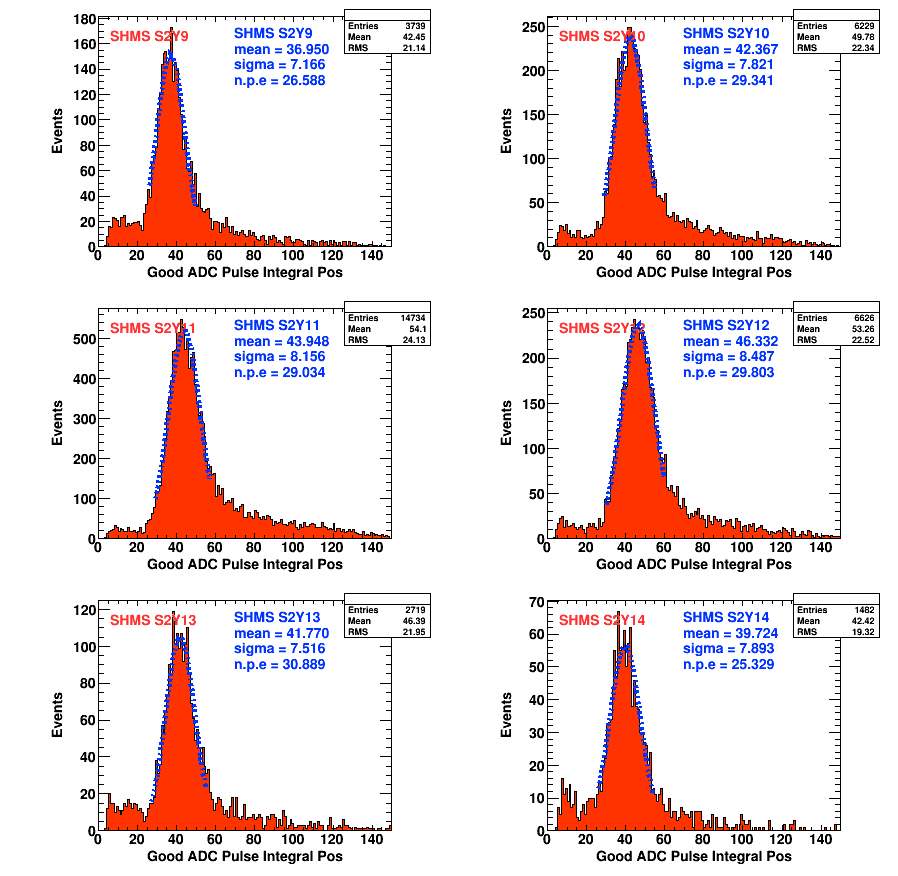}
\caption{Number of photoelectrons response from the quartz plane, negative end PMTs.
\label{fig:neg_npe_b}}
\end{figure}

The performance of the quartz plane (S2Y) was studied with beam during the Hall C commissioning in Fall of 2017. An example plot of the photoelectron response from most bars in the quartz plane is shown in 
Fig.\,\ref{fig:neg_npe_b}.
Only electrons with an incident angle close to 90$^\circ$ were chosen here to eliminate the bias coming from possibly reduced photon collection efficiency due to sub-optimal angles of the photon cones. All PMTs and optical couplings performed satisfactory.

\begin{figure}[htbp]
\includegraphics[width=\linewidth]{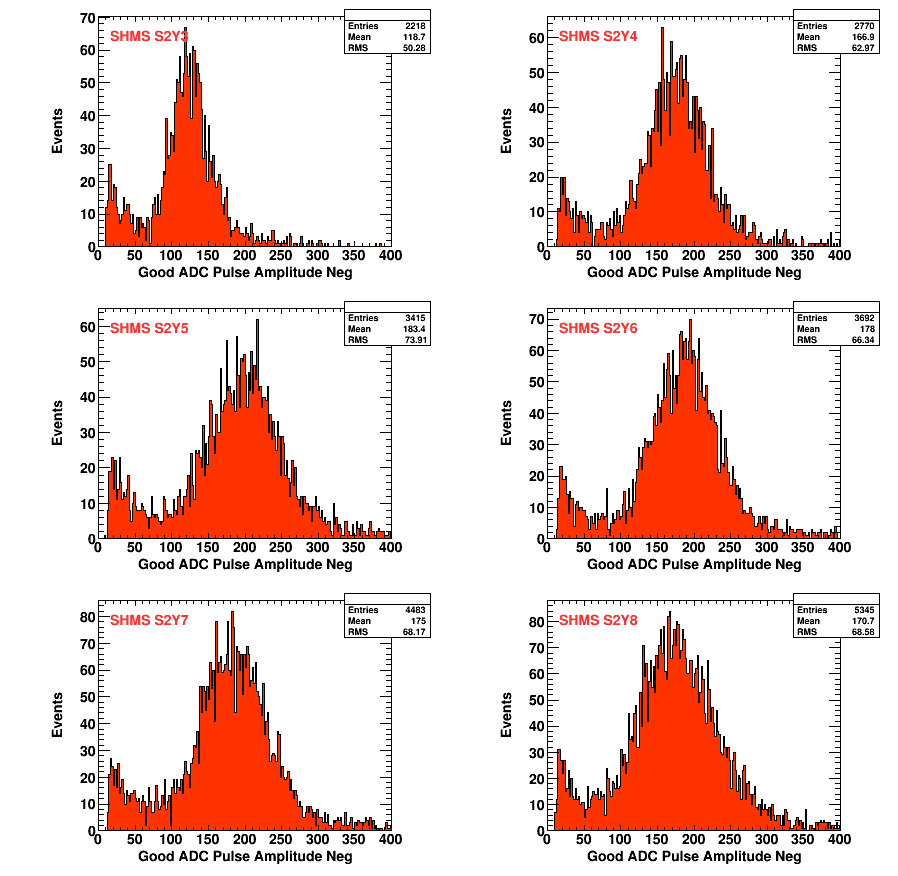}
\caption{PMT pulse amplitude from pions with momenta of 1.96~GeV/c.
\label{fig:amp_pion_lowp}}
\end{figure}

\begin{figure}[htbp]
\includegraphics[width=\linewidth]{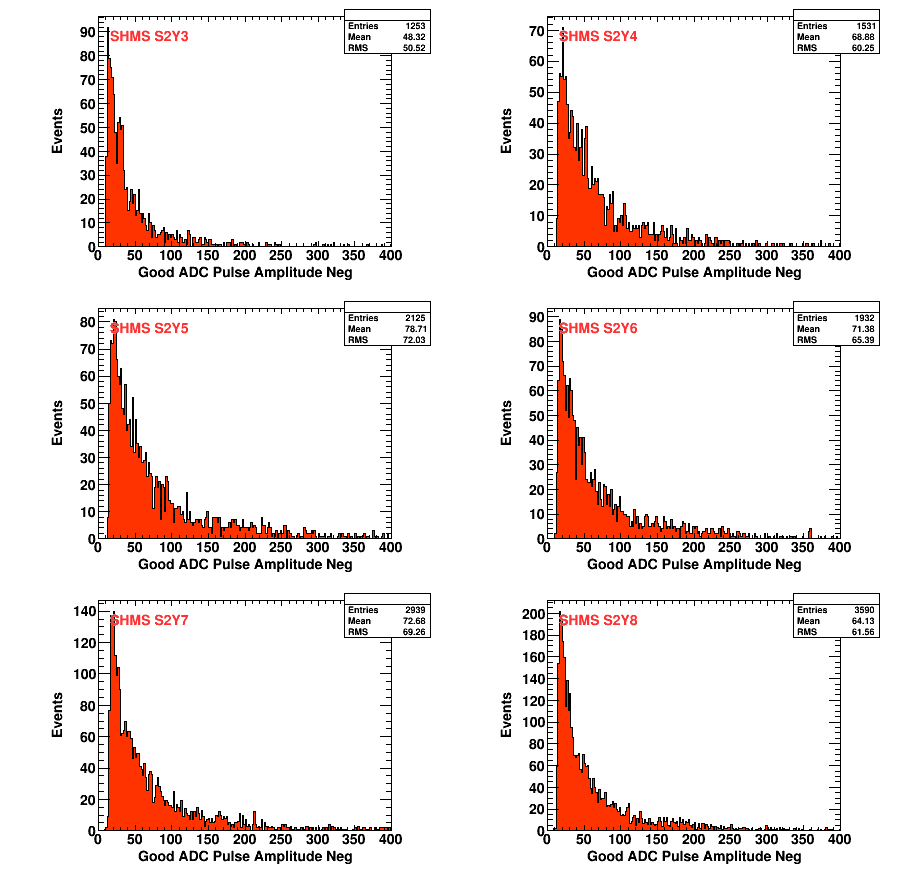}
\caption{PMT pulse amplitude from protons with momenta of 1.96~GeV/c.
\label{fig:amp_proton_lowp}}
\end{figure}

\begin{figure}[htbp]
\includegraphics[width=\linewidth]{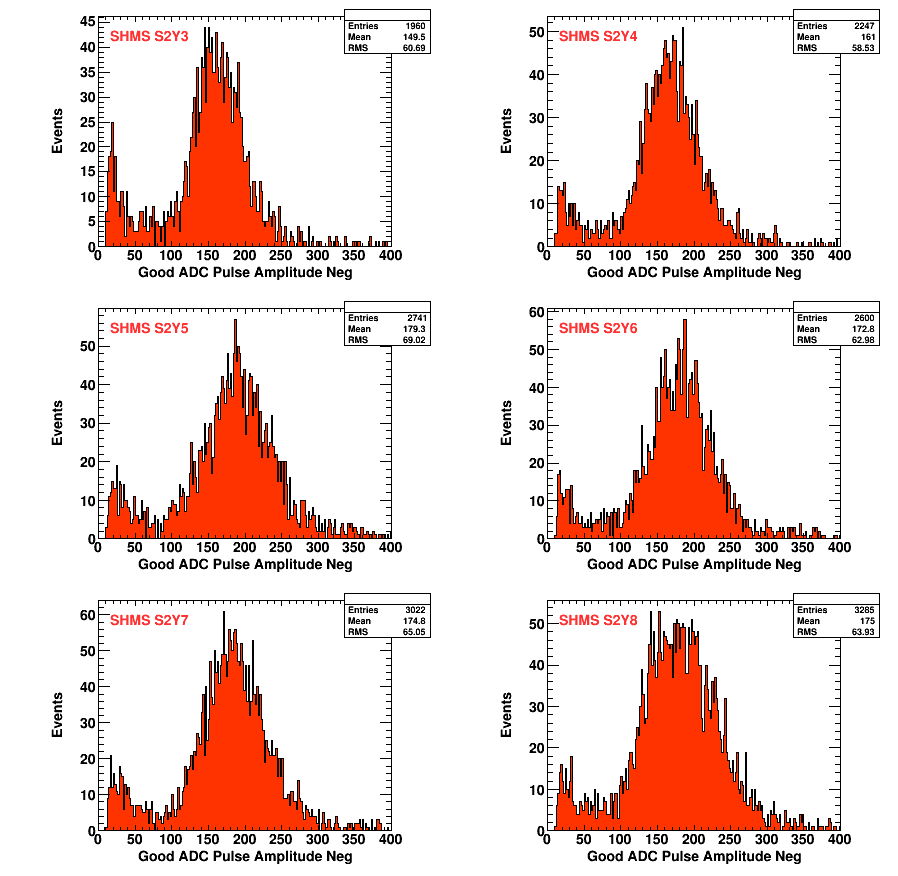}
\caption{PMT pulse amplitude from protons with momenta of 5.05~GeV/c.
\label{fig:amp_proton_highp}}
\end{figure}

Beam data confirmed the expectation that the detection efficiency for low momentum protons, for example, will be smaller than that for pions or electrons simply due to the reduced number of Cherenkov photons that particles close to their firing threshold will produce. This is exemplified by Figs.\,\ref{fig:amp_pion_lowp}, \ref{fig:amp_proton_lowp} and \ref{fig:amp_proton_highp}.

\subsubsection{Drift Chamber Performance}

\begin{figure}[htbp]
\includegraphics[width=\linewidth]{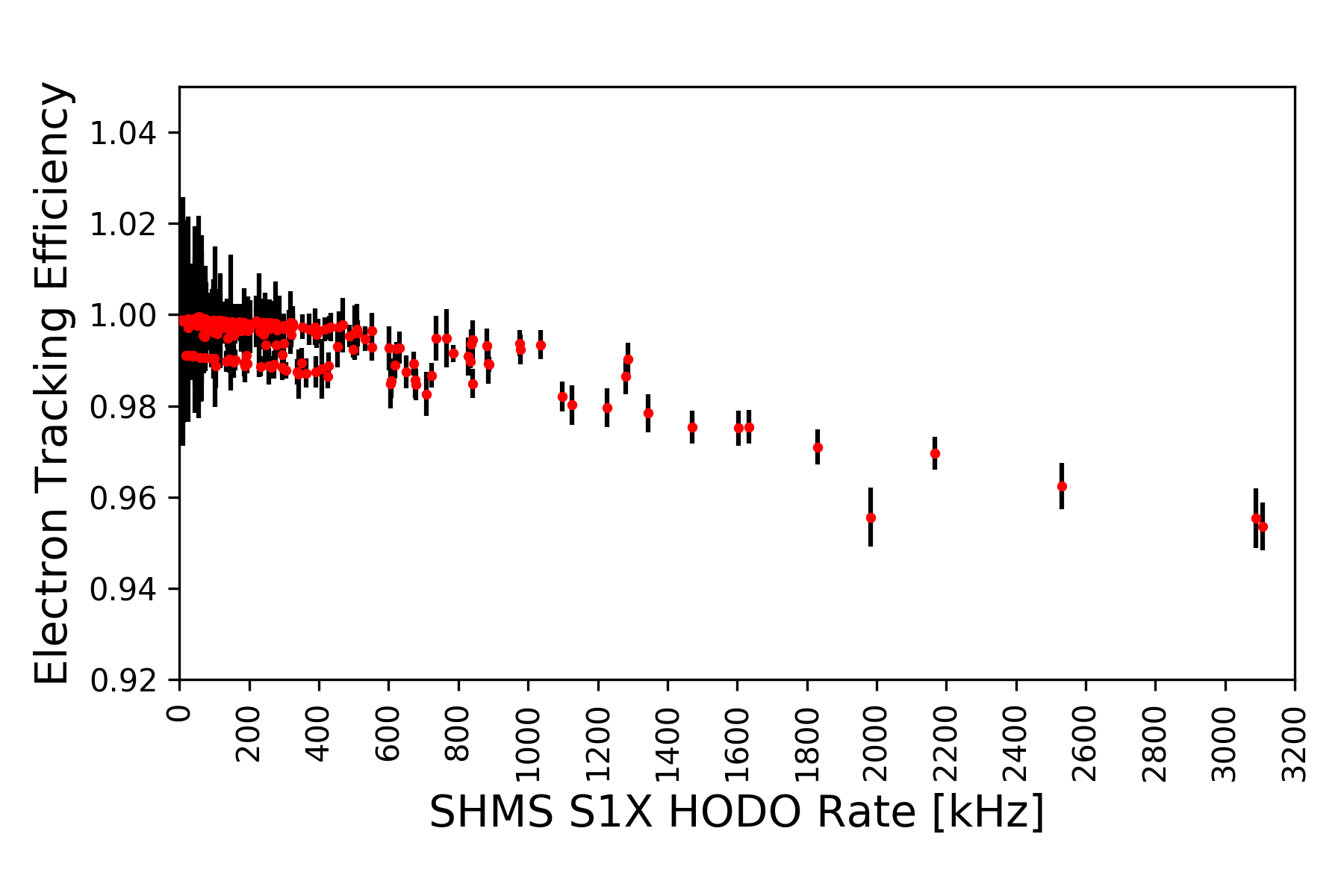}
\caption{\label{fig:SHMS_DC_TrackEff_Elec} The SHMS electron tracking efficiency as a function of the S1X hodoscope trigger rate. 
}
\end{figure}

\begin{figure}[htbp]
\includegraphics[width=\linewidth]{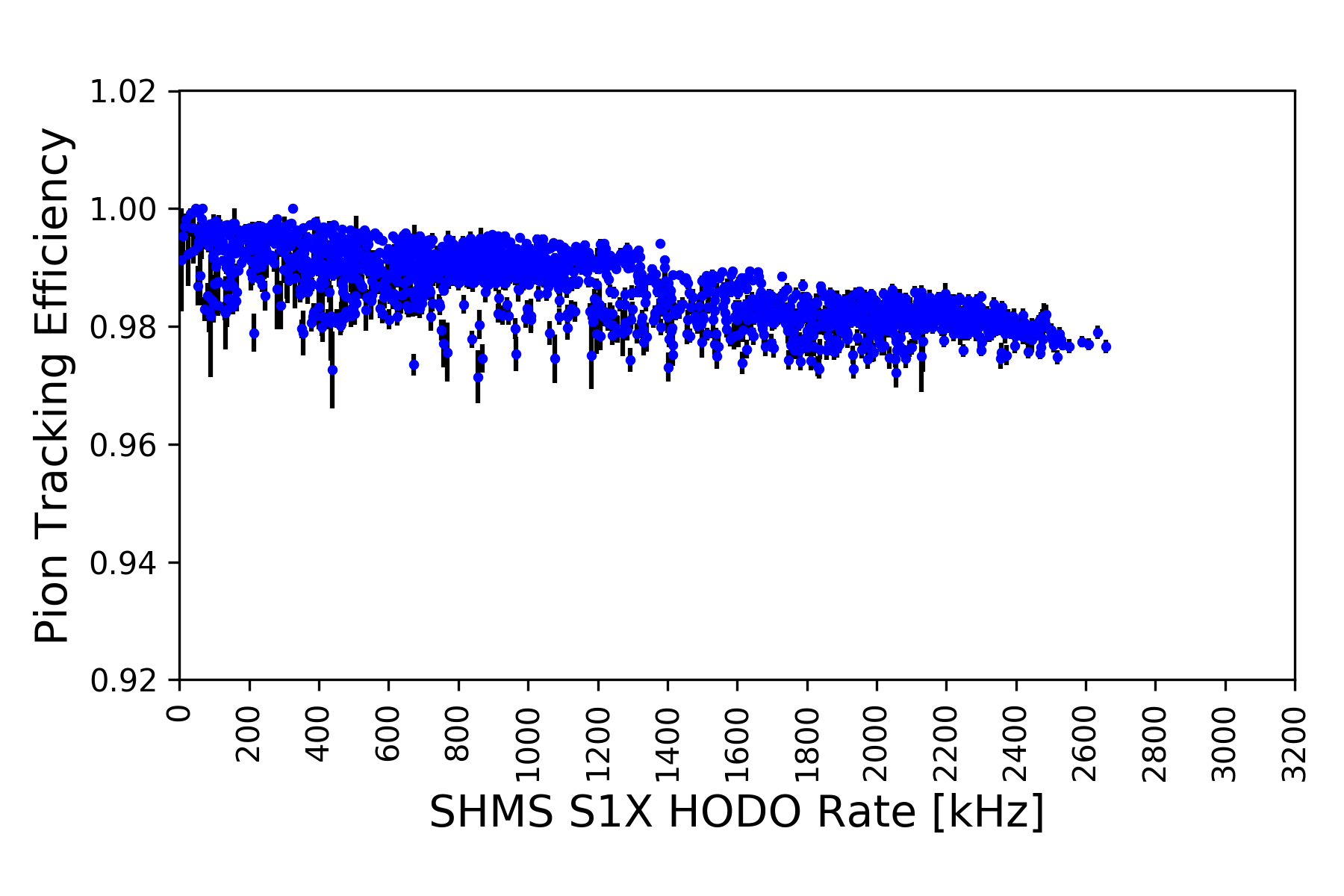}
\caption{\label{fig:SHMS_DC_TrackEff_Pion} The SHMS pion tracking efficiency as a function of the S1X hodoscope trigger rate. 
}
\end{figure}

The SHMS drift chambers have proven to provide reliable tracking for electrons and hadrons across a broad range of momenta. The position resolution depends on how well the wire planes are aligned in the reconstruction software, but per-plane resolutions of 175 $\mu$m are typical. The drift chambers have also performed very well at high rate, with tracking efficiencies exceeding 96\%, even at pre-trigger rates over 2\,MHz. The tracking efficiency as a function of the S1X hodoscope trigger rate (a good proxy for the overall event rate) in the SHMS can be seen in Figs.\,\ref{fig:SHMS_DC_TrackEff_Elec} and \ref{fig:SHMS_DC_TrackEff_Pion}.


\subsubsection{HGC Performance}
\label{Subsec:HGC_Perf}

The performance of the HGC is determined by its capacity to separate particle species based on the detected number of photoelectrons (NPE). In particular, the HGC is a threshold Cherenkov detector and thus identifies particles based on whether a signal is greater than 1.5 NPE. The metrics of performance to be discussed are the detector efficiency and contamination.

Efficiency in this context refers to the ratio of events of a particular particle species selected by all detectors, divided by the number of events selected as that same species without any information from the HGC. For example,  
\begin{equation}
  \eta_\text{HGC} = \frac{\text{$\pi^+$ detected with HGC signal}}{\text{$\pi^+$ detected without HGC signal}},
\end{equation}
where $\eta_\text{HGC}$ represents the efficiency of the HGC for detecting a $\pi^+$. The selection criteria include a single reconstructed track per event, as well as cuts on potential PID information from  timing, reconstructed $\beta$, the calorimeter, aerogel and HGC.   

Contamination refers to the fraction of events identified by non-HGC detectors which should be sub-threshold in the HGC, but which nevertheless yield more than 1.5 NPE in the HGC. For example, if the HGC is configured for $\pi^+$/$K^+$ separation, and the non-HGC detectors have identified a sample of clean $K^+$, then the contamination is defined as the fraction of that clean sample where the HGC saw a light level consistent with a $\pi^+$.

\begin{figure}[htbp]
  \centering
  \includegraphics[width=\linewidth]{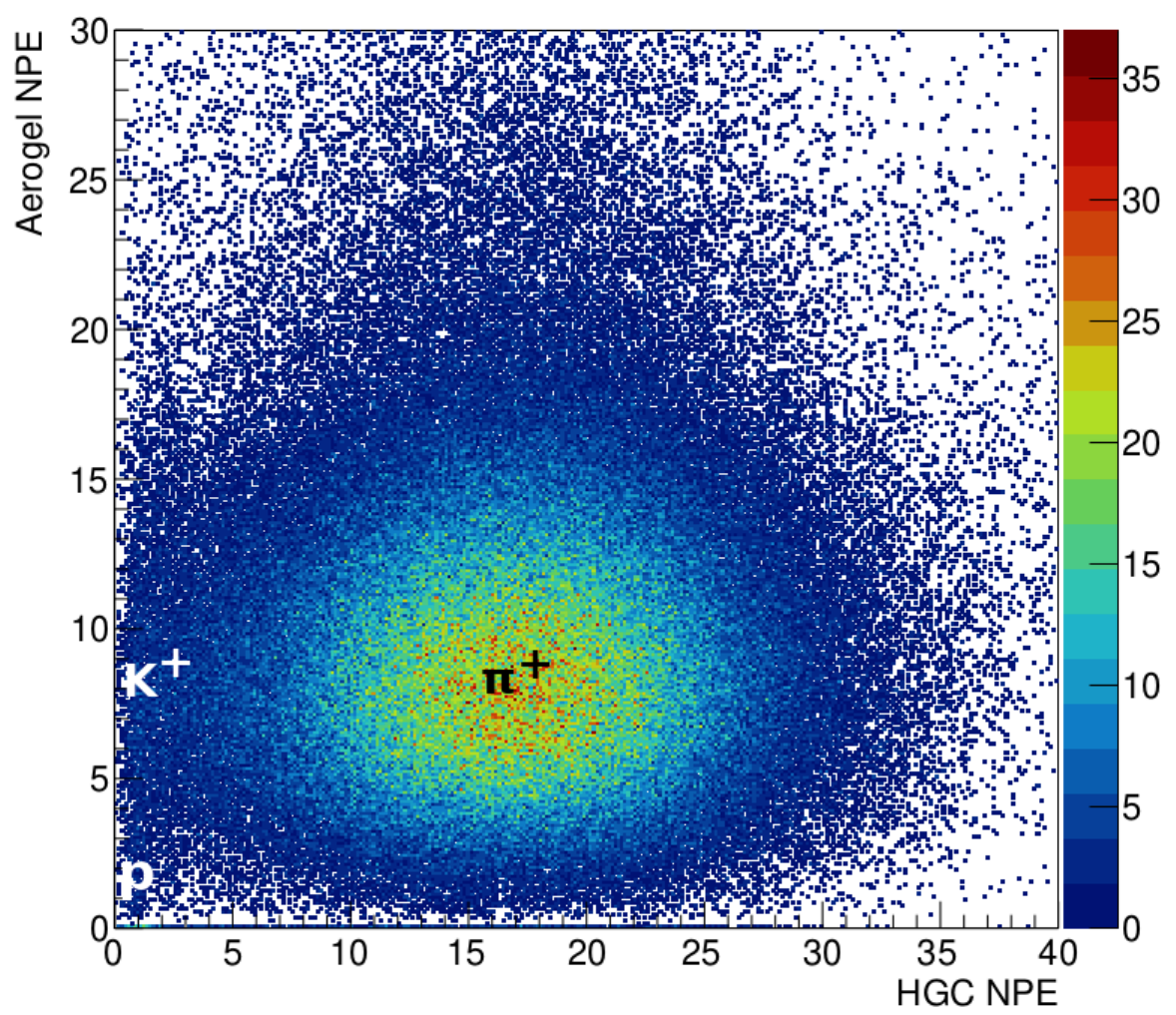}
  \caption[HGC Particle Identification]{Demonstration of the particle identification capability of the Heavy Gas Cherenkov. Pictured is the separation between $\pi^+$, $K^+$ and proton at the 8.186\,GeV beam energy and 6.053\,GeV/c SHMS central momentum. The refractive indexes of HGC and aerogel Cherenkov detectors are 1.00143 and 1.011, respectively.}
  \label{fig:piK}
\end{figure}

\begin{table}[b!] 
  \begin{center}
    \begin{tabular}{|c|c|c|} \hline
      PID Configuration & Efficiency & Contamination \\\hline
      $e^-$/$\pi^-$ & $95.99$\% & $0.01\%$\\\hline
      $\pi^+$/$K^+$ & $98.22$\% & $0.1\%$\\\hline
    \end{tabular} 
    \caption[Heavy Gas Cherenkov Particle Identification]{Summary of the Heavy Gas Cherenkov performance in separating between particle species. The efficiency is quoted for the lighter, above-threshold particle. The contamination is quoted for the heavier, below-threshold particle. Both are based on a photoelectron yield cut greater than 1.5.}
      \label{table:HGCPID}
  \end{center}
\end{table}

Two runs were chosen to show HGC efficiency and contamination, one where the HGC discriminated between $e^-$/$\pi^-$ and the other $\pi^+$/$K^+$. The  $e^-$/$\pi^-$  run featured the HGC filled with CO$_2$ at 1 atm and an SHMS central momentum of -3.0\,GeV/c. Particle identification was established by a cut on the normalized calorimeter energy. 
The $\pi^+$/$K^+$ run had the HGC filled with C$_4$F$_{10}$ at 1\,atm, giving a $\pi$ momentum threshold of 2.8\,GeV/c and a $K$ momentum threshold of 9.4\,GeV/c.
Particle identification was performed by a cut on the aerogel Cherenkov detector and the normalized calorimeter energy. The spectrum obtained for the $\pi^+$/$K^+$ separation is shown in Fig.\,\ref{fig:piK}. This figure illustrates the broad distribution of NPE produced by $\pi^+$
above their momentum threshold. At the lower end of the NPE axis, there is a large number of events producing no light, or just the SPE. These events correspond to $K^+$, since they are below the momentum threshold to produce Cherenkov light. The presence of the SPE is likely due to $\delta$-rays (i.e., knock-on $e^-$).

\begin{figure}[htbp]
\includegraphics[width=1\linewidth]{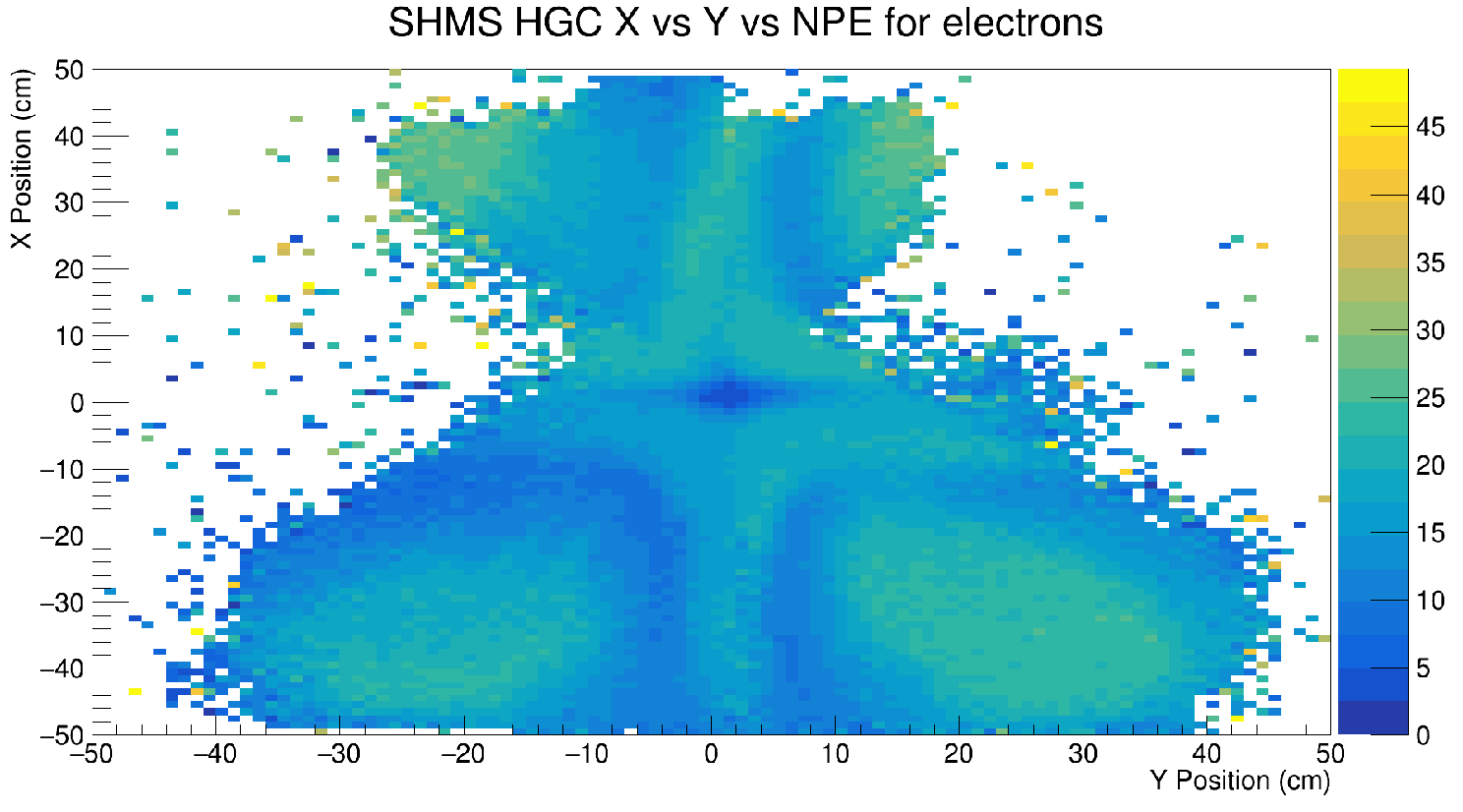}
\caption{\label{fig:SHMS_HGC_Central_Inefficiency} Number of photoelectrons (NPE) as a function of X/Y position at the centre of the HGC for electrons in the SHMS (selected using a high purity cut of $E_{TotNorm} > 0.9$ in the SHMS calorimeter)\cite{TrottaThesis}. A drop in NPE yield can clearly be observed towards the centre of the HGC.
}
\end{figure}

As discussed in Sec\,\ref{subsec:HGC_Design}, there is a region of lower efficiency in the center of the HGC due to the overlap and alignment of the mirrors. This region can be observed in experimental operation, as shown in Fig.\,\ref{fig:SHMS_HGC_Central_Inefficiency} near (0,0) in the focal plane coordinate system. 
A summary of the particle identification efficiency and contamination performance is given in Table\,\ref{table:HGCPID}.

Lastly, measurements of the $\pi$ efficiency across a variety of momentum settings can be used to verify the index of refraction of the Cherenkov media. The relationship between $\pi$ efficiency and momentum is fit with the equation \cite{BrashAerogel}
\begin{equation}
  \label{eq:index}
  \eta_{HGC}=1-e^{-(p-p_o)/\Gamma},
\end{equation}
where $\eta_{HGC}$ is the detector efficiency, $p$ is the momentum of the $\pi$, and $p_o$ and $\Gamma$ are free parameters. Data taken in the range of 2.53\,GeV/c to 5.05\,GeV/c with the HGC filled with C$_4$F$_{10}$ yields an index of refraction of $n=1.001\pm0.002$. This is in agreement with the accepted value of $n=1.00143$ \cite{C4F10Index}. Additional performance details are given in \cite{RyanThesis}.


\subsubsection{Aerogel Performance}
\label{shms_aero_perform}

\begin{figure}[htb!]
    \includegraphics[width=\linewidth]{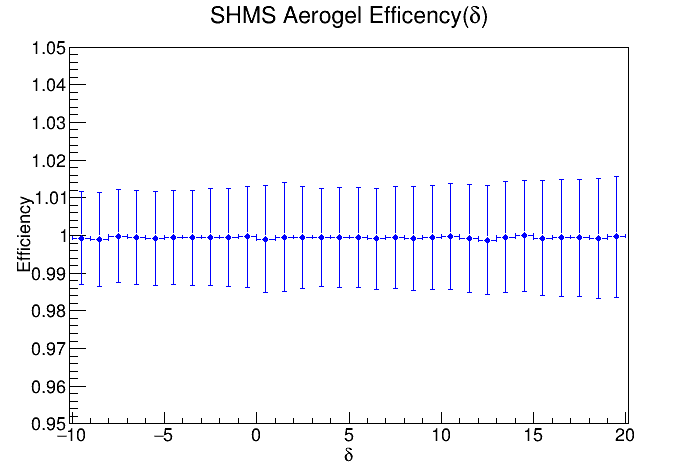}
    \caption{\label{fig:aero_eff_plot}
    The $K^+$ efficiency of the aerogel is plotted over a range of $\delta$. This efficiency is taken at a beam energy of 6.2\,GeV for an SHMS central momentum of 3.486\,GeV/c. The refractive index of the aerogel detector is 1.015.
    }
\end{figure}
\begin{table}[htb!] 
  \begin{center}
    \begin{tabular}{|c|c|c|} \hline
      PID Configuration & Efficiency & Contamination \\\hline
      $K^+$/$p$ & $99.94$\% & $0.1\%$\\\hline
    \end{tabular} 
    \caption[Aerogel Particle Identification]{Aerogel performance for kaon-proton separation when requiring greater than 1.5 photoelectrons. The efficiency and contamination are defined as in Table \ref{table:HGCPID}.}
    \label{table:AeroPID}
  \end{center}
\end{table}

The primary use of the aerogel Cherenkov detector in the SHMS is to distinguish between kaons and protons. A variety of aerogel tile refractive indices are used to cover a range of momenta. A cut of greater than 1.5 photoelectrons is used. Fig.\,\ref{fig:piK} shows the NPE yield of the Heavy Gas Cherenkov as well as the aerogel Cherenkov detector. This figure shows the importance of having both the Heavy Gas and the aerogel Cherenkov detectors as the kaon and proton would be indistinguishable without the aerogel. 

In order to get clean kaon samples, a high detector efficiency in the aerogel is required. The aerogel efficiency is determined by
\begin{equation}
  \label{eq:aero_eff}
  \eta_{\text{aero}}=\frac{\text{$K^+$ detected with aerogel signal}}{\text{$K^+$ detected without aerogel signal}},
\end{equation}
where the detector efficiency is represented by $\eta_{\text{aero}}$. The efficiency of the aerogel detector can be seen in Table\,\ref{table:AeroPID}. It is clear that the aerogel has a very high efficiency, crucially though, this efficiency also runs over the full range of $\delta$ as seen in Fig.\,\ref{fig:aero_eff_plot}. This, plus the ability to change refractive indices, allows for high purity kaon identification over a wide range of kinematics.

\subsubsection{Calorimeter Performance}
\label{shms_calo_perform}

\begin{figure}[hbtp]
\begin{centering}
\includegraphics[width=\linewidth]{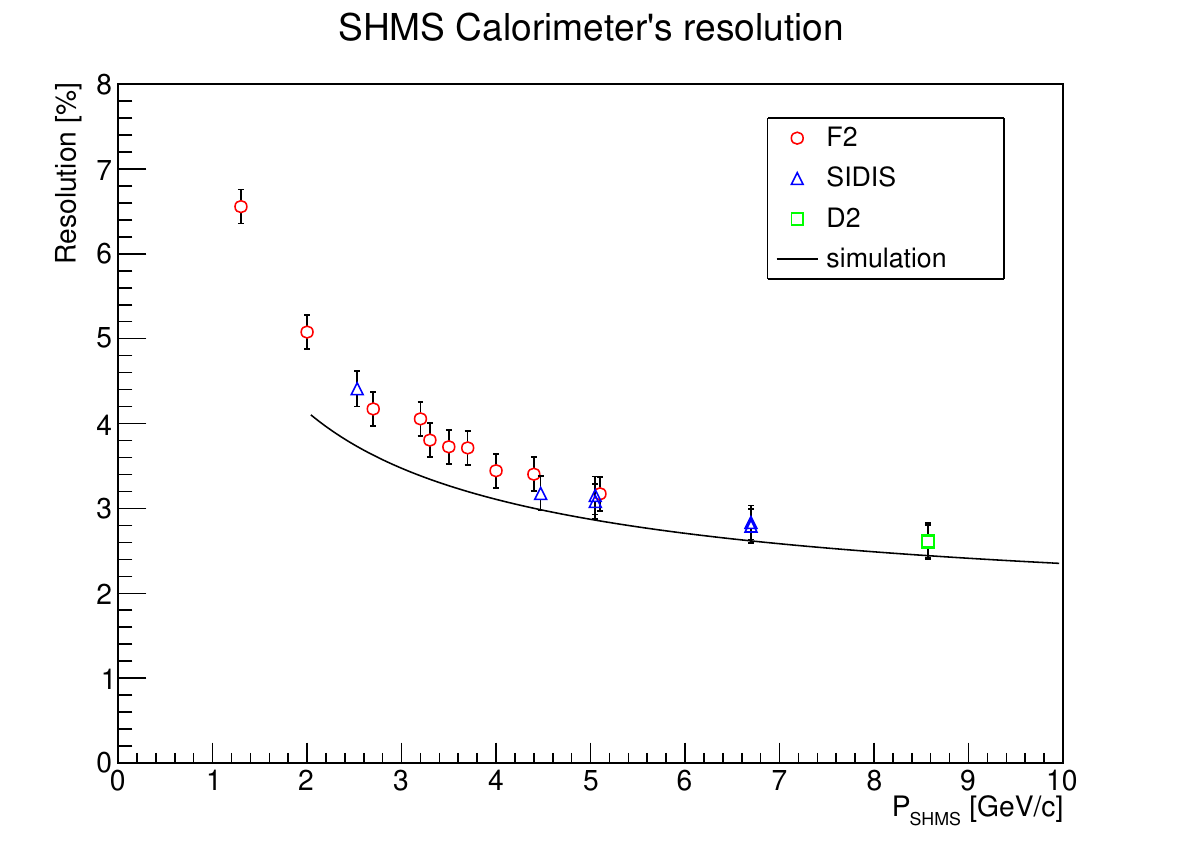}
\caption{Resolution of the SHMS calorimeter from calibrations of runs from the Spring 2018 run period. The solid line is a result from  early simulations.}
\label{fig:shms_calo_resolution}
\end{centering}
\end{figure}

The performance of the SHMS calorimeter under beam conditions was first tested during the 12\,GeV Hall C Key Performance Parameter Run in spring of 2017. As part of the SHMS detector package, the calorimeter was commissioned in the Hall C fall run period of the same year. 
As discussed briefly in Sec.\,\ref{calo_calibr}, $E_{Norm}$ should be $\sim$1 for electrons. This quantity can be utilised for PID selection. In the few GeV/c range, pions and electrons are well separated. The early analyses of the calorimeter data also demonstrate satisfactory performance of the detector in terms of resolution, as demonstrated in Fig.\,\ref{fig:shms_calo_resolution}.

\section{Conclusion}
\label{sec:conclude}

The SHMS has been in service since 2017. Through a range of experiments that utilised a wide variety of running conditions, the SHMS has demonstrated itself to be a reliable and stable spectrometer, both in terms of its ion optics and its detector package.

Numerous experiments have completed and published high profile results since the SHMS was commissioned in 2017. This includes many high profile results on color transparency \cite{jr:DBhetuwal_PRL2020, jr:DBhetuwal_PRC2023}, the EMC effect \cite{jr:AKarki_PRC2023}, deuteron structure \cite{jr:CYero_PRL2020} and proton structure \cite{jr:RLi_Nature2022}. This also extends to detailed studies of the proton's gravitational form factors \cite{jr:BDuran_Nature2023}.
Results on the F2 ratio for D/H\cite{biswas2024newmeasurementsdeuteronproton} and the flavor dependence in charged pion SIDIS\cite{bhatt2024flavordependencechargedpion} were recently submitted for publication.  

Many more high profile scientific results are expected in the near future, with several experimental campaigns now completed and data analysis in advanced stages. Due to the design parameters of the SHMS, it could also be utilised extensively in an upgraded, 20+~GeV Jefferson Lab scenario. Some possible experiments that utilise the SHMS in a 20+~GeV era are outlined in the 22~GeV white paper \cite{PrePrint:JLab22GeV}.

\section*{Acknowledgments}
This material is based upon work supported by the U.S. Department of Energy, Office of Science, Office of Nuclear Physics under contract DE-AC05-06OR23177. This work is supported by the Natural Sciences and Engineering Research Council of Canada (NSERC) SAPIN-2021-00026 and an award from the SAP-RTI program. This work was supported in part by the United States National Science Foundation grants PHY1914034, PHY1039446, PHY2309976, PHY2012430 and the Consortium MRI, PHY0723062.

\section*{References}
\label{}



\bibliographystyle{elsarticle-num} 
\bibliography{bib_items}







\end{document}